\documentclass[aps,groupedaddress,nofootinbib,floatfix,epfs,preprintnumbers]{revtex4}
\usepackage{amsfonts,amscd,amsmath,amssymb,graphicx,color,float,xcolor}
\usepackage[section]{placeins}
\DeclareMathOperator{\arccot}{arccot}
\def\ep{\text{e}}
\def\g{\mathsf{g}}
\def\oh{\frac{1}{2}}
\def\s{\mathsf{s}}
\def\k{\mathsf{k}}
\def\rv{r_v}

\def\r0{r_0}
\def\xo{x_{\text{\tiny 0}}}
\def\ro{r_{\text{\tiny 0}}}
\def\vo{{\text v}_{\text{\tiny 1}}}
\def\vop{{\text v}_{\text{\tiny 1}}^{\text{\tiny +}}}
\def\vom{{\text v}_{\text{\tiny 1}}^{\text{\small -}}}
\def\vn{{\text v}_{\text{\tiny 0}}}
\def\vin{{\text v}_{\text{\tiny 1}}}
\def\vz{{\text v}_{\text{\tiny 0}}^{\text{\tiny 1}}}
\def\vzi{{\text v}_{\text{\tiny 0}}^{\text{\tiny 3}}}
\def\vi{{\text v}_{\text{\tiny 0}}^{\text{\footnotesize i}}}
\def\vzf{{\text v}_{\text{\tiny 01}}^{\text{\tiny 1}}}
\def\vzs{{\text v}_{\text{\tiny 02}}^{\text{\tiny 1}}}
\def\vzt{{\text v}_{\text{\tiny 03}}^{\text{\tiny 1}}}
\def\QQb{\text{\tiny Q}\bar{\text{\tiny Q}}}
\def\QQ{\text{\tiny QQ}}
\def\3Q{\text{\tiny 3Q}}
\def\i{\text{\tiny I}}
\def\ii{\text{\tiny II}}
\def\iii{\text{\tiny III}}

\def\2Q{\text{\tiny QQ}}
\begin{document}
\preprint{LMU-ASC 11/25}
\title{Some Aspects of Three-Quark Potentials (Part II)}
\author{Oleg Andreev}
 \thanks{Also on leave from L.D. Landau Institute for Theoretical Physics}
\affiliation{Arnold Sommerfeld Center for Theoretical Physics, LMU-M\"unchen, Theresienstrasse 37, 80333 M\"unchen, Germany}
\begin{abstract} 
 We continue our investigation of the effective string model for the triply heavy quark system, mimicking that in pure $SU(3)$ gauge theory. We present analytical and numerical studies of the three-quark potential for isosceles and collinear geometries. In the general case, we derive the asymptotic expression of the potential in the infrared limit. Here we also demonstrate the universality of the string tension and interpret the transition between two distinct regimes, occurring when one of the triangle's angles formed by the quarks is equal to $\frac{2}{3}\pi$, as a breaking of permutational symmetry. This symmetry breaking implies the emergence of a heavy quark dressed by gluons, transforming in the two-index antisymmetric representation. Additionally, we discuss various aspects of the $Y$- and $\Delta$-laws, diquarks, and connections to lattice QCD. 
  \end{abstract}
\maketitle
\vspace{0.1cm}
\section{Introduction}
\renewcommand{\theequation}{1.\arabic{equation}}
\setcounter{equation}{0}

 The triply heavy baryons are the last missing members of the lowest-lying baryon multiplets in the quark model. Despite numerous efforts to discover them, they remain a challenge for both experiment and theory \cite{bj}. On the theory side, the potential models have proven to be powerful tools for gaining insight into the structure and properties of charmonium and bottomonium, and it is therefore expected that they would also be useful for studying the baryons.
 
 This is the second of two companion papers in which we continue to develop the string based description of the triply heavy baryons and compute the three-quark potential, inspired by pure $SU(3)$ gauge theory, using the gauge/string duality \cite{a3Q2016}. The motivation for this is as follows. The two most common phenomenological ans\"atze for the potential are:

(1) The $\Delta$-law \cite{cornwall}, which asserts that   

\begin{equation}\label{Delta}
	E_{\3Q}=\oh\sum_{i<j}^3E_{\QQb}(\vert\mathbf{x}_{ij}\vert)
	\,,
\end{equation}
where $E_{\QQb}$ is the heavy quark-antiquark potential, $\mathbf{x}_i$ is the position of the $i$-quark, and $\mathbf{x}_{ij}=\mathbf{x}_i-\mathbf{x}_j$. The Cornell potential is commonly used for $E_{\QQb}$ \cite{cornell}: $E_{\QQb}(\vert\mathbf{x}_{ij}\vert)=-\frac{\alpha_{\QQb}}{\vert\mathbf{x}_{ij}\vert}+\sigma\vert\mathbf{x}_{ij}\vert+2c$, where $\sigma$ is the string tension and $c$ is a normalization constant. This approach effectively replaces a genuine three-body interaction with a sum of pairwise interactions.

(2) In contrast, the $Y$-law incorporates a true three-quark interaction and states that

\begin{equation}\label{YC}
	E_{\3Q}=-\sum_{i<j}^3\frac{\alpha_{\QQ}}{\vert\mathbf{x}_{ij}\vert}\,\,
	+\sigma L_{\text{min}}+3c
	\,,
\end{equation}
where $\alpha_{\QQ}$ is the quark-quark coupling and $L_{\text{min}}$ is the minimal total length of the string network connecting the quarks with a junction at the Fermat point of the triangle formed by the quarks. Since the triply heavy baryons have not yet been observed experimentally, predictions based on either the $\Delta$-law or the $Y$-law remain untested. In this situation, lattice gauge theory continues to be a fundamental tool for testing phenomenological ans\"atze. The baryonic Wilson loops have been studied on the lattice \cite{bali}. It was found that in the infrared (IR) regime, the lattice data fit the $Y$-law rather than the $\Delta$-law \cite{suga,pdf}. However, more recent investigations have revealed an issue even with the $Y$-law \cite{komas}. The lattice data for the subtracted potential $E_{\3Q}-\sigma L_{\text{min}}-3c$ point out that, in contrary to \eqref{YC}, it doesn't vanish in the IR limit. Thus, both ans\"atze are rather oversimplified. Because the potential plays a pivotal role in predicting the baryon properties, there is a strong need for refinement of the phenomenological ans\"atze.

The paper is organized as follows. In Sec.II, we set the framework and recall some preliminary results. Then, in Sec.III we construct and analyze the three-quark potential for quarks located at the vertices of an isosceles triangle. We continue in Sec.IV with a collinear geometry, where the quarks lie on a straight line. Here we also discuss diquarks. In Sec.V, we consider the IR limit for a general quark configuration and explores the asymptotic behavior of the potential. We conclude in Sec.VI with a discussion of the implications of our findings and some open problems. Appendix A introduces our notation and lists some useful formulas. To make the paper more self-contained, we include technical details and derivations in Appendices B and C. Appendices D, E, and F provide supplementary material related to Sec.III-V.

\section{Preliminaries}
\renewcommand{\theequation}{2.\arabic{equation}}
\setcounter{equation}{0}

In our study of the three-quark system, we use the formalism developed in \cite{a3Q2016}. This formalism is general and can be adapted to any model of AdS/QCD. However, for illustration, we perform calculations within one of the simplest models, the so called soft wall metric model. In this model, the background five-dimensional geometry is chosen to be a one-parameter deformation of Euclidean $\text{AdS}_5$ space of radius $R$:

\begin{equation}\label{metric}
ds^2=\ep^{\s r^2}\frac{R^2}{r^2}\Bigl(dt^2+(dx^i)^2+dr^2\Bigr)
\,.
\end{equation}
Here $r$ denotes the fifth (radial) coordinate, $\s$ is a deformation parameter, and $i=1,2,3$. The boundary lies at $r=0$, while the soft wall at $r=1/\sqrt{\s}$. The wall effectively prevents strings from penetrating deep into the bulk. 

Our analysis relies on two key components. The first is a fundamental string governed by the Nambu-Goto action 

\begin{equation}\label{NG}
S_{\text{\tiny NG}}=\frac{1}{2\pi\alpha'}\int d^2\xi\,\sqrt{\gamma^{(2)}}
\,,
\end{equation}
where $\gamma$ is an induced metric, $\alpha'$ is a string parameter, and $\xi^i$ are world-sheet coordinates. The second component is a high-dimensional analogue of the string junction, commonly referred to as the baryon vertex.\footnote{We use this terminology, to distinguish between the four-dimensional string junction and its five-dimensional counterpart.} In the context of AdS/CFT, the vertex is a five brane wrapped on an internal space $\mathbf{X}$ \cite{witten}. From the five-dimensional viewpoint, this object looks point-like. As shown in \cite{a3Q2016}, the action for the baryon vertex written in the static gauge

\begin{equation}\label{baryon-v}
S_{\text{vert}}=\tau_v\int dt \,\frac{\ep^{-2\s r^2}}{r}
\end{equation}
yields the results in very good agreement with lattice QCD calculations of the three-quark potential. Actually, $S_{\text{vert}}$ represents the worldvolume of the brane, assuming $\tau_v={\cal T}_5R\,\text{vol}(\mathbf{X})$ with ${\cal T}_5$ the brane tension.  Unlike AdS/CFT, we treat $\tau_v$ as a free parameter to account for $\alpha'$-corrections as well as the possible effects of other background fields.\footnote{In analogy with AdS/CFT, we expect the presence of Ramond-Ramond background fields on $\mathbf{X}$.}

\begin{figure}[htbp]
\centering
\includegraphics[width=6.5cm]{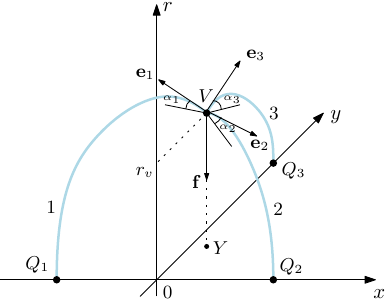}
\caption{{\small A typical baryonic configuration in five dimensions. The heavy quarks are denoted by $Q_i$, the baryon vertex by $V$, its projection onto the boundary by $Y$, and the tangent angles of the strings by $\alpha_i$ ($i=1,\dots,3$). The arrows indicate the forces acting on the vertex, which is located in the bulk at $r=r_v$.}}
\label{barconf}
\end{figure}

A baryonic configuration in five dimensions is constructed from strings in the following way. Heavy quarks are placed at boundary points, with each string ending on a quark. The strings meet at a baryon vertex located in the bulk, as shown in Figure \ref{barconf}. The total action is thus the sum of the Nambu-Goto actions plus the action of the vertex. Explicitly,

\begin{equation}\label{action5d}
S=\sum_{i=1}^3 S_{\text{\tiny NG}}^{(i)}\,+S_{\text{vert}}
\,.	
\end{equation}
For a static configuration, the net force at the vertex must vanish, which leads to the force balance equation 

\begin{equation}\label{fbe}
\mathbf{e}_1+\mathbf{e}_2+\mathbf{e}_3+\mathbf{f}=0
\,,
\end{equation}
where $\mathbf{e}_i$ are string tensions, and $\mathbf{f}$ is a gravitational force acting on the vertex. The explicit formulas for the tensions can be determined using the equation \eqref{e}, and the gravitational force is given by 

 \begin{equation}\label{f}
	\mathbf{f}=\Bigl(0,\,0,\,-\tau_v\partial_{r_v}\,\frac{\ep^{-2\s r_v^2}}{r_v}\Bigr)
	\,.
\end{equation}
Technically, the force balance equation can be derived by extremizing the total action with respect to the position of the baryon vertex.

Although this effective string model\footnote{See also \cite{astbr3Q,T-mu} for related developments.} is not exactly dual to pure $SU(3)$ gauge theory, we pursue it for several compelling reasons: (i)  It offers insights into problems lacking predictions from phenomenology or lattice QCD but tractable within our model framework. (ii) The model's predictions for the quark-antiquark \cite{az1} and three-quark \cite{a3Q2016} potentials show near perfect agreement with lattice results and QCD phenomenology \cite{white,a3Q2016}. (iii) The model permits the derivation of explicit analytic expressions, enhancing its utility and transparency.

\section{Isosceles Triangle Geometry }
\renewcommand{\theequation}{3.\arabic{equation}}
\setcounter{equation}{0}

\begin{figure}[htbp]
\centering
\includegraphics[width=6.5cm]{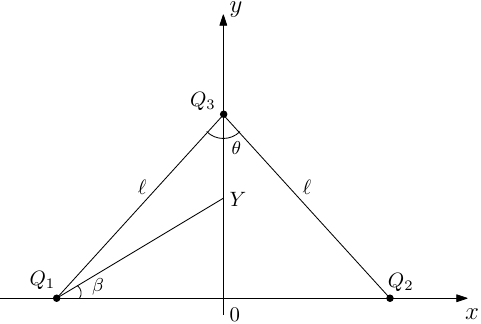}
\caption{{\small An isosceles triangle with leg length $\ell$ and apex angle $\theta$. Here $\beta=\angle{YQ_1Q_2}$.}}
\label{isot}
\end{figure}

We begin with the case in which the quarks are at the vertices of an isosceles triangle, as shown in Figure \ref{isot}.

For future reference, we list the following simple but useful formulas:

\begin{equation}\label{iso-geo}
\ell=\frac{\cos\beta}{\sin\frac{\theta}{2}}\,\vert Q_1Y\vert
\,,\qquad
\cos\bigl(\beta+\tfrac{\theta}{2}\bigr)\vert Q_1Y\vert=
\sin\tfrac{\theta}{2}\vert Q_3Y\vert
\,,
\end{equation}
which are valid for $0\leq\beta\leq (\pi-\theta)/2$.

Because of axial symmetry, $\alpha_1=\alpha_2$, and the point $Y$ lies on the $y$-axis. In this case, the expressions for the string tension vectors take the form  $\mathbf{e}_1=-\g w(\rv)(\cos\beta\cos\alpha_1,\sin\beta\cos\alpha_1,\sin\alpha_1)$, $\mathbf{e}_2=\g w(\rv)(\cos\beta\cos\alpha_1,-\sin\beta\cos\alpha_1,-\sin\alpha_1)$, and $\mathbf{e}_3=\g w(\rv)(0,\cos\alpha_3,-\sin\alpha_3)$, with $w(\rv)=\frac{\ep^{\s\rv^2}}{\rv^2}$. As a result, the force balance equation \eqref{fbe} reduces to two non-trivial components, corresponding to the $y$- and $r$-directions:

\begin{equation}\label{fbe-iso}
	\cos\alpha_3=2\sin\beta\cos\alpha_1
	\,,\qquad
	2\sin\alpha_1+\sin\alpha_3=3\k(1+4v)\ep^{-3v}
	\,.
\end{equation}
Here $v=\s\rv^2$ and $\k=\frac{\tau_v}{3\g}$. It is important to note that for $\beta>(\pi-\theta)/2$, the string configuration does not exist, as the force balance equation has no solution in this case.

\subsection{The basic string configurations}

\begin{figure}[htbp]
\centering
\includegraphics[width=4.75cm]{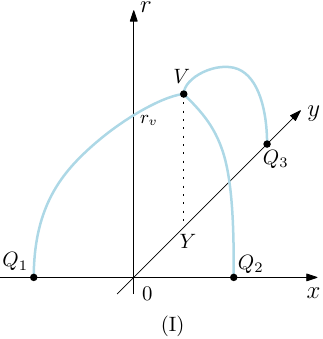}
\hspace{.95cm}
\includegraphics[width=5cm]{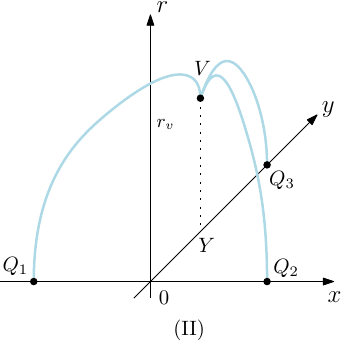}
\hspace{.95cm}
\includegraphics[width=5.80cm]{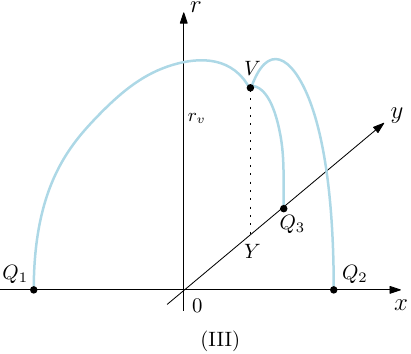}
\caption{{\small Basic string configurations associated to the different signs of the tangent angles. In (I) $\alpha_1\geq 0$ and $\alpha_3\leq 0$. In (II) all the angles are non-positive. In (III) $\alpha_1\leq 0$ and $\alpha_3\geq 0$.}}
\label{conI-III}
\end{figure}

The goal of this subsection is to briefly describe the three string configurations shown in Figure \ref{conI-III}.
These form a basic set that provides a proper framework for constructing the three-quark potential in the isosceles triangle geometry. The Figure also illustrates that, for small $\ell$, configuration (I) typically corresponds to small $\theta$, configuration (II) to $\theta\approx\pi/3$, and configuration (III) to large $\theta$. It may be somewhat surprising that there are only three basic configurations rather than four. This is due to the fact that the configuration with all non-negative tangent angles does not exist for our choice of parameters.

For the purposes of this paper, what we need to do can be summarized as follows. As explained in Appendix B, the lengths $\vert Q_iY\vert$'s and string energies can be expressed in terms of the tangent angles and parameter $v$. For a given angle $\theta$, the second equation in \eqref{iso-geo}, which we call the geometrical constraint, together with the force balance equations \eqref{fbe-iso} allows us to numerically express all the angles in terms of $v$. Since the vertex energy depends only on $v$, we can write the energy of each basic configuration in parametric form as $E=E(v)$ and $\ell=\ell(v)$, where $\ell$ is given by the first equation in \eqref{iso-geo}.
\subsubsection{Configuration I}

Let us begin with configuration (I). The geometrical constraint now takes the form 

\begin{equation}\label{geoc1}
\cos\bigl(\beta+\tfrac{\theta}{2}\bigr){\cal L}^+(\alpha_1,v)=
\sin\tfrac{\theta}{2}{\cal L}^-(\lambda_3,v)
\,,
\end{equation}
where the functions ${\cal L}^\pm$ are defined in Appendix A and $\lambda_3=-\text{ProductLog}(-v\,\ep^{-v}/\cos\alpha_3)$, as follows from Eq.\eqref{lambda}. $\ell$ can be extracted from Eq.\eqref{iso-geo} by substituting $\vert Q_1Y\vert={\cal L}^+(\alpha_1,v)/\sqrt{\s}$, and the energy is obtained from Eqs.\eqref{baryon-v}, \eqref{E+}, and \eqref{E-}. The result is\footnote{The superscript refers to a string configuration. We often omit it to reduce clutter.}  

\begin{equation}\label{El1}
\ell^{\,\i}=\frac{1}{\sqrt{\s}}\,\frac{\cos\beta}{\sin\frac{\theta}{2}}\,
{\cal L}^+(\alpha_1,v)
\,,\qquad
E^{\,\i}=\g\sqrt{\s}
\biggl(2{\cal E}^+(\alpha_1,v)+{\cal E}^-(\lambda_3,v)+3\k\frac{\ep^{-2v}}{\sqrt{v}}\,\biggr)+3c
\,.
\end{equation}
Here $\g=\frac{R^2}{2\pi\alpha'}$ and $c$ is a normalization constant. Combined with the geometrical constraint and the force balance equations, this yields the function $E^{\,\i}(\ell)$ in parametric form.\footnote{The allowed range of $v$ is discussed in subsection B and Appendix D.} 

Now consider the small-$\ell$ behavior of $E^{\,\i}$. This limit corresponds to $v\rightarrow 0$ as follows from the asymptotic behavior of the function ${\cal L}^+$. In this case, the force balance equations and the geometrical constraint reduce to

\begin{gather}
	\cos\alpha_3=2\sin\beta\cos\alpha_1
	\,,\quad
	2\sin\alpha_1+\sin\alpha_3=3\k
	\,, \label{fbe-small}\\
	\cos\bigl(\beta+\tfrac{\theta}{2}\bigr)\sqrt{\cos\alpha_3}\,I(\cos^2\hspace{-2pt}\alpha_1,\tfrac{3}{4},\tfrac{1}{2})
	=
	\sin\tfrac{\theta}{2}\sqrt{\cos\alpha_1}
	\bigl(1+I(\sin^2\hspace{-2pt}\alpha_3,\tfrac{1}{2},\tfrac{3}{4})\bigr)
\,,
\label{El1-small}
\end{gather}
where $I$ denotes the regularized Beta function. Given $\theta$, the three remaining angles can be found from these equations. Taking the limit $v\rightarrow 0$ in Eqs.\eqref{El1}, and using Eqs.\eqref{fL+smallx}, \eqref{fE+smallx} and \eqref{E-y=0}, we get

\begin{equation}\label{El1-small2}
E^{\,\i}=-\frac{\alpha^{\i}}{\ell}+3c+o(1)
\,,\quad\text{with}\quad
\alpha^{\i}=-\frac{\cos\beta}{\sin\frac{\theta}{2}}\,
{\cal L}^+_0(\alpha_1)
\bigl(2{\cal E}^+_0(\alpha_1)+{\cal E}^-_0(\alpha_3)+3\k\bigr)\g
\,.	
\end{equation}
Here the coefficients ${\cal L}_0^+$, ${\cal E}_0^+$ and ${\cal E}^-_0$ are defined in Appendix A. The notation ${\cal E}_0^-(\alpha_3)$ is used for ${\cal E}_0^-(\cos\alpha_3)$. For our parameter values considered, the factor $2{\cal E}^+_0+{\cal E}^-_0+3\k$ is negative, and thus $\alpha^{\i}$ is positive. 

It is also of interest to consider the limit in which quark $Q_3$ is sent to infinity. According to \eqref{L-y=1} this corresponds to $\lambda_3\rightarrow 1$. However, this limit is not allowed, as it violates the geometrical constraint: for fixed $\theta$, the right-hand side of Eq.\eqref{geoc1} diverges, while the left-hand side remains finite. One possible way around this issue would be to consider the limit $v\rightarrow 1$ with $\alpha_1 = \alpha_3 = 0$, in which case both sides diverge. However, $\alpha_1 = \alpha_3 = 0$ does not satisfy the second force balance equation.

\subsubsection{Configuration II}

This configuration is governed by the same action as before. As a result, the force balance equations \eqref{fbe-iso} remain valid. In fact, since $\alpha_1$ now changes the sign from positive to negative, the corresponding expressions for the geometrical constraint, leg length and energy can be obtained simply by replacing ${\cal L}^+$ and ${\cal E}^+$ with ${\cal L}^-$ and ${\cal E}^-$. Thus, we immediately arrive at 

\begin{equation}\label{geoc2}
\cos\bigl(\beta+\tfrac{\theta}{2}\bigr){\cal L}^-(\lambda_1,v)=
\sin\tfrac{\theta}{2}{\cal L}^-(\lambda_3,v)
\,,
\end{equation}
and 

\begin{equation}\label{El2}
\ell^{\,\ii}=\frac{1}{\sqrt{\s}}\,\frac{\cos\beta}{\sin\frac{\theta}{2}}\,
{\cal L}^-(\lambda_1,v)
\,,\qquad
E^{\,\ii}=\g\sqrt{\s}
\biggl(2{\cal E}^-(\lambda_1,v)+{\cal E}^-(\lambda_3,v)+3\k\frac{\ep^{-2v}}{\sqrt{v}}\,\biggr)+3c
\,.
\end{equation}
Together with \eqref{fbe-iso}, these equations provide a parametric representation of $E^{\,\ii}$ as a function of $\ell$.

It is now straightforward to analyze the small-$\ell$ behavior of $E^{\,\ii}(\ell)$. The force balance equations \eqref{fbe-small} apply in this case as well. A similar analysis to the one carried out previously shows that the geometrical constraint becomes

\begin{equation}\label{El2-small}
\cos\bigl(\beta+\tfrac{\theta}{2}\bigr)\sqrt{\cos\alpha_3}
	\bigl(1+I(\sin^2\hspace{-2pt}\alpha_1,\tfrac{1}{2},\tfrac{3}{4})
	\bigr)
	=
	\sin\tfrac{\theta}{2}\sqrt{\cos\alpha_1}
\bigl(1+I(\sin^2\hspace{-2pt}\alpha_3,\tfrac{1}{2},\tfrac{3}{4})\bigr)
\,.
\end{equation}
Given a fixed angle $\theta$, the other angles can be determined from these equations. As before, we replace ${\cal L}^+_0$ and ${\cal E}^+_0$ with ${\cal L}^-_0$ and ${\cal E}^-_0$ in Eq.\eqref{El1-small2}, yielding  

\begin{equation}\label{El2-small2}
E^{\,\ii}=-\frac{\alpha^{ \ii}}{\ell}+3c+o(1)
\,,\quad\text{with}\quad
\alpha^{\ii}=-\frac{\cos\beta}{\sin\frac{\theta}{2}}\,
{\cal L}^-_0(\alpha_1)
\bigl(2{\cal E}^-_0(\alpha_1)+{\cal E}^-_0(\alpha_3)+3\k\bigr)\g
\,.	
\end{equation}
The coefficient $\alpha^{\ii}$ is positive, by the same reasoning used for $\alpha^{\i}$ in Eq.\eqref{El1-small2}.

Now we would like to examine the large-$\ell$ behavior of $E^{\,\ii}$. As before, the basic idea is to let one or both of the $\lambda$ parameters approach $1$. There are two possible cases to consider: $\lambda_1\rightarrow 1$ at fixed $\lambda_3<1$, where only strings 1 and 2 (ending on $Q_1$ and $Q_2$) become infinitely long, and $\lambda_1,\,\lambda_3\rightarrow 1$ , where all the strings become infinitely long. 
  
  We begin with the latter case, which leads to the $Y$-law, or more precisely to the term $\sigma L_{\text{min}}$ within it, as discussed in \cite{a3Q2008}. Our goal here is to determine the leading correction to this result. As the $\lambda$'s approach $1$, it follows immediately from \eqref{v-lambda} that $\cos\alpha_1=\cos\alpha_3=v\,\ep^{1-v}$. Substituting it into the first equation in \eqref{fbe-iso} yields $\beta = \frac{\pi}{6}$. This implies that the point $Y$ coincides with the Fermat point of the triangle, as expected. It also implies that the cosine factor in \eqref{geoc2} becomes $\cos\bigl(\tfrac{\pi}{6}+\tfrac{\theta}{2}\bigr)$. Since it must be positive, the apex angle is constrained by $\theta<\frac{2}{3}\pi$.\footnote{For $\theta=\frac{2}{3}\pi$, this limit is undefined.} It is straightforward to determine the value of $v$ corresponding to $\lambda_1=\lambda_3=1$. With $\alpha_1=\alpha_3$, the second equation in \eqref{fbe-iso} gives $\sin\alpha_1=\k(1+4v)\ep^{-3v}$. Combining this with the expression for $\cos\alpha_1$, leads to the equation
  
\begin{equation}\label{v1}
1-\k^2(1+4v)^2\ep^{-6v}-v^2\ep^{2(1-v)}=0
\,,
\end{equation}
whose solution $\vo$ provides an upper bound on $v$ in the interval $[0,1]$. Notably, $\vo$ is independent of $\theta$.

To get the $Y$-law, we first use the asymptotic formulas \eqref{L-y=1} and \eqref{E-y=1}, obtaining

\begin{equation}\label{El2-large}
\ell=-\frac{\sqrt{3}}{2\sqrt{\s}\sin\frac{\theta}{2}}\,
\ln(1-\lambda_1)+O(1)
\,,\qquad
E^{\,\ii}=-\ep\g\sqrt{\s}
\bigl(2\ln(1-\lambda_1)+\ln(1-\lambda_3)\bigr)+O(1)
\,.
\end{equation}
Next, expressing $1 - \lambda_3$ in terms of $1 - \lambda_1$ via the geometric constraint,\footnote{See also Appendix F.} we get

\begin{equation}\label{El2-large2}
E^{\,\ii}=2\cos(\tfrac{\theta}{2}-\tfrac{\pi}{3})\,\sigma\ell +O(1)
\,.	
\end{equation}
This is the $Y$-law, with $2\cos(\tfrac{\theta}{2}-\tfrac{\pi}{3})\,\ell=\min(2\vert Q_1Y\vert+\vert Q_3Y\vert)$ and $\sigma=\ep\g\s$ the physical string tension \cite{az1}.

To compute the leading correction to the $Y$-law, which is a constant term in the asymptotic expansion for large $\ell$, we consider 

\begin{equation}
\begin{split}
	E^{\,\ii}-2\cos(\tfrac{\theta}{2}-\tfrac{\pi}{3})\,\sigma\ell
	=&
	\g\sqrt{\s}\Bigl(2\bigl({\cal E}^-(\lambda_1,v)-\ep {\cal L}^-(\lambda_1,v)\bigr)+{\cal E}^-(\lambda_3,v)-\ep {\cal L}^-(\lambda_3,v)+3\k\frac{\ep^{-2v}}{\sqrt{v}}\Bigr)+3c\\
	&+
	2\bigl(1-\cos(\beta-\tfrac{\pi}{6})\bigr)\sin\tfrac{\theta}{2}\cos^{-1}\hspace{-3pt}\beta\,\sigma\ell
	\,.
	\end{split}
\end{equation}
Letting $\lambda_1$ approach $1$ and using \eqref{fELy=1}, we find that 

\begin{equation}\label{El2-large3}
E^{\,\ii}=2\cos(\tfrac{\theta}{2}-\tfrac{\pi}{3})\,\sigma\ell+C^{\,\ii}+o(1)
\,,\qquad\text{with}\qquad
C^{\,\ii}=3c-3\g\sqrt{\s}\Bigl({\cal I}(\vo)-\k\frac{\ep^{-2\vo}}{\sqrt{\vo}}\Bigr)
\,,
\end{equation}
where the function ${\cal I}$ is  defined in Appendix A. In the last step, we used the fact that the factor $1-\cos(\beta-\tfrac{\pi}{6})$ behaves like a power law in $1-\lambda_1$, whereas $\ell$ like a logarithm. This result extends the result of \cite{a3Q2016} from the equilateral triangle geometry to the isosceles one. The main conclusions then are that the leading correction is independent of $\theta$ and  includes equal contributions from all the strings, enhanced by the contribution from the baryon vertex. Thus, in this limit, all the strings enter on equal footing.

To study the other limit, we proceed as follows. The tangent angle $\alpha_1$ is still given by $\cos\alpha_1=v\,\ep^{1-v}$, but this no longer holds for $\alpha_3$. To determine it, let us consider the geometric constraint. Since the right hand side remains finite, we immediately obtain $\beta=\tfrac{\pi}{2}-\tfrac{\theta}{2}$. In other words, in this limit $\beta$ approaches the base angle. Substituting it into the first equation in \eqref{fbe-iso} gives $\cos\alpha_3=2\cos\tfrac{\theta}{2}\,v\ep^{1-v}$. Then, from the second equation, we find

\begin{equation}\label{v1m}
2\sqrt{1-v^2\ep^{2(1-v)}}
+
\sqrt{1-2(1+\cos\theta)\,v^2\ep^{2(1-v)}}
+
3\k(1+4v)\ep^{-3v}=0
\,,
\end{equation}
which shows explicit dependence on the apex angle $\theta$. Accordingly, the solution $\vom$ is $\theta$-dependent and establishes an upper bound on $v$. The allowed range of $\theta$ is governed by the two conditions: $\lambda_3=1$ and $\alpha_3=0$, as discussed in Appendix F.

Using \eqref{L-y=1} and \eqref{E-y=1}, we find that to leading order in $1-\lambda_1$ the expressions for $\ell$ and $E^{\,\ii}$ simplify to   

\begin{equation}\label{El2-2large}
\ell=-\frac{1}{\sqrt{\s}}\ln(1-\lambda_1)+O(1)
\,,\qquad
E^{\,\ii}=-2\ep\g\sqrt{\s}	\ln(1-\lambda_1)+O(1)
\,.
\end{equation}
Then it follows immediately that

\begin{equation}\label{El2-2large2}
E^{\,\ii}=2\sigma\ell+O(1)
\,,
\end{equation}
with the same string tension as before. Notably, in this limit the leading term is independent of the apex angle $\theta$. 

To compute the constant term, we consider the difference between $E^{\,\ii}$ and the leading term appearing in the large-$\ell$ expansion

\begin{equation}\label{El2-2large3}
	E^{\,\ii}-2\sigma\ell= \g\sqrt{\s}
	\Bigl(2\bigl({\cal E}^-(\lambda_1,v)-\ep{\cal L}^-(\lambda_1,v)\bigr)+{\cal E}^-(\lambda_3,v)+3\k\frac{\ep^{-2v}}{\sqrt{v}}
	+2\ep\,\frac{\sin\tfrac{\theta}{2}-\cos\beta}{\cos(\beta+\tfrac{\theta}{2})}
	\,
	{\cal L}^-(\lambda_3,v)\Bigr)+3c
	\,.
\end{equation}
Taking the limit $\lambda_1\rightarrow 1$, we obtain 

\begin{equation}\label{El2-2large4}
	E^{\,\ii}-2\sigma\ell= 
	\g\sqrt{\s}\Bigl(
	-2{\cal I}(\vom)+{\cal E}^-(\lambda_3,\vom)-2\ep\cos\tfrac{\theta}{2}{\cal L}^-(\lambda_3,\vom)+3\k\frac{\ep^{-2\vom}}{\sqrt{\vom}}
	\Bigr)+3c
	\,.
	\end{equation}
In the last step we used the fact that $\lim_{\beta\rightarrow\tfrac{\pi-\theta}{2}}\bigl(\sin\tfrac{\theta}{2}-\cos\beta\bigr)/\cos(\beta+\tfrac{\theta}{2})=-\cos\tfrac{\theta}{2}$. This yields the asymptotic expansion

\begin{equation}\label{El2-2large5}
{E}^{\,\ii}=2\sigma\ell+\tilde{C}^{\,\ii}+o(1)
\,,\qquad\text{with}\qquad
\tilde{C}^{\,\ii}
=
3c
-
\g\sqrt{\s}\Bigl(2{\cal I}(\vom)-{\cal E}^-(\lambda_3,\vom)
+2\ep\cos\tfrac{\theta}{2}{\cal L}^-(\lambda_3,\vom)
-3\k\frac{\ep^{-2\vom}}{\sqrt{\vom}}\Bigr)
\,.
\end{equation}
The key new feature here is that the constant term depends explicitly on the apex angle $\theta$. This has a simple explanation. In the limit as $\lambda_1\rightarrow 1$, the two strings become infinitely long, while the third remains finite in size. Its length, as well as the energy, depends on the apex angle that directly affects the constant term but not the leading one.

\subsubsection{Configuration III}

Finally, consider configuration $III$. A straightforward way to proceed is as follows. Since $\alpha_3$ changes its sign from negative to positive, the expressions for the geometrical constraint, leg length, and energy can be obtained from those of subsection 2 by respectively replacing ${\cal L}^-$ and ${\cal E}^-$ with ${\cal L}^+$ and ${\cal E}^+$. The force balance equations, however, remain unchanged. Thus, we can rewrite \eqref{geoc2} and \eqref{El2} as   

\begin{equation}\label{geoc3}
\cos\bigl(\beta+\tfrac{\theta}{2}\bigr){\cal L}^-(\lambda_1,v)=
\sin\tfrac{\theta}{2}{\cal L}^+(\alpha_3,v)
\,
\end{equation}
and  
\begin{equation}\label{El3}
\ell^{\,\iii}=\frac{1}{\sqrt{\s}}\,\frac{\cos\beta}{\sin\frac{\theta}{2}}\,
{\cal L}^-(\lambda_1,v)
\,,\qquad
E^{\,\iii}=\g\sqrt{\s}
\biggl(2{\cal E}^-(\lambda_1,v)+{\cal E}^+(\alpha_3,v)+3\k\frac{\ep^{-2v}}{\sqrt{v}}\,\biggr)+3c
\,,
\end{equation}
and parametrically define $E^{\,\iii}$ as a function of $\ell$. 

As before, when studying the small-$\ell$ behavior, we take the limit $v\rightarrow 0$, in which the geometrical constraint becomes

\begin{equation}\label{El3-small}
\cos\bigl(\beta+\tfrac{\theta}{2}\bigr)\sqrt{\cos\alpha_3}\
	\bigl(1+I(\sin^2\alpha_1,\tfrac{1}{2},\tfrac{3}{4})\bigr)
	\bigr)
	=
	\sin\tfrac{\theta}{2}\sqrt{\cos\alpha_1}
\,I(\cos^2\alpha_3,\tfrac{3}{4},\tfrac{1}{2})\bigr)
\,.
\end{equation}
Of course, the force balance equations \eqref{fbe-small} still hold in this limit. The asymptotic expansion can be obtained by replacing ${\cal E}^-_0(\alpha_3)$ in \eqref{El2-small2} by ${\cal E}^+_0(\alpha_3)$, with the result 

\begin{equation}\label{El3-small2}
E^{\,\iii}=-\frac{\alpha^{\iii}}{\ell}+3c+o(1)
\,,\quad\text{with}\quad
\alpha^{\iii}=-\frac{\cos\beta}{\sin\frac{\theta}{2}}\,
{\cal L}^-_0(\alpha_1)
\bigl(2{\cal E}^-_0(\alpha_1)+{\cal E}^+_0(\alpha_3)+3\k\bigr)\g
\,.	
\end{equation}
Again, the coefficient $\alpha^{\iii}$ is positive for our parameter values. 

Likewise, we can obtain the formulas for large $\ell$ corresponding to the limit $\lambda_1\rightarrow 1$ at fixed $\alpha_3\geq0$. Two facts are essential here. First, the tangent angle $\alpha_3$ changes its sign when going from configuration $II$ to $III$, implying that Eq.\eqref{v1m} becomes  

\begin{equation}\label{v1p}
2\sqrt{1-v^2\ep^{2(1-v)}}
-
\sqrt{1-2(1+\cos\theta)\,v^2\ep^{2(1-v)}}
+
3\k(1+4v)\ep^{-3v}=0
\,.
\end{equation}
The solution $\vop$ to this equation is $\theta$-dependent and establishes an upper bound on the parameter $v$. One might intuitively expect the range of $\theta$ to be set by the two conditions: $\alpha_3=0$ and $\alpha_3=\frac{\pi}{2}$. While this is true for the latter case, leading to $\theta=\pi$,  it does not hold for the former. This is because $\vop$ is a double-valued function of $\theta$.\footnote{For more on this, see Appendix D.} Second, in this limit, strings 1 and 2 become infinitely long, while string 3 remains finite and contributes only to the subleading constant term in the expansion in $\ell$. This allows us to deduce directly from \eqref{El2-2large5} that  

\begin{equation}\label{El3-large4}
E^{\,\iii}=2\sigma\ell+C^{\,\iii}+o(1)
\,,\qquad\text{with}\qquad
C^{\,\iii}
=
3c
-
\g\sqrt{\s}\Bigl(
2{\cal I}(\vop)-{\cal E}^+(\alpha_3,\vop)
+
2\ep\cos\tfrac{\theta}{2}{\cal L}^+(\alpha_3,\vop)
-
3\k\frac{\ep^{-2\vop}}{\sqrt{\vop}}
\Bigr)
\,.
\end{equation}
Here the coefficient of the leading term is also $2$, and the constant term depends on $\theta$.

\subsection{The simplest examples}

Having understood the basic string configurations, we can now proceed to compute the three-quark potential. But before doing so, we need to specify the model parameters. For the purposes of this paper, it is natural to use those from \cite{a3Q2016}, which were obtained by fitting lattice QCD data for pure $SU(3)$ gauge theory to the string model under consideration. Accordingly, we set $\g = 0.176$, $\s = 0.44\,\text{GeV}^2$, and $\k=-0.083$.\footnote{Note that string configurations with all non-negative tangent angles are not allowed due to the negative value of $\k$.} These values will be used in all subsequent estimates unless otherwise specified. Here, we do not attempt to compute the three-quark potential for all values of $\theta$. Instead, our goal is to illustrate the above ideas through a few examples, which we hope will clarify the general approach. A more complete analysis is presented in Appendix D.
\subsubsection{$\theta<\frac{2}{3}\pi$}

Let us first recall how the potential is defined at $\theta=\frac{\pi}{3}$, which corresponds to the equilateral triangle geometry \cite{a3Q2016}. In this case, symmetry plays a pivotal role, allowing one, in particular to express the tangent angle analytically in terms of the parameter. As a result, the potential is given parametrically by  

\begin{equation}\label{Elpi/3}
\ell=\ell^{\,\ii}(v)\,,
\qquad 
E_{\3Q}=E^{\,\ii}(v)
\,,
\quad\text{for}\quad
0\leq v\leq \vo
\,.
\end{equation}
Here the upper bound corresponds to the solution of Eq.\eqref{v1}, and numerically, $\vo=0.978$.

Now consider $\theta=\frac{\pi}{6}$, a typical example with an acute apex angle. As explained in Appendix D, configuration $I$ dominates at small $\ell$, and is then replaced by configuration $II$ for larger values of $\ell$. Consequently, the potential is defined parametrically by two piecewise functions 

\begin{equation}\label{Elpi/6}
\ell=\begin{cases}
\ell^{\,\i} (v)& \text{for}\quad 0\,\leq v\leq \vz\,,\\ 
\ell^{\,\ii}(v)& \text{for}\quad \vz\leq v\leq \vo\,,
\end{cases}
\qquad
E_{\3Q}=\begin{cases}
E^{\,\i} (v)& \text{for}\quad 0\,\leq v\leq \vz\,,\\
E^{\,\ii}(v)& \text{for}\quad \vz\leq v\leq \vo\,,
\end{cases}
\end{equation}
where $\vz$ is the solution to the equation $\alpha_1(v)=0$.\footnote{The upper index in $\vi$ refers to a tangent angle, and the lower to its value.} At this value of $v$ the tangent angle $\alpha_1$ changes the sign from positive to negative, signaling a transition between the string configurations. The value of $\vz$ can be found numerically from Eqs.\eqref{fbe-iso} and \eqref{geoc1} which take the form 

\begin{equation}\label{v01}
	\sin\beta=\oh\cos\alpha_3\,,
	\qquad
	\cos\alpha_3=\sqrt{1-9\k^2(1+4v)^2\ep^{-6v}}\,,
	\qquad
	\cos\bigl(\beta+\tfrac{\pi}{12}\bigr){\cal L}^+(0,v)=
\tfrac{\sqrt{3}-1}{2\sqrt{2}}\,{\cal L}^-(\lambda_3,v)
\,.
\end{equation}
For our parameter set, this yields $\vz=0.929$. Importantly, $\vz<\vo$, as required for consistency.  

Our next example is $\theta=\frac{\pi}{2}$, corresponding to the right triangle geometry. The analysis closely follows that of the previous case, with the only essential difference being that configuration $III$ is now relevant at small $\ell$. As a result, the potential is given parametrically by

\begin{equation}\label{Elpi/2}
\ell=\begin{cases}
\ell^{\,\iii} (v)& \text{for}\quad 0\,\leq v\leq \vzi\,,\\ 
\ell^{\,\ii}(v)& \text{for}\quad \vzi\leq v\leq \vo\,,
\end{cases}
\qquad
E_{\3Q}=\begin{cases}
E^{\,\iii} (v)& \text{for}\quad 0\,\leq v\leq \vzi\,,\\
E^{\,\ii}(v)& \text{for}\quad \vzi\leq v\leq \vo\,.
\end{cases}
\end{equation}
Here $\vzi$ is determined by the  equations

\begin{equation}\label{v03}
	\sin\beta=\oh\cos^{-1}\hspace{-2pt}\alpha_1\,,
	\qquad
	\cos\alpha_1=\sqrt{1-\tfrac{9}{4}\k^2(1+4v)^2\ep^{-6v}}\,,
	\qquad
	\cos\bigl(\beta+\tfrac{\pi}{4}\bigr){\cal L}^-(\lambda_1,v)
=
\tfrac{1}{\sqrt{2}}\,{\cal L}^+(0,v)
\,.
\end{equation}
Numerically, $\vzi = 0.967$. This value is also consistent with the previously obtained $\vo$.

\begin{figure}[h]
\centering
\includegraphics[width=8.25cm]{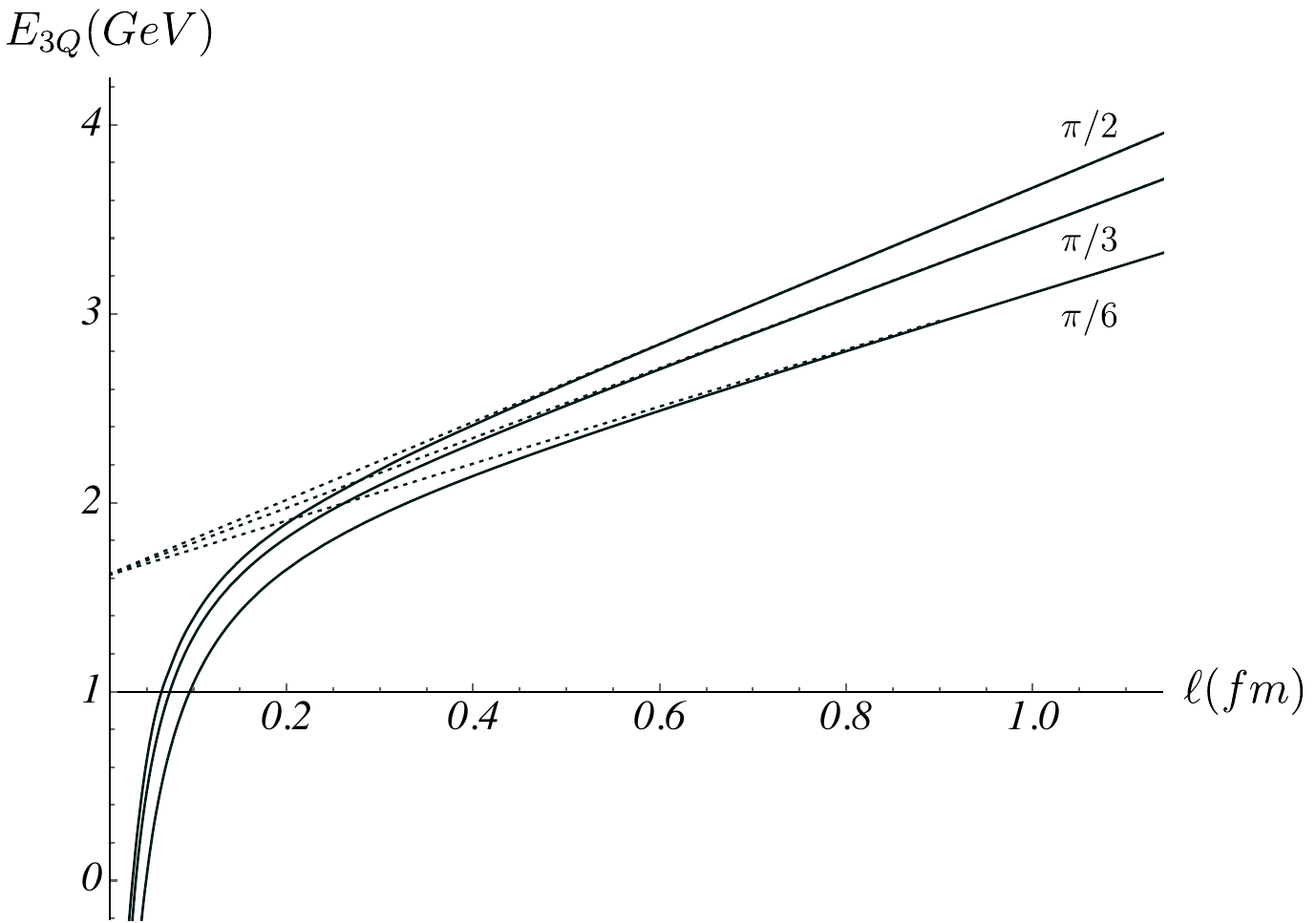}
\caption{{\small $E_{\3Q}$ vs $l$ for several values of $\theta$ not exceeding $\frac{2}{3}\pi$. The dotted lines represent the asymptotic behavior \eqref{largeELI}. We set $c=623\,\text{MeV}$, here and in all subsequent Figures.}}
\label{E3QI}
\end{figure}

In Figure \ref{E3QI} we present the results for the three-quark potential obtained for these three values of $\theta$. 
All of them agree with the large-$\ell$ asymptotics 

\begin{equation}\label{largeELI}
	E_{\3Q}(\ell)=2\cos(\tfrac{\theta}{2}-\tfrac{\pi}{3})\,\sigma\ell+C_{\3Q}+o(1)
	\,,
\end{equation} 
where $C_{\3Q}=C^{\,\ii}$. In other words, for large $\ell$ the potential becomes linear, with the slope dictated by the $Y$-law and the universal constant term (i.e., independent of $\theta$). In Sec.V, we will show that this asymptotic behavior holds for any triangle geometry provided no angle exceeds $\frac{2\pi}{3}$. Another observation is that approaching to the asymptotic form depends on the value of $\theta$.

As first noted in \cite{a3Q2016}, the constant terms appearing in the large- and small-$\ell$ expansions are not equal. This feature sets the five-dimensional string model apart from phenomenological ans\"atze. It is noteworthy that the difference between these constants is well-defined (scheme-independent), and a straightforward estimate yields

\begin{equation}\label{Cuv-Cir}
	3c-C_{\3Q}=267\,\text{MeV}
	\,.
\end{equation}
Thus, the constant term in the large-$\ell$ expansion is smaller than that in the small-$\ell$ expansion. 

It is also instructive to make an estimate in relation to the $\Delta$-ansatz. Assuming that $E_{\QQb}$ is given by \eqref{EQQb}, the difference between the constant terms in the large-$\ell$ expansions is estimated to be 

\begin{equation}\label{3QvsQQb}
	\tfrac{3}{2}C_{\QQb}-C_{\3Q}=-3\g\sqrt{\s}\Bigl(\int^1_{\sqrt{\vo}}\frac{du}{u^2}\,\ep^{u^2}\sqrt{1-u^4\ep^{2(1-u^2)}}+\k\frac{\ep^{-2\vo}}{\sqrt{\vo}}\,\Bigr)=4\,\text{MeV}\,.
	\end{equation}
This suggests that the baryon vertex has only a minimal impact on the configuration of three infinitely long strings. In practical terms, this implies that at large interquark separations the difference between the $Y$- and $\Delta$-laws manifests mainly through their differing slopes rather than through constant terms.\footnote{See, for example, Figure $3$ in \cite{a3Q2016}.} In the next subsection, we will see that the situation is reversed for $\theta>\frac{2}{3}\pi$. It is also worth noting that the Cornell ansatz gives $\tfrac{3}{2}C_{\QQb}=3c$, and thus leads directly to the estimate in \eqref{Cuv-Cir}.

Finally, let us examine the behavior at small $\ell$. As already noted in subsection A, the potential behaves for $\ell\rightarrow 0$ as 

\begin{equation}\label{ELsmall}
E_{\3Q}=-\frac{\alpha_{\3Q}}{\ell}+3c+O(\ell)
\,,\qquad	
\alpha_{\3Q}=
	\begin{cases}
		\alpha^{\i} &\text{for}\quad\theta=\frac{\pi}{6}\,,\\
		\alpha^{\ii} &\text{for}\quad\theta=\frac{\pi}{3}\,,\\
		\alpha^{\iii} &\text{for}\quad\theta=\frac{\pi}{2}
		\,.
	\end{cases}
\end{equation}
It is of physical interest to investigate the pairwise interaction between quarks at small separations. A natural question arises: how accurate is the pairwise quark interaction ($\Delta$-law) in describing this regime? We are now in a position to partially answer this. To this end, it is useful to define the quark-quark coupling as    

\begin{equation}\label{reduced-alpha}
\alpha_{\QQ}=\biggl(2+\frac{1}{2\sin\tfrac{\theta}{2}}\biggr)^{-1}
	\alpha_{\3Q}
	\,.
\end{equation}
The second term inside the parentheses accounts for the length of the triangle's base.

Now, let us present some estimates for the small-$\ell$ behavior. The results are summarized in Table  \ref{estimates1}.
\begin{table*}[htb]
\renewcommand{\arraystretch}{2}
\centering 	\scriptsize
\begin{tabular}{lccccccr}				
\hline
$\theta$  ~~~~&~~~~~ $\cos\alpha_1$ ~&~~~ $\cos\alpha_3$ ~&~~~$\sin\beta$  ~~~&~~~~ $\alpha_{\3Q}$ ~~&~ $\alpha_{\QQ}/\alpha_{\QQb}$ 
\rule[-3mm]{0mm}{8mm}
\\
\hline \hline
$\pi/6$ & 0.951  & 0.500 & 0.263 & 0.484 & 0.488  \\
$\pi/3$ & 0.997 & 0.997 & 0.500 & 0.375 & 0.495\\
$\pi/2$  & 0.921 & 0.848 & 0.460 & 0.336 & 0.491  \\
 \hline \hline
\end{tabular}
\caption{ \small Estimates for the angles and couplings in the small $\ell$ limit. Here $\alpha_{\QQb}$ is the Coulomb coefficient of the quark-antiquark potential (see Appendix C).}
\label{estimates1}
\end{table*}
The two main conclusions follow from this Table. First, as seen from the fourth column, the $Y$-point generally does not coincide with the Fermat point $F$ of the triangle.\footnote{As shown in subsection A, $Y$ approaches $F$ in the limit $\ell \rightarrow \infty$.} The only exception is the case $\theta = \frac{\pi}{3}$, where the two points coincide due to symmetry.  Second, the last column shows that the coupling $\alpha_{\QQ}$ depends weakly on $\theta$, and that the ratio $\alpha_{\QQ}/\alpha_{\QQb}$ is very close to the phenomenologically expected value of $\oh$. Since the discrepancy among the $\alpha_{\QQ}$ values does not exceed $1.5\%$, the assumption of pairwise quark interaction at small separations appears to be quite reasonable.
\subsubsection{$\theta>\frac{2}{3}\pi$}

We now turn to additional examples. For our purposes, it suffices to consider the cases where the apex angle takes the values $\frac{3}{4}\pi$, $\frac{5}{6}\pi$, and $\pi$. The last of these, corresponding to the collinear geometry, was previously discussed in \cite{a3Q2016}. The remaining cases are similar in structure, as explained in Appendix D. The potential is given parametrically by

\begin{equation}\label{Elpi}
	\ell=\ell^{\,\iii}(v)\,,\qquad
	E_{\3Q}=E^{\,\iii}(v)\,,\qquad   
	\text{for}\qquad
	0\leq v\leq \vop
	\,,
\end{equation} 
where $\vop$ is the solution of Eq.\eqref{v1p}.

\begin{figure}[H]
\centering
\includegraphics[width=8.5cm]{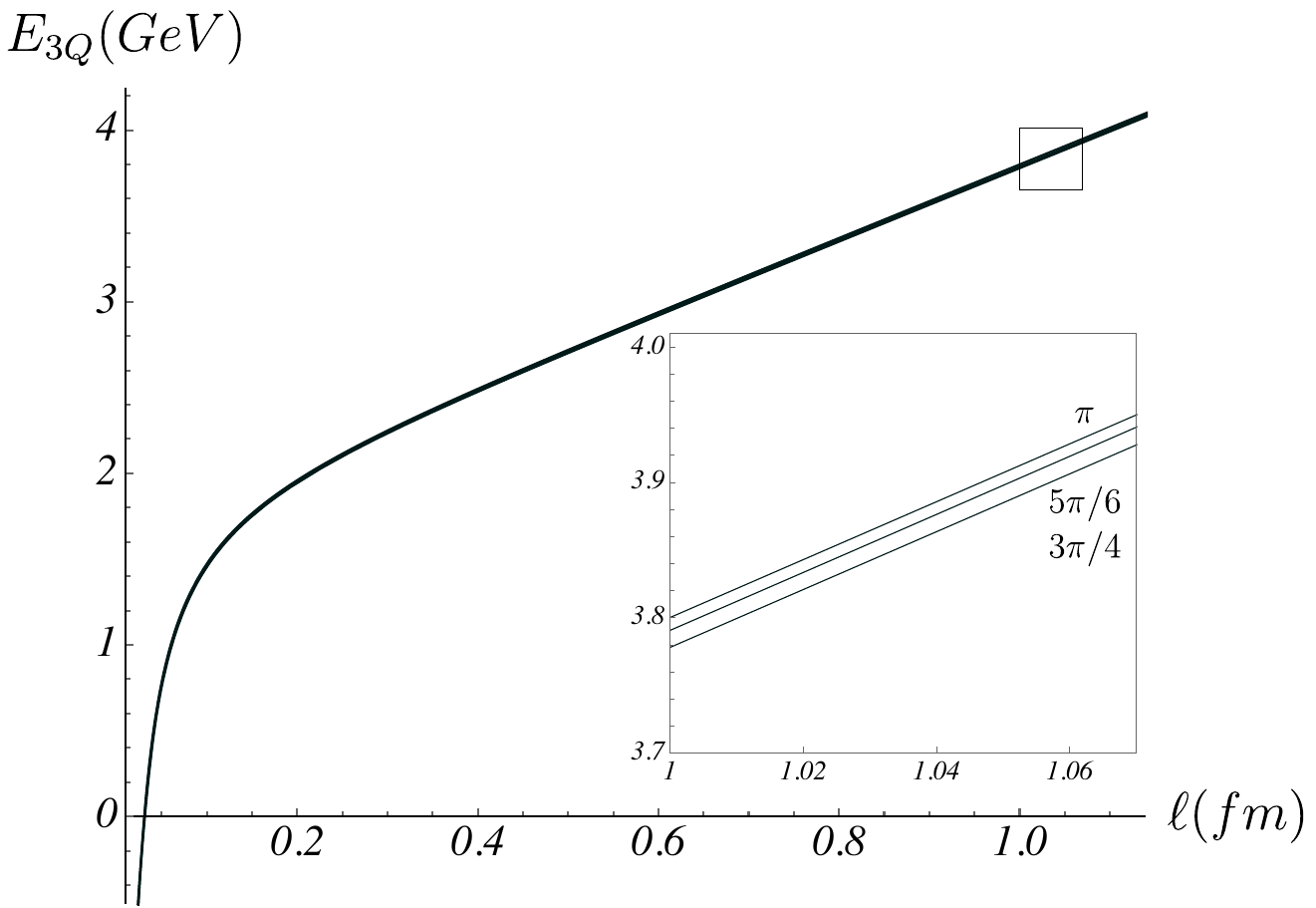}
\caption{{\small $E_{\3Q}$ vs $l$ for several values of $\theta$ exceeding $\frac{2}{3}\pi$.}}
\label{E3QII}
\end{figure}

In Figure \ref{E3QII} we plot the three-quark potential for these values of $\theta$. For large $\ell$, all the curves become linear with 
\noindent the same slope but different constant terms. This behavior follows from the general asymptotic form of the potential

\begin{equation}\label{largeELII}
	E_{\3Q}(\ell)=2\sigma\ell+C_{\3Q}+o(1)
	\,,
\end{equation} 
with $C_{\3Q}=C^{\,\iii}$. It is instructive to analyze this in more detail using the formulas derived in subsection A. The results are summarized in Table \ref{estimatesElpi}. Several conclusions can be drawn immediately:
\begin{table*}[htbp]
\renewcommand{\arraystretch}{2}
\centering 	\scriptsize
\begin{tabular}{lccccccr}				
\hline
$\theta$  ~~~~~~~~& $\vop$~~&~~~~$\cos\alpha_3$~~~~&~~~$3c-C_{\3Q}$  ~~~&~~$C_{\3Q}-\frac{3}{2}C_{\QQb}$  
\rule[-3mm]{0mm}{8mm}
\\
\hline \hline
$3\pi/4$ &  0.614  &  0.691 &  232\,\text{MeV} &  31\,\text{MeV}  \\
$5\pi/6$ &  0.532  &  0.440 &  217\,\text{MeV} &  46\,\text{MeV}  \\
$\pi$    &  0.484  &  0     &  206\,\text{MeV}  &  57\,\text{MeV}  \\
 \hline \hline
\end{tabular}
\caption{ \small Estimates of $C_{\3Q}$ for different values of $\theta$.}
\label{estimatesElpi}
\end{table*}
First, the constant term $C_{\3Q}$ increases with $\theta$ reaching its maximum at $\theta = \pi$. Second, for all cases, $C_{\3Q}$ remains less than $3c$, as also shown in Figure \ref{CIR}. Third, at $\theta=\pi$ the difference between the $\Delta$-law and $E_{\3Q}$ is visible due to different constant terms, while their slopes remain the same. 

 Next, we examine the behavior of the potential at small $\ell$. In all the above examples, it is described by configuration III, leading to 

\begin{equation}\label{ELsmall2}
E_{\3Q}=-\frac{\alpha_{\3Q}}{\ell}+3c+O(\ell)
\,,\qquad
\text{with}
\qquad	
\alpha_{\3Q}=\alpha^{\iii}
\,.	
\end{equation}
As before, it is useful to define the quark-quark coupling using Eq.~\eqref{reduced-alpha}. The corresponding estimates for the Coulomb coefficients are presented in Table \ref{estimatesELpi2}. The conclusions are similar to those from the earlier examples:
\begin{table*}[htbp]
\renewcommand{\arraystretch}{2}
\centering 	\scriptsize
\begin{tabular}{lccccccr}				
\hline
$\theta$  ~~~~&~~~ $\cos\alpha_1$ ~&~~~ $\cos\alpha_3$ ~&~~~$\sin\beta$  ~~~&~~~~ $\alpha_{\3Q}$ ~~~&~~~ $\alpha_{\QQ}/\alpha_{\QQb}$ 
\rule[-3mm]{0mm}{8mm}
\\
\hline \hline
$3\pi/4$ & 0.815  & 0.416 & 0.255 & 0.308 & 0.480  \\
$5\pi/6$ & 0.796  & 0.275 & 0.173 & 0.304 & 0.477\\
$\pi$    & 0.781  & 0     & 0     & 0.300 & 0.475  \\
 \hline \hline
\end{tabular}
\caption{ \small Estimates for the angles and couplings in the small $\ell$ limit.}
\label{estimatesELpi2}
\end{table*}
 In general, the point $Y$ does not coincide with the Fermat point, except in the special case $\theta = \pi$. The ratio $\alpha_{\QQ}/\alpha_{\QQb}$ remains close to $\frac{1}{2}$ across the considered range of $\theta$. Also noteworthy is that the deviation of this ratio reaches its minimum and maximum at the special angles $\theta = \frac{1}{3}\pi$ and $\pi$, respectively, but never exceeds about $4\%$.
 
\section{Collinear Geometry}
\renewcommand{\theequation}{4.\arabic{equation}}
\setcounter{equation}{0}

With the experience gained from the equilateral triangle geometry, particularly the case $\theta = \pi$, we now turn to a generic case of collinear geometry. For convenience, we place the heavy quarks along the $x$-axis at $x_1 = -\ell$, $x_2 = L$, and $x_3 = 0$, as illustrated in Figure \ref{coli}. These points serve as the endpoints of strings that meet at a baryon vertex located in the interior. This geometry significantly simplifies the force balance equation \eqref{fbe}, reducing it to only two components: the $x$- and $r$-components:

\begin{equation}\label{fbe-col}
\cos\alpha_1-\cos\alpha_2-\cos\alpha_3=0
\,,\qquad
\sin\alpha_1+\sin\alpha_2+\sin\alpha_3=3\k(1+4v)\ep^{-3v}
\,.
\end{equation}
The string configuration for this type of geometry was previously studied in \cite{a3Q2016}, where parametric equations for the three-quark potential were derived. However, the resulting equations involve two parameters, making it difficult to analyze the potential in general. To facilitate the discussion, we consider a slice of the two-parameter space by imposing the condition\footnote{Note that the symmetric case $\eta = 1$ corresponds to the case $\theta = \pi$ discussed in Sec.III.}

\begin{equation}\label{eta}
	\ell=\eta\hspace{0.1pt}L
	\,,
\end{equation}
where $\eta$ is fixed and lies in the interval $[0,1]$. Accordingly, we restrict ourselves to the case $\ell \leq L$. The potential for the opposite case can be easily obtained by interchanging the roles of $\ell$ and $L$. In analogy with Sec.III, we refer to Eq.\eqref{eta} as the geometrical constraint. As before, it reduces the number of parameters by one, making practical calculations, especially at small quark separations, more tractable.
\subsection{Basic configurations}

\begin{figure}[htpb]
\centering
\includegraphics[width=6.5cm]{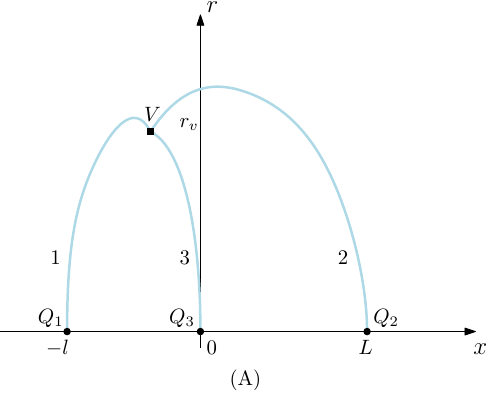}
\hspace{2cm}
\includegraphics[width=6.5cm]{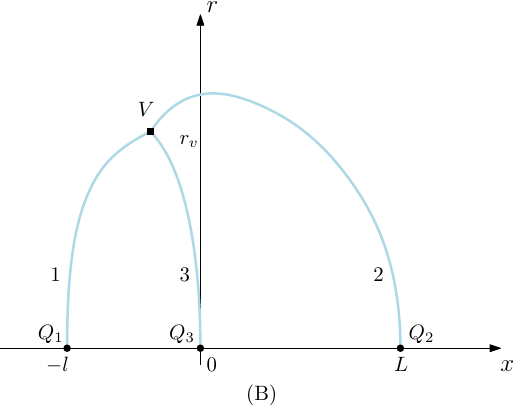}
\caption{{\small Two basic configurations corresponding to different sign of the tangent angle of string 1. In (A), $\alpha_1\leq 0$, and in (B), $\alpha_1\geq 0$.}}
\label{coli}
\end{figure}

For our purposes in this paper, we first consider two basic configurations, as those shown in Figure \ref{coli}. The number  
of possible configurations is limited by the condition $\ell\leq L$.
 \subsubsection{Configuration A}

We begin with configuration A, characterized by $\alpha_1<0$, $\alpha_2<0$, and $\alpha_3>0$. It can be regarded as a deformation of the symmetric case considered in \cite{a3Q2016}.   

Using Eqs.\eqref{l+} and \eqref{l-}, the geometrical constraint can be written as

\begin{equation}\label{gc-colA}
	{\cal L}^-(\lambda_1,v)+{\cal L}^+(\alpha_3,v)
	=\eta
	\bigl(
	{\cal L}^-(\lambda_2,v)-{\cal L}^+(\alpha_3,v)
	\bigr)
	\,.
\end{equation}
Similarly, we can derive expressions for $\ell$ and the configuration energy. The latter follows from Eqs.\eqref{baryon-v}, \eqref{E+}, and \eqref{E-}, leading to

\begin{equation}\label{lEA}
	\ell^{\text{\,\tiny A}}=\frac{1}{\sqrt{\s}}\Bigl({\cal L}^-(\lambda_1,v)+{\cal L}^+(\alpha_3,v)\Bigr)
	\,,
	\qquad
	E^{\text{\,\tiny A}}=\g\sqrt{\s}\Bigl(
	{\cal E}^-(\lambda_1,v)+{\cal E}^-(\lambda_2,v)+{\cal E}^+(\alpha_3,v)+3\k\frac{\ep^{-2v}}{\sqrt{v}}\,
	\Bigr)
	+3c
	\,,
	\end{equation}
	where the superscript $\text{\small A}$ serves as a reminder of the specific string configuration under consideration. The force balance equations and geometrical constraint allow us to express the tangent angles in terms of $v$. As a result, the energy is given in parametric form by $E^{\text{\,\tiny A}}=E^{\text{\,\tiny A}}(v)$ and $\ell^{\text{\,\tiny A}}=\ell^{\text{\,\tiny A}}(v)$, with the range for $v$ to be discussed shortly.
	
We are now ready to analyze the small-$\ell$ behavior of $E^{\text{\,\tiny A}}$. The analysis is based on the observation that small $\ell$ corresponds to small $v$, as follows from the asymptotics \eqref{fL+smallx} and \eqref{L-y=0}. Taking the limit $v\rightarrow 0$ in Eqs.\eqref{fbe-col} and \eqref{gc-colA}, we get 

\begin{equation}\label{fbe-colA}
\cos\alpha_1-\cos\alpha_2-\cos\alpha_3=0
\,,\qquad
\sin\alpha_1+\sin\alpha_2+\sin\alpha_3=3\k
\,
\end{equation}
and
\begin{equation}\label{gc-colA0}
\eta=
\frac{\cos^{-\oh}\hspace{-.9mm}\alpha_1
\bigl[1+I\bigl(\sin^2\hspace{-.9mm}\alpha_1,\tfrac{1}{2},\tfrac{3}{4}\bigr)\bigr]
+
\cos^{-\oh}\hspace{-.9mm}\alpha_3\,I\bigl(\cos^2\hspace{-.9mm}\alpha_3,\tfrac{3}{4},\oh\bigr)
}
{\cos^{-\oh}\hspace{-.9mm}\alpha_2
\bigl[1+I\bigl(\sin^2\hspace{-.9mm}\alpha_2,\oh,\tfrac{3}{4}\bigr)\bigr]
-
\cos^{-\oh}\hspace{-.9mm}\alpha_3\,I\bigl(\cos^2\hspace{-.9mm}\alpha_3,\tfrac{3}{4},\oh\bigr)}
\,.	
\end{equation}
At a given $\eta$, the solution to these equations determines the limiting values of the tangent angles. We then expand $\ell^{\text{\,\tiny A}}$ and $E^{\text{\,\tiny A}}$ near $v=0$ and reduce the two equations to a single one, with the result 

\begin{equation}\label{Ecola-small}
E^{\text{\,\tiny A}}=-\frac{\alpha^{\text{\tiny A}}}{\ell}+3c+o(1)
\,,\quad\text{with}\quad
\alpha^{\text{\tiny A}}=-
\bigl({\cal L}^-_0(\alpha_1)+{\cal L}_0^+(\alpha_3)\bigr)
\bigl({\cal E}^-_0(\alpha_1)+{\cal E}^-_0(\alpha_2)
+
{\cal E}^+_0(\alpha_3)+3\k\bigr)\g
\,.	
\end{equation}
The second factor in $\alpha^{\text{\tiny A}}$ ensures its positivity for our parameter set.

In studying the large-$\ell$ behavior, we take both $\lambda$'s to approach $1$. In that limit, the tangent angles of the side strings go to $\cos\alpha_1=\cos\alpha_2=v\ep^{1-v}$, so that $\alpha_1=\alpha_2$. Solving the force balance equations under this condition gives: $\sin\alpha_1=(3\k(1+4v)\ep^{-3v}-1)/2$ and $\alpha_3=\pi/2$. The compatibility of the two expressions for $\alpha_1$ implies that 

\begin{equation}\label{colav1}
2\sqrt{1-v^2\ep^{2(1-v)}}
-
1
+
3\k(1+4v)\ep^{-3v}=0
\,.
\end{equation}
This is precisely Eq.\eqref{v1p} evaluated at $\theta = \pi$. Importantly, its solution $\vop$ is independent of $\eta$ and provides an upper bound on the parameter $v$, which thus varies from $0$ to $\vop$.

The leading terms in the expansion of $\ell^{\text{\,\tiny A}}$ and $E^{\text{\,\tiny A}}$ can be found using the asymptotics \eqref{L-y=1} and \eqref{E-y=1}. So, 

\begin{equation}\label{Ecola-large}
\ell^{\text{\,\tiny A}}=-\frac{1}{\sqrt{\s}}\ln(1-\lambda_1)+O(1)
\,,\qquad
E^{\text{\,\tiny A}}=-\ep\g\sqrt{\s}\bigl(\ln(1-\lambda_1)+\ln(1-\lambda_2)\bigr)+O(1)
	\,.
	\end{equation}
Then using the geometric constraint, we express $1-\lambda_2$ in terms of $1-\lambda_1$ as $1-\lambda_2=(1-\lambda_1)^{\frac{1}{\eta}}$ and arrive at
\begin{equation}\label{Ecola-large2}
E^{\text{\,\tiny A}}=(1+\eta^{-1})\sigma\ell+O(1)=
\sigma(\ell+L)+O(1)
\,,
\end{equation}
which matches the result first obtained in \cite{a3Q2016}. 

To go further, we compute the next term in this expansion. Consider the difference

\begin{equation}\label{Ecola-large3}
E^{\text{\,\tiny A}}-\sigma(\ell+L)=\g\sqrt{\s}
\Bigl({\cal E}^-(\lambda_1,v)-\ep{\cal L}^-(\lambda_1,v)+{\cal E}^-(\lambda_2,v)-\ep{\cal L}^-(\lambda_2,v)+{\cal E}^+(\alpha_3,v)+3\k\frac{\ep^{-2v}}{\sqrt{v}}
\Bigr)+3c
\,.	
\end{equation}
After taking the limit $\lambda_1\rightarrow 1$ and using \eqref{QL+} and \eqref{fELy=1}, we find

\begin{equation}\label{Ecola-large4}
E^{\text{\,\tiny A}}-\sigma(\ell+L)
=
\g\sqrt{\s}
\Bigl(-2{\cal I}(\vop)+{\cal Q}(\vop)+3\k\frac{\ep^{-2\vop}}{\sqrt{\vop}}
\Bigr)+3c
\,.
\end{equation}
Thus, the large-$\ell$ expansion takes the form

\begin{equation}\label{Ecola-large5}
E^{\text{\,\tiny A}}=\sigma (\ell+L)+C^{\text{\,\tiny A}}+o(1)
\,,\qquad\text{with}\qquad
C^{\text{\,\tiny A}}=3c-\g\sqrt{\s}\Bigl(2{\cal I}(\vop)-{\cal Q}(\vop)-3\k\frac{\ep^{-2\vop}}{\sqrt{\vop}}\Bigr)
\,.
\end{equation}
Here, $C^{\text{\,\tiny A}}$ is independent of $\eta$ and coincides with $C^{\text{\tiny III}}$ evaluated at $\theta = \pi$. We conclude that the constant term is universal for all collinear geometries.

Finally, it is worth noting that neither the limit $\lambda_1 \to 1$ at fixed $\lambda_2$, nor the limit $\lambda_2 \to 1$ at fixed $\lambda_1$, is allowed due to the geometrical constraint.

\subsubsection{Configuration B}

We now briefly discuss configuration B. The only difference from configuration A is that the tangent angle $\alpha_1$ is non-negative. As a result, the corresponding expressions can be obtained from those of configuration A by replacing, for string 1, the functions ${\cal L}^-$ and ${\cal E}^-$ with ${\cal L}^+$ and ${\cal E}^+$. After doing so, we get 

\begin{equation}\label{gc-colB}
	{\cal L}^+(\alpha_1,v)+{\cal L}^+(\alpha_3,v)
	=
	\eta
	\bigl({\cal L}^-(\lambda_2,v)-{\cal L}^+(\alpha_3,v)\bigr)
	\,,
\end{equation}
\begin{equation}\label{lEB}
	\ell^{\text{\,\tiny B}}=\frac{1}{\sqrt{\s}}\Bigl({\cal L}^+(\alpha_1,v)+{\cal L}^+(\alpha_3,v)\Bigr)
	\,,
\qquad
E^{\text{\,\tiny B}}=\g\sqrt{\s}\Bigl(
	{\cal E}^+(\alpha_1,v)+{\cal E}^-(\lambda_2,v)+{\cal E}^+(\alpha_3,v)+3\k\frac{\ep^{-2v}}{\sqrt{v}}\,
	\Bigr)
	+3c
	\,.
\end{equation}
The force balance equations \eqref{fbe-col} also apply to configuration B. Together with the geometrical constraint, they allow us to express the tangent angles in terms of $v$. Accordingly, the energy as a function of $\ell$ can be written in parametric form as $E^{\text{\,\tiny B}}=E^{\text{\,\tiny B}}(v)$ and $\ell^{\text{\,\tiny B}}=\ell^{\text{\,\tiny B}}(v)$. 

As usual, when studying the small-$\ell$ behavior, we consider the limit $v\rightarrow 0$. In this limit, the force balance equations reduce to \eqref{fbe-colA}, while the geometrical constraint becomes

\begin{equation}\label{gc-col2s}
\eta=
\frac{\cos^{-\oh}\hspace{-.9mm}\alpha_1
I\bigl(\cos^2\hspace{-.9mm}\alpha_1,\tfrac{3}{4},\tfrac{1}{2}\bigr)
+
\cos^{-\oh}\hspace{-.9mm}\alpha_3\,I\bigl(\cos^2\hspace{-.9mm}\alpha_3,\tfrac{3}{4},\oh\bigr)
}
{\cos^{-\oh}\hspace{-.9mm}\alpha_2
\bigl(1+I\bigl(\sin^2\hspace{-.9mm}\alpha_2,\oh,\tfrac{3}{4}\bigr)\bigr)
-
\cos^{-\oh}\hspace{-.9mm}\alpha_3\,I\bigl(\cos^2\hspace{-.9mm}\alpha_3,\tfrac{3}{4},\oh\bigr)}
\,.	
\end{equation}
The derivation of this formula proceeds analogously to that of \eqref{gc-colA0}. The corresponding asymptotic expansion follows directly from Eq.\eqref{Ecola-small}, where we substitute ${\cal L}_0^-(\alpha_1),,{\cal E}_0^-(\alpha_1) \rightarrow {\cal L}_0^+(\alpha_1),,{\cal E}_0^+(\alpha_1)$. So,

\begin{equation}\label{Ecolb-small}
E^{\text{\,\tiny B}}=-\frac{\alpha^{\text{\tiny B}}}{\ell}+3c+o(1)
\,,\quad\text{with}\quad
\alpha^{\text{\tiny B}}=-
\bigl({\cal L}^+_0(\alpha_1)+{\cal L}_0^+(\alpha_3)\bigr)
\bigl({\cal E}^+_0(\alpha_1)+{\cal E}^-_0(\alpha_2)
+
{\cal E}^+_0(\alpha_3)+3\k\Bigr)\g
\,.	
\end{equation}
The coefficient $\alpha^{\text{\,\tiny B}}$ is positive, for the same reasons discussed below Eq.\eqref{Ecola-small}.

Finally, we note that the limit $\lambda_2\rightarrow 1$ is not allowed by the geometrical constraint. Thus, configuration B is irrelevant at large $\ell$.

\subsection{Some examples}

The goal here is to describe some simple examples. Additional details are provided in Appendix E.

The symmetric case $\eta=1$ was already considered in \cite{a3Q2016} and in Sec.III. Therefore, the first example we consider here is $\eta=\frac{1}{2}$. In this case, as explained in Appendix E, the potential is determined by a single string configuration, namely configuration A. The corresponding expressions are

\begin{equation}\label{Elcol2}
	\ell=\ell^{\text{\,\tiny A}}(v)\,,\qquad
	E_{\3Q}=E^{\text{\,\tiny A}}(v)\,,\qquad 
	\text{for}\qquad
	0\leq v\leq \vop
	\,,
\end{equation} 
where $\vop$ is the solution to Eq.\eqref{colav1} (see also Table II).

The next cases are $\eta=\frac{1}{3}$ and $\eta=\frac{1}{10}$. As shown in Appendix E, for these values of $\eta$ the potential is given parametrically as a piecewise function by  

\begin{equation}\label{Elcol3}
\ell=\begin{cases}
\ell^{\text{\,\tiny B}} (v)& \text{for}\quad 0\,\leq v\leq \vz\,\,,\\ 
\ell^{\text{\,\tiny A}}(v)& \text{for}\quad \vz\leq v\leq \vop ,
\end{cases}
\qquad
E_{\3Q}=\begin{cases}
E^{\,\text{\tiny B}} (v)& \text{for}\quad 0\,\leq v\leq \vz\,\,,\\
E^{\,\text{\tiny A}}(v)& \text{for}\quad \vz\leq v\leq \vop .
\end{cases}
\end{equation}
Here $\vz$ is a solution to the equation $\alpha_1(v)=0$ and $\vop$ is the same as before. Numerically, we find $\vz=0.070$ for $\eta=\frac{1}{3}$ and $\vz=0.118$ for $\eta=\frac{1}{10}$.

\begin{figure}[htbp]
\centering
\includegraphics[width=8.25cm]{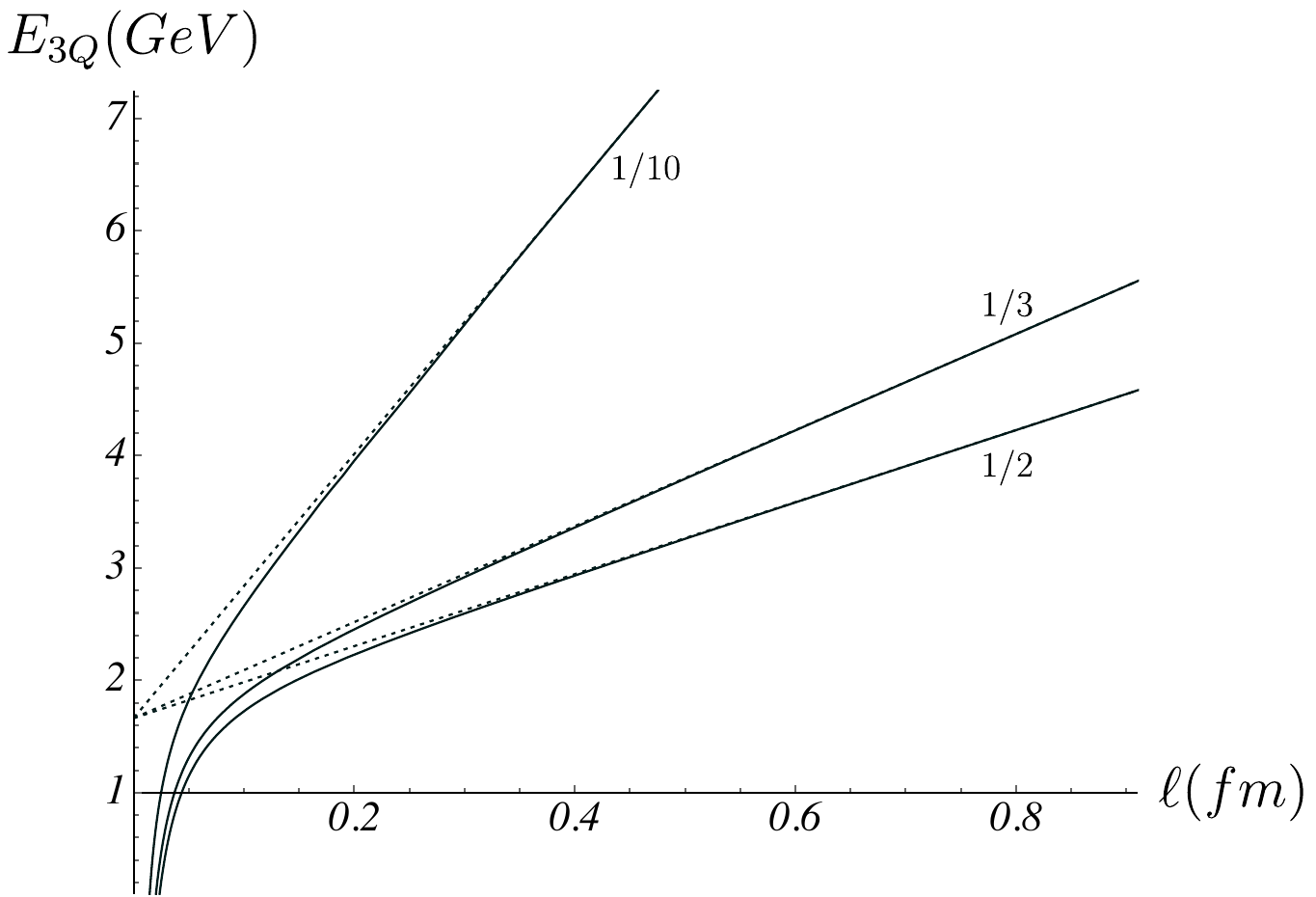}
\hspace{1.25cm}
\includegraphics[width=8cm]{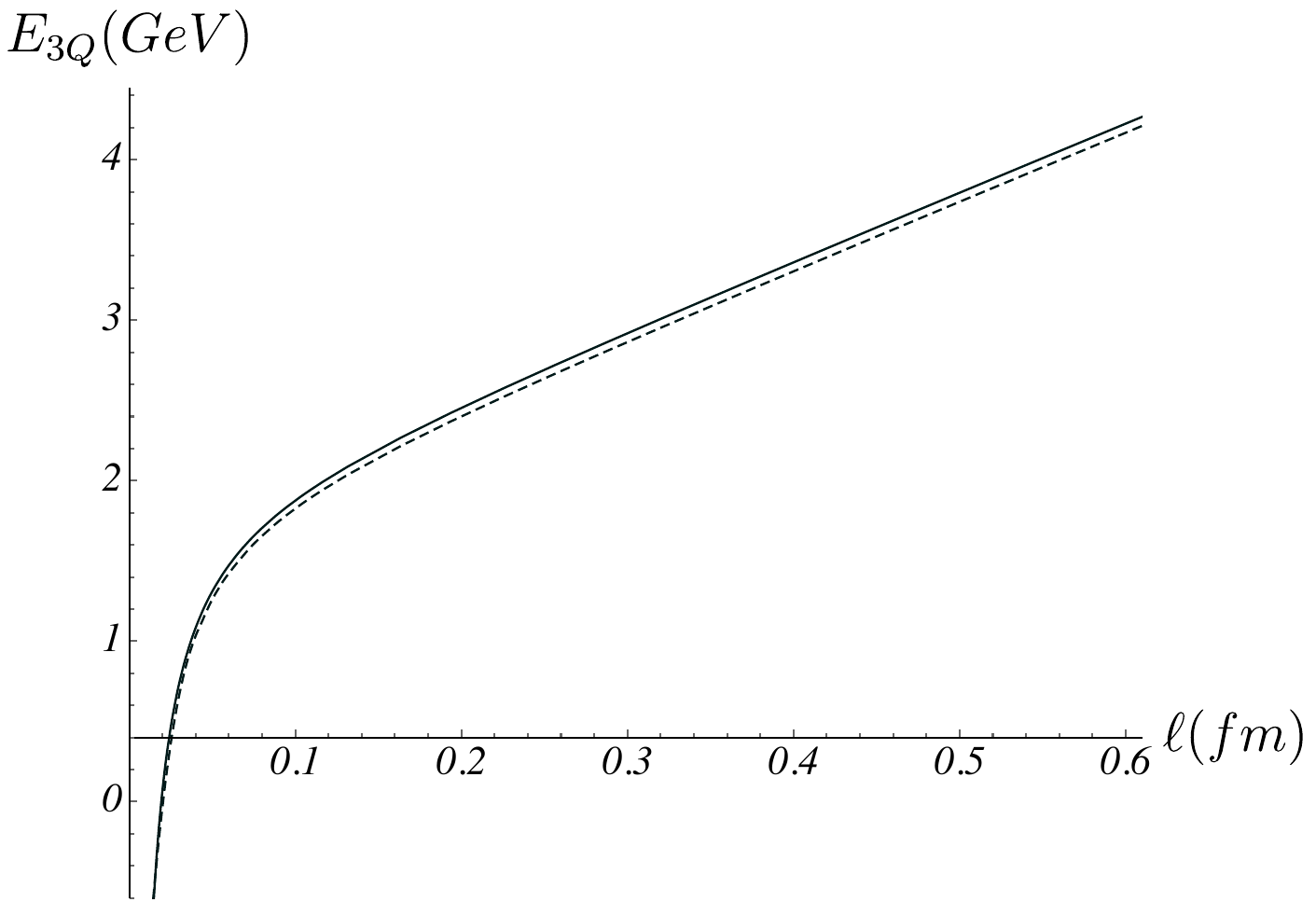}
\caption{{\small $E_{\3Q}$ vs $\ell$ for $\eta=\frac{1}{10},\frac{1}{3},\frac{1}{2}$ (left) and for $\eta=\frac{1}{3}$ (right). The dotted lines show the asymptotic behavior \eqref{Elc-large} and the dashed line represents the $\Delta$-law.}}
\label{Elcoll}
\end{figure}

As an illustration, the left panel of Figure \ref{Elcoll} shows the results of numerical computations. For large values of $\ell$, the curves 
asymptotically approach straight lines with the same constant term but different slopes, in agreement with Eq.\eqref{Ecola-large5}. Explicitly, this can be written as 

\begin{equation}\label{Elc-large}
	E_{\3Q}(\ell)=(1+\eta^{-1})\sigma\ell+C_{\3Q}+o(1)
	\,,
\end{equation} 
where $C_{\3Q}=C^{\,\text{\tiny A}}$. Notably, the $\Delta$-law yields the same leading term. Since it also provides a good approximation at small $\ell$, one may expect that the difference between the genuine three-quark potential and the $\Delta$-law is small.  To explore this, we compare the two in the right panel of the Figure, which plots the potential derived from the $\Delta$-law. The visible difference arises from the constant terms in the large-$\ell$ expansions and amounts to approximately $57\,\text{MeV}$, as indicated in Table II. This difference could potentially be determined with reasonable accuracy through numerical simulations.

Now let us discuss the behavior of the potential at small $\ell$. In this case, the corresponding asymptotics can be read off from Eqs.\eqref{Ecola-small} and \eqref{Ecolb-small}, yielding 

\begin{equation}\label{Elcol-small}
E_{\3Q}=-\frac{\alpha_{\3Q}}{\ell}+3c+O(\ell)
\,,\qquad	
\alpha_{\3Q}=
	\begin{cases}
		\alpha^{\,\text{\tiny A}} &\text{for}\quad\eta=\frac{1}{2}\,,\\
		\alpha^{\,\text{\tiny B}} &\text{for}\quad\eta=\frac{1}{3}\,,\frac{1}{10}
		\,.
\end{cases}
\end{equation}
The corresponding expression for the quark-quark coupling is

\begin{equation}\label{reduced-alphacol}
	\alpha_{\QQ}=\frac{1+\eta}{(1+\eta)^2+\eta}\alpha_{\3Q}
	\,.
\end{equation}
For a given value of $\eta$, the tangent angles and couplings are obtained numerically using the formulas in subsection A. The results are summarized in Table \ref{estimates4}. 
\begin{table*}[htbp]
\renewcommand{\arraystretch}{2}
\centering 	\scriptsize
\begin{tabular}{lccccccr}				
\hline
$\eta$  ~~~~~~~~& $\cos\alpha_1$~~&~~~~$\cos\alpha_3$~~~~&~~~$\alpha_{\3Q}$  ~~~&~~$\alpha_{\QQ}/\alpha_{\QQb}$  
\rule[-3mm]{0mm}{8mm}
\\
\hline \hline
$1/2$  &  0.999     &  0.677      &  0.221    &  0.4761    \\
$1/3$  &  0.986     &  0.822      &  0.190    &  0.4759    \\
$1/10$ &  0.936     &  0.918      &  0.142    &  0.4714     \\
 \hline \hline
\end{tabular}
\caption{ \small Estimates of the tangent angles and couplings in the small $\ell$ limit.}
\label{estimates4}
\end{table*}
At this point, a simple but important observation is worth mentioning. Unlike $C_{\3Q}$, the coupling $\alpha_{\QQ}$ depends on $\eta$. This suggests that $\alpha_{\QQ}$ has a more intricate dependence on the geometry of the triangle. 

\subsection{A remark on diquarks}

Having developed a technique for calculating the three-quark potentials, we can now shed some light on diquarks in the triply heavy baryons.\footnote{For a recent review on diquarks, see \cite{diquarks} and references therein.} A commonly used approximation is to treat a diquark as nearly point-like, so that it behaves similarly to an antiquark. This effectively reduces the genuine three-body interaction to a two-body one.  In the case of triply heavy baryons, this approximation implies that the three-quark potential can be approximated by a quark-antiquark potential. Of course, diquarks are not truly point-like objects. In what follows, we examine how accurate this approximation is within the framework of the current model.

Consider the collinear configuration of the heavy quarks shown in Figure \ref{coli}. As before, we are interested in the case $\ell \leq L$, but with one important modification: we now fix the length $\ell$ and allow $L$ to vary. The geometric constraint is then given by

\begin{equation}\label{gcd}
\ell=\ell^{\text{\,\tiny A}}\,,\quad\text{or}\quad
\ell=\ell^{\text{\,\tiny B}}
\,
\end{equation}
depending on whether the system is describe by configuration A or B. This constraint, combined with the force balance equations, allows us to write the potential $E_{\3Q}(L)$ in parametric form with a parameter $v$. Explicitly, 

\begin{equation}\label{Ed01}
L=\begin{cases}
L^{\text{\,\tiny A}} (v)& \text{for}\quad \text{v}_\ast \leq v\leq \vn\,,\\ 
L^{\text{\,\tiny B}}(v)& \text{for}\quad \vn\leq v\leq \vin\,,
\end{cases}
\qquad
E_{\3Q}=\begin{cases}
E^{\,\text{\tiny A}} (v)& \text{for}\quad \text{v}_\ast \leq v\leq\vn\,,\\
E^{\,\text{\tiny B}}(v)& \text{for}\quad \vn\leq v\leq \vin\,\,,
\end{cases}
\end{equation}
with $L^{\text{\,\tiny A,\tiny B}}=\frac{1}{\sqrt{\s}}\bigl({\cal L}^-(\lambda_2,v)-{\cal L}^+(\alpha_3,v)\bigr)$. The bounds on $v$ are determined by the Equations: $L(v)=\ell$ for $\text{v}_\ast$, $\alpha_1(v)=0$ for $\vn$, and $\lambda_2(v)=1$ for $\vin$. The second corresponds to a transition between the configurations, while the third to the limit of infinitely long string ending on $Q_2$. In contrast to the constraint \eqref{eta}, the constraint \eqref{gcd} does not prevent us from taking the limit $\lambda_2\rightarrow 1$ ($L\rightarrow\infty$) for configuration B.

\begin{figure}[htbp]
\centering
\includegraphics[width=7.8cm]{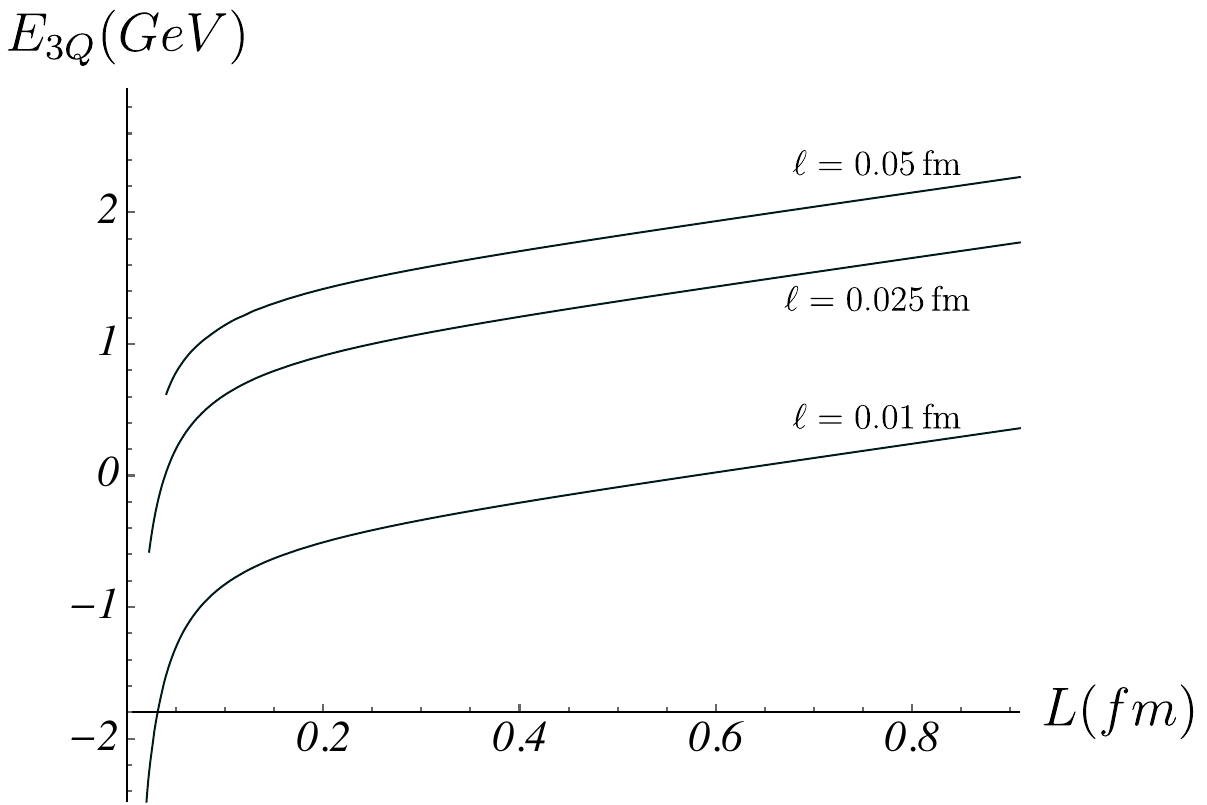}
\hspace{1.5cm}
\includegraphics[width=7cm]{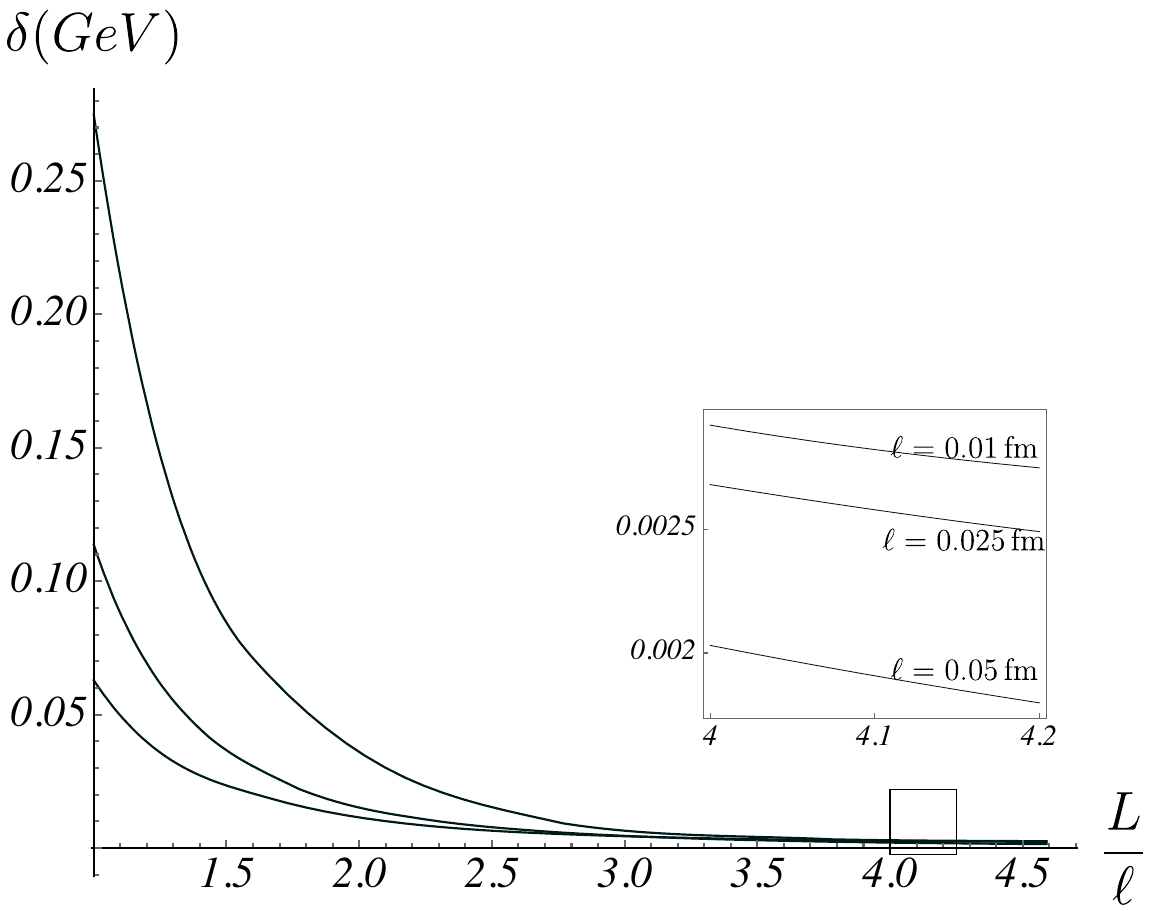}
\caption{{\small $E_{\3Q}$(left) and  $\delta$ (right) as a function of $L$ .}}
\label{Ediq}
\end{figure}

To proceed further, we consider $\ell$ to be small enough assuming that the diquark $[Q_1Q_3]$ can be regarded as nearly point-like. Our first goal is to examine the three-quark potential as a function of $L$. The corresponding plots are shown in the left panel of Figure \ref{Ediq}. As seen, $E_{\3Q}$ behaves linearly at large $L$ and becomes nearly Coulomb-like  
 at small $L$. This is qualitatively consistent with heavy quark-diquark symmetry, which in the limit $\ell \to 0$ gives \cite{a3Q2016}

\begin{equation}\label{Eqd}
	E_{\3Q}(\ell,L)=E_{\QQb}(L+\ell/2)+E_{\QQ}(\ell)
	\,,
	\qquad\text{with}\qquad
	E_{\QQ}(\ell)=-\frac{\alpha_{\QQ}}{\ell}+c+\sigma_0\ell+o(\ell)
	\,,
\end{equation}
 where $E_{\QQb}$ is the heavy quark-antiquark potential defined in Appendix C, $\alpha_{\QQ}=0.466\,\alpha_{\QQb}$, and $\sigma_0=0.785\,\sigma$. 

Our next goal is to quantify the difference between the potentials. To this end, we define a function of $L$

\begin{equation}\label{delta}
\delta=E_{\QQb}(L+\ell/2)-E_{\3Q}(\ell,L)+E_{\QQ}(\ell)
\,,
\end{equation}
as motivated by heavy quark-diquark symmetry.\footnote{Alternatively, the constant term in \eqref{delta} can be fixed by imposing $\delta=0$ as $L\rightarrow\infty$.}  Importantly, $\delta$ doesn't depend on the normalization constant $c$. Our results for $\delta$ are shown in the right panel of the Figure as a function of $L/\ell$. As expected, $\delta$ decreases with increasing $L$. Quantitatively, the value of $\delta$ drops below $3\,\text{MeV}$ for $L\gtrsim 4\ell$, suggesting that the diquark approximation is reasonable for phenomenological applications in the nonperturbative region.\footnote{However, it was reported that the accuracy of the diquark approximation diminishes  with increasing temperature \cite{bakry}.}  

 We conclude with a couple of brief remarks. First, the above analysis is applicable to the triply heavy baryons. Extending it to baryons containing light quarks, or to multiquark hadrons, remains a task for future work. Second, the analysis provides  at least an approximate estimate of finite-size effects, which are important for understanding the domain of validity of heavy quark-diquark symmetry \cite{wise}
 
\section{The infrared behavior of the three-quark potential}
\renewcommand{\theequation}{5.\arabic{equation}}
\setcounter{equation}{0}

We aim here to show that the main features of the infrared behavior of the three-quark potential hold generally, and are not limited to the specific examples discussed in Sec.III-IV.

For the analysis that follows, it is useful to decompose the vectors $\mathbf{e}$ in \eqref{fbe} as $\mathbf{e}=(\vec\ep, \ep_r)$, where $\ep_r$ is the radial component, and the remaining components are grouped into $\vec\ep$. If so, then $\vert\vec{\ep}\vert=\g w(r_v)\cos\alpha$ and $\ep_r=-\g w(r_v)\sin\alpha$, as follows from \eqref{e}. The force balance equation is written in component form as 

\begin{equation}\label{fbeg}
\vec\ep_1+\vec\ep_2+\vec\ep_3=0\,,\qquad
\sin\alpha_1+\sin\alpha_2+\sin\alpha_3=3\k(1+4v)\ep^{-3v}
\,.	
\end{equation}

\subsection{The leading correction to the $Y$-law}

\begin{figure}[htbp]
\centering
\includegraphics[width=6.25cm]{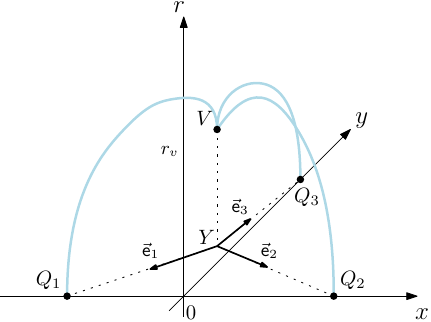}
\caption{{\small A string configuration relevant in the IR limit, with the negative tangent angles. The arrows on the $xy$-plane indicate the vectors $\vec\ep$.}}
\label{3Q5dY}
\end{figure}

Consider a triangle $Q_1Q_2Q_3$ whose largest angle doesn't exceed $\frac{2}{3}\pi$. In \cite{a3Q2008}, it was shown that the $Y$-law, more precisely the $\sigma L_{\text{min}}$ term, emerges in the limit of infinitely long strings, or equivalently, when all $\lambda_i\rightarrow 1$. This regime is referred to as the infrared  limit. The corresponding string configuration is shown in Figure \ref{3Q5dY}. We assume also that the largest angle is 
not very close to $\frac{2}{3}\pi$, so that the IR limit is described by a single string configuration. As explained in Appendix F, this assumption holds for our choice of parameters. The goal of the present section is to compute the leading correction to the $Y$-law. 

A string becomes infinitely long as its $\lambda$ parameter approaches $1$, as explained in Appendices A and B. In this limit, the tangent angle at the baryon vertex is given by $\cos\alpha=v\ep^{1-v}$, and it is the same for all the strings shown in the Figure. Because of this, the vectors $\vec{\ep}$ all have the same length. This implies that the angles between them must be $\frac{2}{3}\pi$ to satisfy the first equation in \eqref{fbeg}, and therefore, the point $Y$ coincides with the Fermat point $F$ of the triangle.\footnote{Note that this argument doesn't extend to triangles with angles greater or equal to $\frac{2}{3}\pi$, because in such cases, $F$ coincides with one of the triangle's vertices, and the corresponding string cannot be infinitely long.} Thus, we recover the $Y$-law. Moreover, from the second equation in \eqref{fbeg}, we find that $\sin\alpha=\k(1+4v)\ep^{-3v}$. Combining this with the above expression for $\cos\alpha$ leads directly to Eq.\eqref{v1} whose solution is $\vo$. The above derivation of the $Y$-law is more physical than that of \cite{a3Q2008}, where it was derived by solving a variational problem.

We now provide more detail. Using the formulas from Appendix B, we can show that 

\begin{equation}\label{Y}
\vert YQ_i\vert=\frac{1}{\sqrt{\s}}{\cal L}^-(\lambda_i,v)\,,\qquad
E_{\3Q}=\g\sqrt{\s}\Bigl(\sum_{i=1}^3{\cal E}^-(\lambda_i,v)+3\k\frac{\ep^{-2v}}{\sqrt{v}}\Bigr)+3c
\,.
\end{equation}
Taking the limit $\lambda_i\rightarrow 1$ and keeping only the leading terms, we get 

\begin{equation}\label{Y-large}
\vert FQ_i\vert=-\frac{1}{\sqrt{\s}}\ln(1-\lambda_i)+O(1)\,,\qquad
E_{\3Q}=-\g\ep\sqrt{\s}
\sum_{i=1}^3\ln(1-\lambda_i)\,+O(1)
\,.
\end{equation}
Here, we used the asymptotic expansions \eqref{L-y=1} and \eqref{E-y=1}, along with the fact that $Y\rightarrow F$. From this, it follows immediately that 

\begin{equation}\label{Y-large2}
E_{\3Q}=\sigma\sum_{i=1}^3\vert FQ_i\vert
+O(1)
\,
\end{equation}
which is exactly the $Y$-law, where $\sigma$ is the string tension and $\sum_{i=1}^3\vert FQ_i\vert$ is the minimal total length of the strings. It is noteworthy that the tension is the same as that in the $Q\bar Q$ system reviewed in Appendix C, indicating its universality.

To derive the leading correction to the $Y$-law, consider the difference 

\begin{equation}\label{Y-large3}
E_{\3Q}-\sigma\sum_{i=1}^3\vert YQ_i\vert
=\g\sqrt{\s}\Bigl(\sum_{i=1}^3{\cal E}^-(\lambda_i,v)-\ep{\cal L}^-(\lambda_i,v)\,
+
3\k\frac{\ep^{-2v}}{\sqrt{v}}\Bigr)+3c
\,.
\end{equation}
After taking the limit $\lambda_i\rightarrow 1$, we find that 

\begin{equation}\label{Y-large4}
E_{\3Q}-\sigma\sum_{i=1}^3\vert FQ_i\vert
=3\g\sqrt{\s}\Bigl(-{\cal I}(\vo)
+
\k\frac{\ep^{-2\vo}}{\sqrt{\vo}}\Bigr)+3c
\,.
\end{equation}
In the last step we used \eqref{fELy=1} evaluated at $x=\vo$. This gives the final result 

\begin{equation}\label{Y-large5}
E_{\3Q}=\sigma\sum_{i=1}^3\vert FQ_i\vert+C_{\3Q}+o(1)
\,,
\qquad
C_{\3Q}=3c-3\g\sqrt{\s}\Bigl({\cal I}(\vo)
-
\k\frac{\ep^{-2\vo}}{\sqrt{\vo}}\Bigr)
\,,
\end{equation}
with $C_{\3Q}$ a geometry independent constant. The conclusion is that the leading correction to the $Y$-law is universal for all triangles with angles less than $\frac{2}{3}\pi$. Note that the asymptotic expression \eqref{El2-large3} is a special case of \eqref{Y-large5}. 

We conclude with two remarks. First, we did not impose any geometrical constraints. The IR limit exists regardless of how the $\lambda$'s tend to $1$. This enhances the robustness of the derivation. Of course, one may impose such constraints and take the limit along specific paths. In Appendix F, we show how to do this in the case of fixed angles. Second, it follows from \eqref{Y-large5} that the difference between the constant terms in the UV and IR limits is scheme-independent. Specifically, 

 \begin{equation}\label{Cuv-ir}
3c-C_{\3Q}=3\g\sqrt{\s}\Bigl({\cal I}(\vo)
-
\k\frac{\ep^{-2\vo}}{\sqrt{\vo}}\Bigr)
\,.
\end{equation}
 For our parameter set, the right hand side is positive, with its value given in \eqref{Cuv-Cir}. This implies that $3c>C_{\3Q}$, or, in other words, the constant term in the IR limit is smaller than that in the UV limit.
\subsection{What happens when a triangle has an angle greater than $\frac{2}{3}\pi$?}
 
\begin{figure}[htbp]
\centering
\includegraphics[width=8cm]{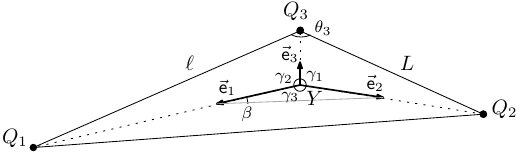}
\caption{{\small An obtuse triangle with side lengths $\ell$ and $L$. The $\gamma$'s are angles between the vectors $\vec\ep\,$'s, and $\beta$ is a base angle of the isosceles triangle formed by the vectors $\vec\ep_1$ and $\vec\ep_2$.}}
\label{3Q120}
\end{figure}

We now consider the opposite situation from before: the IR behavior of the three-quark potential in the case when one of the internal angles of the triangle exceeds $\frac{2}{3}\pi$. Without loss of generality, we assume this to be $\theta_3$, as shown in Figure \ref{3Q120}. We analyze the IR limit in which two of the strings become 
infinitely long, while the third remains finite. As depicted, the infinitely extended strings are those ending on $Q_1$ and $Q_2$, which corresponds to taking $\lambda_1,\,\lambda_2 \rightarrow 1$.

The string configuration is similar to that of Figure \ref{3Q5dY}, but here we must account for two possible choices for the string ending on $Q_3$. So, we have 

\begin{equation}\label{El120}
\vert YQ_i\vert=\frac{1}{\sqrt{\s}}{\cal L}^-(\lambda_i,v)\,,
\quad i=1,2\,,
\quad
\vert YQ_3\vert=\frac{1}{\sqrt{\s}}
{\cal L}^{\pm}(A,v)\,,
\quad
E_{\3Q}=\g\sqrt{\s}\Bigl(\sum_{i=1}^2{\cal E}^-(\lambda_i,v)+{\cal E}^{\pm}(A,v)
+
3\k\frac{\ep^{-2v}}{\sqrt{v}}\Bigr)+3c
\,,
\end{equation}
where $A=\alpha_3$ for the plus sign and $A=\lambda_3$ for the minus sign. Letting $\lambda_1,\,\lambda_2\rightarrow 1$ and using the asymptotic formulas of Appendix A, we find to leading order 

\begin{equation}\label{El1202}
\vert YQ_i\vert=-\frac{1}{\sqrt{\s}}\ln(1-\lambda_i)+O(1)\,,\quad i=1,2\,,
\qquad
E_{\3Q}=-\ep\g\sqrt{\s}\sum_{i=1}^2\ln(1-\lambda_i)
+O(1)
\,.
\end{equation}
We have omitted $\vert YQ_3\vert$ here as it does not contribute to the leading order. Since $\ell=\vert YQ_1\vert+O(1)$ and $L=\vert YQ_2\vert+O(1)$ in the limit where $\vert YQ_1\vert$ and $\vert YQ_2\vert$ go to infinity, we can rewrite \eqref{El1202} as  

\begin{equation}\label{El1203}
E_{\3Q}=\sigma(\ell+L)+O(1)
\,.
\end{equation}
This result aligns precisely with the predictions of four-dimensional string models in the IR limit. From the perspective of five-dimensional models, it complements the result of \cite{a3Q2008}, thereby filling a gap in the understanding of the IR behavior of the three-quark potential.

Before deriving the leading correction to \eqref{El1203}, we briefly examine the force balance equations in the IR limit, proceeding in three steps. First, note that $\cos\alpha_1=\cos\alpha_2=v\,\ep^{1-v}$ as $\lambda_1,\lambda_2\rightarrow 1$ and, as a consequence, $\vert\vec\ep_1\vert=\vert\vec\ep_2\vert$. This allows us to relate the angle $\beta$ to $\gamma_3$ via $\beta = (\pi - \gamma_3)/2$. Second, to determine $\gamma_3$, we compare leading terms in the expression $\vec{Q_3Q_1}\cdot\vec{Q_3Q_2}=(\vec{Q_3Y}+\vec{YQ_1})\cdot(\vec{Q_3Y}+\vec{YQ_2})$ as $\vert YQ_1\vert$ and $\vert YQ_2\vert$ become large. This leads to $\theta_3=\gamma_3$. Third, we decompose the $\vec\ep\,$'s into components parallel and perpendicular to $\vec\ep_3$. Since the left hand side of the first equation in \eqref{fbeg} has only one non-trivial component along $\vec\ep_3$, the force balance equations reduce to

\begin{equation}\label{fbe120}
	\cos\alpha_3=2\cos\frac{\theta_3}{2}v\ep^{1-v}
	\,,\qquad
	2\sin\alpha_1+\sin\alpha_3=3\k(1+4v)\ep^{-3v}
	\,.
\end{equation}
Combining these with the expression for $\cos\alpha_1$, we obtain either Eq.\eqref{v1m} or Eq.\eqref{v1p}, depending on whether $\alpha_3$ is negative or positive.  

To compute the leading correction to \eqref{El1203}, consider

\begin{equation}\label{EL1204}
E_{\3Q}-\sigma(\ell+L)
=\g\sqrt{\s}\Bigl(\sum_{i=1}^2{\cal E}^-(\lambda_i,v)-\ep{\cal L}^-(\lambda_i,v)\,
+
{\cal E}^{\pm}(A,v)+2\ep\cos\gamma_1{\cal L}^{\pm}(A,v)
+
3\k\frac{\ep^{-2v}}{\sqrt{v}}\Bigr)+3c
\,.
\end{equation}
In deriving this, we used the cosine rule for large $\vert YQ_1\vert$ and $\vert YQ_2\vert$, which gives $\ell=\vert YQ_1\vert-\vert YQ_3\vert\cos\gamma_2+o(1)$ and $L=\vert YQ_2\vert-\vert YQ_3\vert\cos\gamma_1+o(1)$, along with the fact that $\gamma_1=\gamma_2$. The latter follows from the first equation in \eqref{fbeg} if $\vert\vec\ep_1\vert=\vert\vec\ep_2\vert$. After taking the limit $\lambda_1,\,\lambda_2\rightarrow 1$, we get 

\begin{equation}\label{EL1205}
E_{\3Q}-\sigma(\ell+L)
=\g\sqrt{\s}\Bigl(
-2{\cal I}(\text{v})
+
{\cal E}^{\pm}(A,\text{v})-2\ep\cos\frac{\theta_3}{2}{\cal L}^{\pm}(A,\text{v})
+
3\k\frac{\ep^{-2\text{v}}}{\sqrt{\text{v}}}\Bigr)+3c+o(1)
\,,
\end{equation}
 where $\text{v}=\vop$ for the plus sign and $\text{v}=\vom$ for the minus sign. In the last step, we used the relation $\gamma_1=\pi-\frac{\theta_3}{2}$. An interesting observation here is that the right hand side depends only on the obtuse angle $\theta_3$, and  not on $\ell$ or $L$. The result for collinear geometry in Sec.IV is a special case of this general behavior.
 
So far we have considered the IR limit without referring to the parameter values. More specifically, as seen in Figure \ref{v1pm}, the solution $\vom$ exists only in a tiny interval near $\theta_3=\frac{2}{3}\pi$. This implies that outside this interval, the IR limit is governed by a single string configuration with $\alpha_3>0$. The potential thus behaves as
 
 \begin{equation}\label{EL1206}
E_{\3Q}=\sigma\bigl(\vert Q_1Q_3\vert+\vert Q_2Q_3\vert\bigr)
+
C_{\3Q}+o(1)
\,,
\quad
C_{\3Q}
=
3c
-
\g\sqrt{\s}\Bigl(
2{\cal I}(\vop)-{\cal E}^+(\alpha_3,\vop)
+
2\ep\cos\tfrac{\theta_3}{2}{\cal L}^+(\alpha_3,\vop)
-
3\k\frac{\ep^{-2\vop}}{\sqrt{\vop}}
\Bigr)
\,.
\end{equation}
Unlike the $Y$-law, the constant term is not universal, as it depends explicitly on the value of the obtuse angle. Note also that the asymptotic expressions \eqref{El3-large4} and \eqref{Elc-large} are special cases of \eqref{EL1206}. 

\begin{figure}[htbp]
 \includegraphics[width=6.5cm]{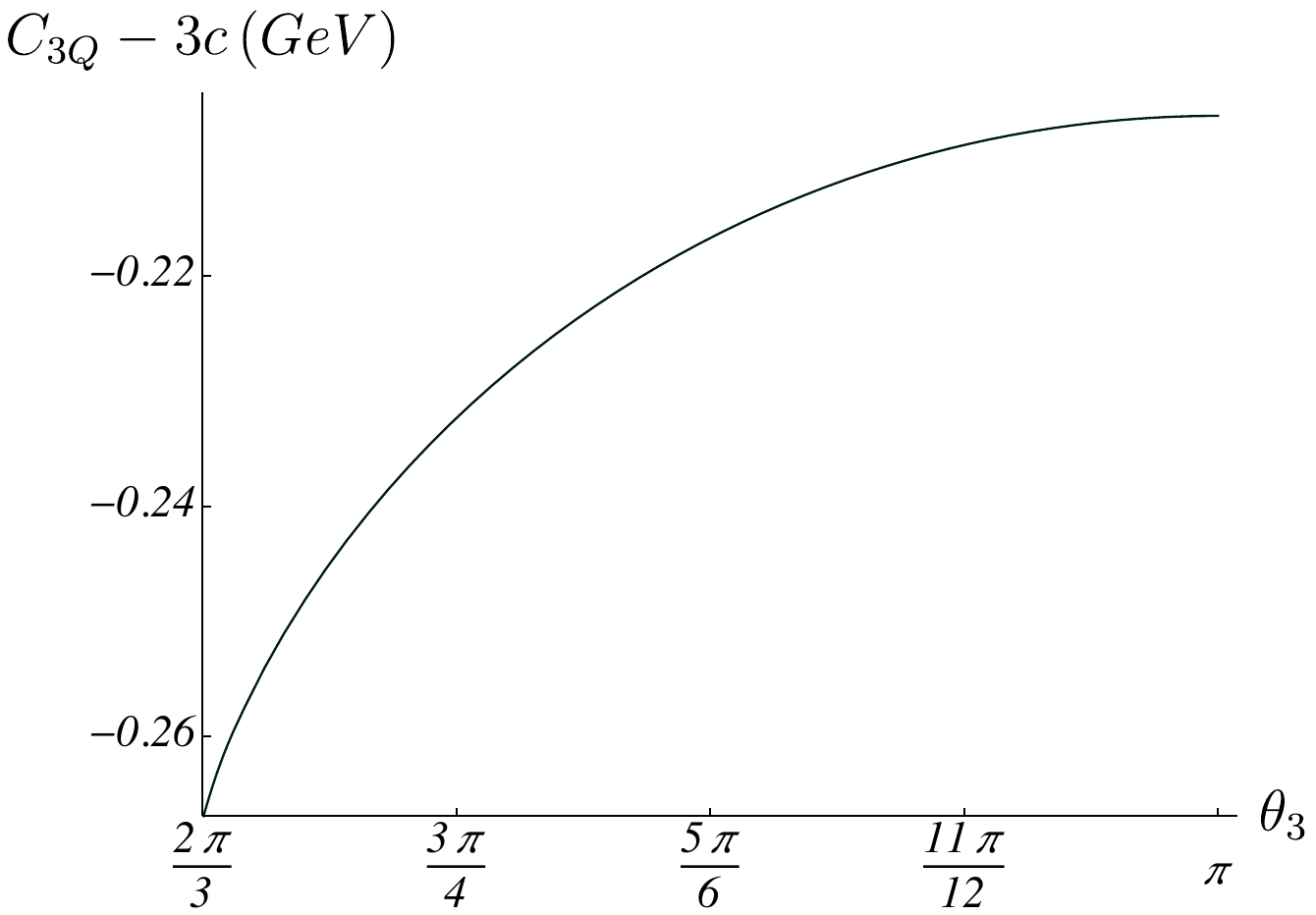}
\caption{{\small The value of $C_{\3Q}$ as a function of the largest angle of the triangle $Q_1Q_2Q_3$. }}
\label{CIR}
\end{figure}

A few remarks are in order. First, it is instructive to analyze the $\theta$-dependence of $C_{\3Q}$ in detail. As shown in Figure \ref{CIR}, $C_{\3Q}$ increases with $\theta_3$ and reaches its maximum at $\theta_3 = \pi$.
Second, as before, we have not imposed any geometrical constraints. The IR limit exists independently of the manner in which $\lambda_1$ and $\lambda_2$ approach $1$ that enhances the robustness of the derivation. Of course, one could impose constraints to simplify the analysis. In Appendix E we examine the case where the ratio $\ell/L$ is held fixed. Third, unlike in the four-dimensional string models, the point $Y$ does not coincides with the vertex $Q_3$ playing the role of the Fermat point in this case. Even in the IR limit, it remains at a finite distance from $Q_3$ unless $\theta_3=\pi$. Nonetheless, the leading terms in the potential are the same in both types of models, as they come from the two infinitely long strings. The third string being finite contributes only to subleading corrections.

\section{Conclusions}

We conclude our discussion with several remarks. 

(1) It is widely accepted that at small quark separations, excluding extremely small ones where quarks behave almost freely due to asymptotic freedom, the three-quark potential is well described by the $\Delta$-law, while at larger separations, it follows the $Y$-law. In \cite{a3Q2016}, it was shown that for the equilateral triangle geometry the current model smoothly interpolates between these two laws. In this paper, we have provided further evidence supporting this behavior by examining the isosceles and collinear geometries. We also demonstrated that in the current model the string tension is universal (geometry independent). 

However, some challenging questions remain. First, in general, the point $Y$ (the projection of a baryon vertex on the boundary) does not coincide with the Fermat point $F$, as might be expected from the four-dimensional string models. The only exception is the IR limit, where $Y=F$. Why is this the case? The explanation lies in the interpretation of the five-dimensional model as an effective four-dimensional one, but with strings having different tensions. When the tensions of three strings differ, the first equation in Eq.\eqref{fbeg} no longer implies $Y=F$. As explained in Sec.V, only in the IR limit their tensions become equal, leading to $Y=F$. By the same reasoning, a fat string in QCD can be viewed as a collection of thin strings with different tensions \cite{aFlux}. Second, the model in question is an effective string model based on the three-quark interaction in five dimensions. Given that, why does it well approximate the $\Delta$-law in four dimensions, which involves only pairwise (two-quark) interactions? We do not have a satisfactory answer to the second question. 

(2) As discussed in Sec.V, in the IR limit the three-quark potential exhibits two different behaviors depending on the value of the largest angle in the triangle formed by the heavy quarks.\footnote{We assume that none of the angles is very close to $\frac{2}{3}\pi$.} The first regime occurs when the largest angle is less than $\frac{2}{3}\pi$. In this case the asymptotic behavior is given by \eqref{Y-large5}, where the leading term represents the $Y$-law and the subleading term is a universal constant. The second regime occurs when the largest angle exceeds $\frac{2}{3}\pi$. In this case, the asymptotic expression is provided by \eqref{EL1206}, with the subleading term depending on the geometry of the quark configuration. The reason for such a difference is that in the first regime all the strings become infinitely long and, therefore, are symmetric under permutations of the quarks, while in the second regime only two strings are infinitely long, reducing the symmetry to permutations of just two quarks.

\begin{figure}[htbp]
\centering
\includegraphics[width=4.25cm]{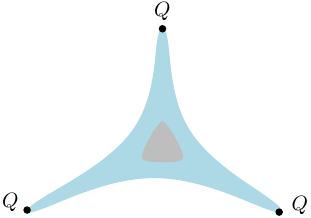}
\hspace{2.5cm}
\includegraphics[width=2.75cm]{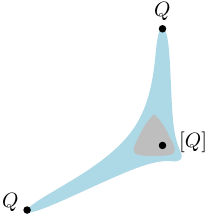}
\caption{{\small Schematic illustration of gluon flux tubes in baryons. Left: The $Y$-shaped flux structure with a junction shown in gray, as observed in lattice QCD \cite{flux}. Right: The flux configuration in the case of large angles, exceeding $\frac{2}{3}\pi$. $[Q]$ denotes a heavy quark dressed by gluons.}}
\label{gflux}
\end{figure}

This symmetry breaking has an important implication, namely the emergence of a heavy object transforming in the two-index antisymmetric representation of gauge group, which we  denote by $[Q]$. In perturbative QCD, a quark $Q_\alpha$ transforms in the fundamental representation of $SU(3)$ or, equivalently, in its two-index antisymmetric representation $Q^{\alpha\beta}=\varepsilon^{\alpha\beta\gamma}Q_{\gamma}$. To gain insight into the strong coupling regime, it is useful to draw an analogy with baryonic Wilson loops. Each of loops is constructed from three path-ordered Wilson exponentials joined at two junctions involving the epsilon tensors. In the QCD string picture, these exponentials correspond to gluon flux-tubes (fat strings), while the junctions correspond to dense gluon cores (string junctions), as illustrated in Figure \ref{gflux} on the left. A quark, when inside the core, becomes
a composite object dressed by gluons, as in the Figure on the right, and due to the epsilon tensor transforms in the two-index antisymmetric representation. As we saw in Sec.V, when the largest angle exceeds $\frac{2}{3}\pi$, one of the quarks does approach the baryon vertex in the IR limit, while the other recede. Using $[Q]$, the leading term in Eq.\eqref{EL1206} can be immediately recovered as coming from two strings stretched between the quark sources. Notably, a similar construction is possible with an antiquark. In the context of tetraquarks, a $[Q][\bar Q]$ pair called the $Q\bar Q$ seeding was suggested in \cite{isgur} and recently found in the current model \cite{aQQqq}. 

  However, the transition between the two regimes is puzzling. The results in Appendix F suggest that within a very narrow angular range near $\frac{2}{3}\pi$, all basic string configurations may contribute to the three-quark potential. But it is unclear how to make this precise. An alternative idea would be to investigate this range using closed string constructions \cite{zk} or lattice simulations.
 
 (3) The issue of the subtracted potential \cite{komas} can be addressed within the current model. The key point is that the constant terms differ in the UV and IR limits.\footnote{This discrepancy was first found in the case of the equilateral triangle geometry \cite{a3Q2016}.} If, therefore, $E_{\3Q}-\sigma L_{\text{min}}-3c$ does not vanish in the IR limit, $E_{\3Q}-\sigma L_{\text{min}}-C_{\3Q}$ does. As seen from Figure 17 of \cite{komas}, the minimum discrepancy (in absolute value) arises in the collinear geometry, while the maximum in several geometries where the largest angle does not exceed $\frac{2}{3}\pi$.\footnote{Due to limited data, it is not clear what happens with the obtuse-wide geometry considered in \cite{komas}. Our model predicts that its discrepancy value lies between the two extrema.} This agrees with the result shown in Figure \ref{CIR} as the discrepancy originates from the difference $C_{\3Q}-3c$. There is another important implication. We saw in Sec.IV that in the case of the collinear geometry, the potential $E_{\3Q}$ deviates from the $\Delta$-law. The deviation arises because the difference $C_{\3Q}-\frac{3}{2}C_{\QQb}$ is non-zero. Note that such a difference identically vanishes in the UV limit, where $3c-\frac{3}{2}\cdot 2c=0$.

It is worth mentioning another issue related to lattice QCD: the quark-quark coupling $\alpha_{\QQ}$. It has been known that the lattice data are compatible with $\alpha_{\QQ}\simeq\oh\alpha_{\QQb}$ \cite{suga}. The results of Secs. III and IV indicate that this coupling exhibits a weak dependence on geometry (quark positions). With continued progress, high-precision simulations should soon be feasible, allowing for direct comparison with our predictions.

(4) In summary, understanding the properties of baryons and the interactions between quarks within them is of primary importance. The $QQQ$ system presents a rich and complex landscape of physics, with many pressing questions that remain unanswered. Advancing our theoretical understanding and its applications to hadron phenomenology will require a concerted effort from the high-energy physics community. We hope this study provides important insights and inspiration for future research.

\begin{acknowledgments}
We would like to thank P. de Forcrand, Z. Komargodski, M. Shifman, P.Weisz, and S. Zhong for useful discussions and correspondence concerning this topic. This work was conducted as part of Program FFWR-2024-0011 at the Landau Institute.
\end{acknowledgments}
\appendix
\section{Notation and useful formulas}
\renewcommand{\theequation}{A.\arabic{equation}}
\setcounter{equation}{0}

Throughout the paper, we denote heavy quarks by $Q$ and a baryon vertex in five-dimensional space by $V$. We locate the quarks on the boundary of the five-dimensional space at $r=0$, and the vertex in the interior at $r=\rv$. For convenience, we introduce a dimensionless variable: $v=\s\rv^2$, which ranges from 0 to 1 and indicates the proximity of the vertex to the soft-wall located at 1 in these units. 

To express the resulting formulas compactly, we make use of the following basic functions \cite{astbr3Q}:

\begin{equation}\label{fL+}
{\cal L}^+(\alpha,x)=\cos\alpha\sqrt{x}\int^1_0 du\, u^2\, \ep^{x (1-u^2)}
\Bigl[1-\cos^2{}\hspace{-1mm}\alpha\, u^4\ep^{2x(1-u^2)}\Bigr]^{-\frac{1}{2}}
\,,
\qquad
0\leq\alpha\leq\frac{\pi}{2}\,,
\qquad 
0\leq x\leq 1
\,.
\end{equation}
${\cal L}^+$ is a non-negative function that vanishes when $\alpha=\frac{\pi}{2}$ or $x=0$. It exhibits a singularity at $(0,1)$. If $\alpha$ approaches $\alpha_0$ as $x$ approaches $0$, then the behavior of ${\cal L}^+$ for small $x$ is 

\begin{equation}\label{fL+smallx}
{\cal L}^+(\alpha,x)=\sqrt{x}\bigl({\cal L}^+_0(\alpha_0)+O(x)\bigr)
\,,
\qquad\text{with}\qquad
{\cal L}^+_0(\alpha_0)=\frac{1}{4}
\cos^{-\oh}\hspace{-.9mm}\alpha_0\,B\bigl(\cos^2\hspace{-.9mm}\alpha_0;\tfrac{3}{4},\tfrac{1}{2}\bigr)
\,.
\end{equation}
Here $B(z;a,b)$ is the incomplete beta function;

 \begin{equation}\label{fL-}
{\cal L}^-(y,x)=\sqrt{y}
\biggl(\,
\int^1_0 du\, u^2\, \ep^{y(1-u^2)}
\Bigl[1-u^4\,\ep^{2y(1-u^2)}\Bigr]^{-\frac{1}{2}}
+
\int^1_
{\sqrt{\frac{x}{y}}} 
du\, u^2\, \ep^{y(1-u^2)}
\Bigl[1-u^4\,\ep^{2y(1-u^2)}\Bigr]^{-\frac{1}{2}}
\,\biggr)
\,,
\quad
0\leq x\leq y\leq 1
\,.
\end{equation}
 This function is non-negative and vanishes at the origin. It becomes singular at $y=1$, where  

\begin{equation}\label{L-y=1}
{\cal L}^-(y,x)=-\ln(1-y)+O(1)
\quad
\text{for fixed $x$}
\,.
\end{equation}
If $y=x/\rho$ with $0<\rho\leq 1$ as $x\rightarrow 0$, then the small-$x$ behavior of ${\cal L}^-$ is 

\begin{equation}\label{L-y=0}
	{\cal L}^-(y,x)=
	\sqrt{x}\bigl({\cal L}^-_0(\rho)+O(x)\bigr)
	\,,\qquad\text{with}\qquad
	{\cal L}^-_0(\rho)=\frac{1}{4}\rho^{-\oh}{\cal B}\bigl(1-\rho^2;\tfrac{1}{2},\tfrac{3}{4}\bigr)
\,.
\end{equation}
Here ${\cal B}(z;a,b)=B(a,b)+B(z;a,b)$. The ${\cal L}$ functions are related by 

\begin{equation}\label{L-L+}
{\cal L}^-(y,y)={\cal L}^+(0,y)
\,,
\qquad
{\cal L}^-(y,0)=2{\cal L}^+(0,y)
	\,;
\end{equation}

\begin{equation}\label{fE+}
{\cal E}^+(\alpha,x)=
\frac{1}{\sqrt{x}}
\int^1_0\,\frac{du}{u^2}\,\biggl(\ep^{x u^2}
\Bigl[
1-\cos^2{}\hspace{-1mm}\alpha\,u^4\ep^{2x (1-u^2)}
\Bigr]^{-\frac{1}{2}}-1-u^2\biggr)
\,,
\qquad
0\leq\alpha\leq\frac{\pi}{2}\,,
\qquad 
0\leq x\leq 1
\,.
\end{equation}
The function ${\cal E}^+$ is singular at $x=0$ and at the point $(0,1)$. If $\alpha$ approaches $\alpha_0$ as $x\rightarrow 0$, then the small-$x$ behavior of ${\cal E}^+$ is 

\begin{equation}\label{fE+smallx}
{\cal E}^+(\alpha,x)=\frac{1}{\sqrt{x}}\bigl({\cal E}^+_0(\alpha_0)+O(x)\bigr)
\,,
\qquad\text{with}\qquad
{\cal E}_0^+(\alpha_0)=
\frac{1}{4}
\cos^{\oh}\hspace{-.9mm}\alpha_0\,B\bigl(\cos^2\hspace{-.9mm}\alpha_0;-\tfrac{1}{4},\tfrac{1}{2}\bigr)
\,;
\end{equation}

\begin{equation}\label{fE-}
{\cal E}^-(y,x)=\frac{1}{\sqrt{y}}
\biggl(
\int^1_0\,\frac{du}{u^2}\,
\Bigl(\ep^{y u^2}\Bigl[1-u^4\,\ep^{2y(1-u^2)}\Bigr]^{-\frac{1}{2}}
-1-u^2\Bigr)
+
\int^1_{\sqrt{\frac{x}{y}}}\,\frac{du}{u^2}\,\ep^{y u^2}
\Bigl[1-u^4\,\ep^{2y(1-u^2)}\Bigr]^{-\frac{1}{2}}
\biggr) 
\,,
\,\,\,
0\leq x\leq y\leq 1
\,.
\end{equation}
The function ${\cal E}^-$ is singular at $(0,0)$ and at $y=1$. More precisely, as $y$ approaches $1$ with $x$ held fixed, it behaves like

\begin{equation}\label{E-y=1}
	{\cal E}^-(y,x)=-\ep\ln(1-y)+O(1)
	\,.
\end{equation}
If $y=x/\rho$ with $0<\rho\leq 1$ as $x\rightarrow 0$, then the small-$x$ behavior of ${\cal E}^-$ is 
\begin{equation}\label{E-y=0}
	{\cal E}^-(y,x)=
	\frac{1}{\sqrt{x}}
	\Bigl({\cal E}^-_0(\rho)+O(x)\Bigr)
	\,,\qquad\text{with}\qquad
	{\cal E}^-_0(\rho)=\frac{1}{4}\rho^{\oh}{\cal B}\bigl(1-\rho^2;\tfrac{1}{2},-\tfrac{1}{4}\bigr)
\,.
\end{equation}

The ${\cal E}$ functions are related by 

\begin{equation}
{\cal E}^-(y,y)={\cal E}^+(0,y)
\,,
\qquad
{\cal E}^-(y,x)=\frac{1}{\sqrt{x}}+2{\cal E}^+(0,y)+O(\sqrt{x})
\quad\text{as}\quad x\rightarrow 0\,;
\end{equation}

\begin{equation}\label{Q}
{\cal Q}(x)=\sqrt{\pi}\text{erfi}(\sqrt{x})-\frac{\ep^x}{\sqrt{x}}
\,.
\end{equation}
Here $\text{erfi}(x)$ denotes the imaginary error function. The functions ${\cal Q}(x)$ and ${\cal E}^+(\alpha,x)$ are related by 

\begin{equation}\label{QL+}
{\cal E}^+(\tfrac{\pi}{2},v)={\cal Q}(v)
\,.	
\end{equation}
Also, a useful fact is that its small-$x$ behavior is given by 

\begin{equation}\label{Q0}
{\cal Q}(x)=-\frac{1}{\sqrt{x}}\bigl(1-x+O(x^2)\bigr)
\,;
\end{equation}

\begin{equation}\label{I}
	{\cal I}(x)=
	I_0
	-
	\int_{\sqrt{x}}^1\frac{du}{u^2}\ep^{u^2}\Bigl[1-u^4\ep^{2(1-u^2)}\Bigr]^{\frac{1}{2}}
	\,,
\quad\text{with}\quad 
I_0=\int_0^1\frac{du}{u^2}\Bigl(1+u^2-\ep^{u^2}\Bigl[1-u^4\ep^{2(1-u^2)}\Bigr]^{\frac{1}{2}}\Bigr)
\,,
\qquad
0< x\leq 1
\,.
\end{equation}
Numerically, $I_0=0.751$. This function is increasing and vanishes at $x=0.278$. It is related to the functions ${\cal L}^-$ and $ {\cal E}^-$ via

\begin{equation}\label{fELy=1}
	{\cal I}(x)=\ep\,{\cal L}^-(y,x)-{\cal E}^-(y,x)
	\quad\text{as}\quad y\rightarrow 1\quad\text{with}\quad x\quad\text{fixed}
	\,.
	\end{equation}

\section{A static Nambu-Goto string with fixed endpoints}
\renewcommand{\theequation}{B.\arabic{equation}}
\setcounter{equation}{0}

The purpose of this Appendix is to describe some facts about a static Nambu-Goto string in the curved geometry \eqref{metric}, which are essential for constructing string configurations in Sec.III-V. Much of this material is not novel and can be found in \cite{a3Q2016}, whose notation we primarily adopt. 

\begin{figure}[htbp]
\centering
\includegraphics[width=5.25cm]{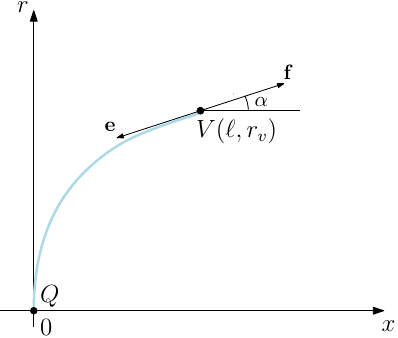}
\hspace{3cm}
\includegraphics[width=5.25cm]{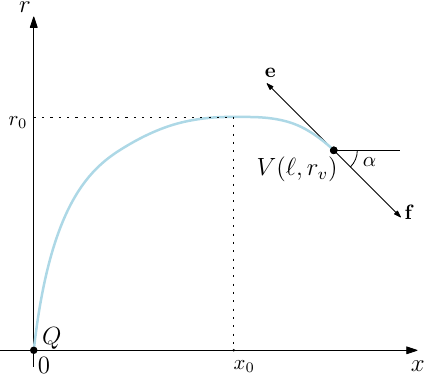}
\caption{{\small A string stretched between two points. $\alpha$ denotes the tangent angle, and the arrows denote two forces acting at point $V$. Here $\rv<1/\sqrt{\s}$. Left: The case $\alpha>0$. Right: The case $\alpha<0$.}}
\label{ngs}
\end{figure}

Consider a static string stretched between two fixed points, $Q(0,0)$ and $V(\ell,\rv)$, in the $xr$-plane, as depicted in Figure \ref{ngs}. In static gauge, where $\xi^1=t$ and $\xi^2=x$, the boundary conditions for $r(x)$ are 

\begin{equation}\label{string-bc}
r(0)=0
\,,\qquad
r(\ell)=\rv
\,.
\end{equation}
The Nambu-Goto action takes the form

\begin{equation}\label{NG2}
S=T\g\int_{0}^{\ell} dx\,w(r)\sqrt{1+(\partial_x r)^2}
\,,
\qquad
w(r)=\frac{\ep^{\s r^2}}{r^2}
\,.
\end{equation}
For convenience, we use the shorthand notation $\g=\frac{R^2}{2\pi\alpha'}$, $T=\int dt$, and $\partial_x r=\frac{\partial r}{\partial x}$. Since the integrand does not depend explicitly on $x$, the equation of motion admits a first integral

\begin{equation}\label{Int}
I=\frac{w(r)}{\sqrt{1+(\partial_x r)^2}}\,.
\end{equation}
At point $V$, it can be written as

\begin{equation}\label{I-PB}
I=w(\rv)\cos\alpha
\,,
\end{equation}
where $\tan\alpha=\partial_xr\vert_{x=\ell}$ and $\alpha\in[-\frac{\pi}{2},\frac{\pi}{2}]$.

In general, the tangent angle $\alpha$ may be positive or negative. For $\alpha>0$, the function $r(x)$ describing a string profile is monotonically increasing on the interval $[0,\ell]$. Conversely, for $\alpha<0$, the situation is more intricate. The function $r(x)$ is increasing on $[0,x_{\text{\tiny 0}}]$ and decreasing on $[x_{\text{\tiny 0}},\ell]$, reaching a local maximum at $x=\xo$. Both cases are depicted in Figure \ref{ngs}. 

 For our purposes in this paper, point $Q$ is associated with an infinitely heavy quark, while point $V$ with a baryon vertex. Therefore, it makes sense to consider forces acting on point $V$ to maintain equilibrium. The force balance equation is simply 

\begin{equation}\label{fbeNG}
\mathbf{e}+\mathbf{f}=0
\,,
\end{equation}
where $\mathbf{e}$ is the string tension, and $\mathbf{f}$ is an external force. It is straightforward to compute the $r$-component of $\mathbf{e}$ which arises from the boundary term in the variation of the action. Indeed, with $\delta r\vert_{x=\ell}=\delta\rv$, we find that 

\begin{equation}\label{NG3}
\delta S=T\g \frac{w(r)\partial_x r}{\sqrt{1+(\partial_x r)^2}}\delta\rv
\,
\end{equation}
that implies $\mathbf{e}_r=-T^{-1}\delta S/\delta\rv=-\g w(\rv)\sin\alpha$. Similarly, in static gauge $\xi^1=t$ and $\xi^2=r$, the boundary term provides the $x$-component of $\mathbf{e}$: $\mathbf{e}_x=-\g w(\rv)\cos\alpha$.\footnote{Alternatively, one can choose a gauge in which the boundary terms provide both components \cite{a3Q2016}.} Putting both components together, we have 

\begin{equation}\label{e}
\mathbf{e}=-\g w(\rv)(\cos\alpha,\sin\alpha)
\,.
\end{equation}
 A simple but important observation is that the magnitude of the tension is determined by a radial coordinate, specifically $\lvert\mathbf{e}\rvert=\g w(r)$. From this point of view, the physical string tension $\sigma$ is equal to $\lvert\mathbf{e}\rvert$ evaluated at $r=1/\sqrt{\s}$, corresponding to the position of the soft wall. 
 
\renewcommand \thesubsubsection {\arabic{subsubsection}}
\subsubsection{The case $\alpha\geq 0$}

Let us examine both cases more systematically, beginning with $\alpha\geq 0$. First, we express $I$ in terms of $\alpha$ and $\rv$. This leads to the differential equation $w(\rv)\cos\alpha=w(r)/\sqrt{1+(\partial_x r)^2}$ which can be integrated over the variables $x$ and $r$. Using the boundary conditions \eqref{string-bc} then gives 

\begin{equation}\label{l+}
\ell
=
\cos\alpha\sqrt{\frac{v}{\s}}\int^1_0 du\, u^2\, \ep^{v (1-u^2)}
\Bigl(1-\cos^2{}\hspace{-1mm}\alpha\, u^4\ep^{2v(1-u^2)}\Bigr)^{-\frac{1}{2}}
=
\frac{1}{\sqrt{\s}}{\cal L}^+(\alpha,v)
\,,
\end{equation}
where $v=\s\rv^2$, and the function ${\cal L}^+$ is defined in Appendix A.

To compute the string energy, we reduce the integral over $x$ in $S$ to an integral over $r$ using the differential equation. Since the resulting expression diverges at $r=0$, we regularize it by imposing a lower cutoff $\epsilon$. In this way, we get 

\begin{equation}\label{E+reg}
E_{R}=\frac{S_R}{T}=
\g\sqrt{\frac{\s}{v}}\int^1_{\sqrt{\tfrac{\s}{v}}\epsilon}\,\frac{du}{u^2}\,\ep^{v u^2}
\Bigl[1-\cos^2{}\hspace{-1mm}\alpha\,u^4\,\ep^{2v(1-u^2)}\Bigr]^{-\frac{1}{2}}
\,.
\end{equation}
$E_R$ behaves for $\epsilon\rightarrow 0$ as  

\begin{equation}\label{E+R}
E_R=\frac{\g}{\epsilon}+E+O(\epsilon)\,.
\end{equation}
So, subtracting the $\tfrac{1}{\epsilon}$ term and letting $\epsilon=0$, we obtain a finite expression for the energy

\begin{equation}\label{E+}
E
=
\g\sqrt{\frac{\s}{v}}
\int^1_0\,\frac{du}{u^2}\,\biggl(\ep^{v u^2}
\Bigl(1-\cos^2{}\hspace{-1mm}\alpha\,u^4\,\ep^{2v(1-u^2)}\Bigr)^{-\frac{1}{2}}-1-u^2\biggr)
+c
=
\g\sqrt{\s}\,{\cal E}^+(v,\alpha)+c
\,.
\end{equation}
Here the function ${\cal E}^+$ is defined in \eqref{fE+}, and $c$ is a normalization constant. 

It is worth noting that for $\alpha=\frac{\pi}{2}$, the above expressions reduce to

\begin{equation}\label{Evert}
\ell=0\,,
\qquad
E=\g\sqrt{\s}{\cal Q}(v)+c\,,
\end{equation}
where the function ${\cal Q}$ is given in Appendix A. In this special case, the string is stretched entirely along the radial direction. 

\subsubsection{The case $\alpha\leq 0$}

The analysis from the previous subsection generalizes straightforwardly to the case $\alpha\leq 0$. A key point, relevant to all expressions below, is that the string configuration involves two segments: one over the interval $[0,\xo]$, where $r(x)$ increases, and another over the interval $[\xo,\ell]$, where $r(x)$ decreases (see the Figure). 

First, we define the first integral at $r=\ro$ so that $I=w(\ro)$, and then integrate the differential equation over both intervals. As a result, we find

\begin{equation}\label{l-}
\ell
=
\sqrt{\frac{\lambda}{\s}}\biggl[
\int^1_0 du\, u^2\, \ep^{\lambda(1-u^2)}
\Bigl(1-u^4\,\ep^{2\lambda(1-u^2)}\Bigr)^{-\frac{1}{2}}
+
\int^1_
{\sqrt{\frac{v}{\lambda}}} 
du\, u^2\, \ep^{\lambda(1-u^2)}
\Bigl(1-u^4\,\ep^{2\lambda(1-u^2)}\Bigr)^{-\frac{1}{2}}
\biggr]
=
\frac{1}{\sqrt{\s}}{\cal L}^-(\lambda,v)
\,.
\end{equation}
Here $\lambda=\s\ro^2$, and the function ${\cal L}^-$ is given by \eqref{fL-}. It is important to note that $\lambda$, $v$, and $\alpha$ are not independent. From the first integral it follows that 

\begin{equation}\label{v-lambda}
\frac{\ep^{\lambda}}{\lambda}=\frac{\ep^{v}}{v}\cos\alpha
\,
\end{equation}
which allows us to express $\lambda$ in terms of $v$ and $\alpha$ as

\begin{equation}\label{lambda}
\lambda=-\text{ProductLog}(-v\,\ep^{-v}/\cos\alpha)\,,
\end{equation}
where $\text{ProductLog}(z)$ is the principal solution for $w$ in the equation $z=w\,\ep^w$ \cite{wolf}.

As before, the string energy is computed by first replacing the integral over $x$ in $S$ to that over $r$ and then imposing the short-distance cutoff on $r$. A direct calculation gives

\begin{equation}\label{e-}
E_{R}=\g\sqrt{\frac{\s}{\lambda}}
\biggl[
\int^1_{\sqrt{\tfrac{\s}{\lambda}}\epsilon}\,\frac{du}{u^2}\,\ep^{\lambda u^2}
\Bigl[1-u^4\,\ep^{2\lambda (1-u^2)}\Bigr]^{-\frac{1}{2}}
+
\int^1_{\sqrt{\frac{v}{\lambda}}}\,\frac{du}{u^2}\,\ep^{\lambda u^2}
\Bigl(1-u^4\,\ep^{2\lambda (1-u^2)}\Bigr)^{-\oh}
\biggr] 
\,.
\end{equation}
To obtain a finite result from $E_R$, we subtract the $\frac{1}{\epsilon}$ term and take the limit $\epsilon\rightarrow 0$. After doing so, we have

\begin{equation}\label{E-}
E
= 
\g\sqrt{\frac{\s}{\lambda}}
\biggl[
\int^1_0\,\frac{du}{u^2}\,
\Bigl(\ep^{\lambda u^2}\Bigl(1-u^4\,\ep^{2\lambda (1-u^2)}\Bigr)^{-\frac{1}{2}}
-1-u^2\Bigr)
+
\int^1_{\sqrt{\frac{v}{\lambda}}}\,\frac{du}{u^2}\,\ep^{\lambda u^2}
\Bigl(1-u^4\,\ep^{2\lambda (1-u^2)}\Bigr)^{-\frac{1}{2}}\biggr] 
+c
=
\g\sqrt{\s}\,{\cal E}^-(\lambda,v)
+c
\,,
\end{equation}
where $c$ is the same normalization constant introduced earlier.

\section{Some details on the quark-antiquark potential}
\renewcommand{\theequation}{C.\arabic{equation}}
\setcounter{equation}{0}

This Appendix provides a brief summary of key results concerning the heavy quark-antiquark potential, which represents the ground state energy of a static quark-antiquark pair. These results are relevant for our discussion in Sec. III and IV. For standard explanations, see \cite{az1}, whose conventions we follow unless stated otherwise. 

\begin{figure}[htbp]
\centering
\includegraphics[width=5.25cm]{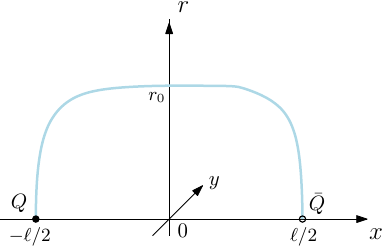}
\caption{{\small A string configuration in five dimensions. Here, $\ell$ denotes the distance between the two quark sources, and $\ro$ is the radial coordinate of the turning point at $x=0$.}}
\label{conQQb}
\end{figure}

The corresponding string configuration is shown in Figure \ref{conQQb}. 
It involves a Nambu-Goto string attached to the heavy quark sources on the boundary of five-dimensional space. Notably, a gravitational force pulls the string into the interior of an AdS-like space \cite{malda}. 

For the background geometry \eqref{metric}, the relation between the string energy and quark separation is expressed parametrically as

\begin{equation}\label{EQQb}
\ell= \frac{2}{\sqrt{\s}}{\cal L}^+(0,\lambda)
 \,,
\quad
E_{\QQb}=2\g\sqrt{\s}\,{\cal E}^+(0,\lambda)+2c
\,.
\end{equation}
Here $\lambda=\s\ro^2$ is a parameter varying from $0$ to $1$, and $c$ is the same normalization constant as in Appendix B. The functions ${\cal L}^+$ and ${\cal E}^+$ are defined in Appendix A. 

The small-$\ell$ behavior of $E_{\QQb}$ is given by 

\begin{equation}\label{EQQb-small}
E_{\QQb}(\ell)=-\frac{\alpha_{\QQb}}{\ell}+2c+\boldsymbol{\sigma}_{\QQb}\ell +o(\ell)
\,,
\end{equation}
with 
\begin{equation}\label{alpha-QQb}
\alpha_{\QQb}=(2\pi)^3\Gamma^{-4}\bigl(\tfrac{1}{4}\bigr)\g
	\,,\qquad
	\boldsymbol{\sigma}_{\QQb}=\oh(2\pi)^{-2}\Gamma^{4}\bigl(\tfrac{1}{4}\bigr)\g\s
	\,.
\end{equation}
In contrast, the large-$\ell$ behavior is given by 

\begin{equation}\label{EQQb-large}
E_{\QQb}(\ell)=\sigma\ell+C_{\QQb}+o(1)
\,,\qquad
\text{with}
\qquad
C_{\QQb}=2c-2\g\sqrt{\s}I_0
\,.
\end{equation}
The factor $I_0$ is defined in Appendix A. It is important to note that the coefficients $\boldsymbol{\sigma}_{\QQb}$ and $\sigma$ differ, with their ratio given numerically by $\boldsymbol{\sigma}_{\QQb}/\sigma=0.805$. Furthermore, the difference between the constant terms in the small- and large-$\ell$ expansions is $2c-C_{\QQb}=175\,\text{MeV}$, indicating that the constant term in the small-$\ell$ expansion is larger.

\section{A short account of the three-quark potential for the isosceles triangle geometry}
\renewcommand{\theequation}{D.\arabic{equation}}
\setcounter{equation}{0}

In general, any string configuration describing the three-quark potential specific to the isosceles triangle geometry can be constructed from the basic configurations of Sec.III. However, when the configuration involves multiple transitions between these basic configurations, the construction becomes more intricate. In this Appendix, we describe the potential, assuming that the apex angle is not in the near vicinity of $\frac{2}{3}\pi$.
\subsection{The small $\ell$ limit}

To gain some insights, we begin by examining the behavior of string configurations at small $\ell$. As shown in Appendix E of \cite{a3Q2016}, in the diquark limit (small $\theta$), the potential is described by configuration I. This remains valid as long as the parameter $v$ is less than $\vz$, where $\vz$ is the solution to the equation $\alpha_1(v) = 0$. We want to find the value of $\theta$ such that the equation has the trivial solution $\vz=0$. In other words, the interval $[0,\vz]$ shrinks to a point, rendering configuration I irrelevant. Substituting $\alpha_1=0$ into Eqs.\eqref{fbe-small} and \eqref{El1-small}, we find that  

\begin{equation}\label{theta1}
\theta^1=2\arccot\biggl(\sqrt{\frac{1-9\k^2}{3+9\k^2}}
\,\Bigl[1+2(1-9\k^2)^{-\frac{3}{4}}
\bigl(1+I(9\k^2,\tfrac{1}{2},\tfrac{3}{4})\bigr)
\Bigr]\biggr)
\,.
\end{equation}
A straightforward estimates gives $\theta^1=0.301\pi$. Thus, at small $\ell$ configuration I is relevant when $\theta<\theta^1$.

A similar analysis applies to configuration III, which is relevant for large $\theta$ \cite{a3Q2016}. Here, we look for a value of $\theta$ such that the equation $\alpha_3(v) = 0$ has the trivial solution $\vzi = 0$. Substituting $\alpha_3=0$ into Eqs.\eqref{fbe-small} and \eqref{El3-small}, we get

\begin{equation}\label{theta3}
\theta^3=2\arccot\biggl(
(3-9\k^2)^{-\oh}
\Bigl[1+2(1-\tfrac{9}{4}\k^2)^{\frac{3}{4}}
\bigl(1+I(\tfrac{9}{4}\k^2,\tfrac{1}{2},\tfrac{3}{4})\bigr)^{-1}
\Bigr]\biggr)
\,.
\end{equation}
Numerically, $\theta^3=0.351\pi$. We conclude that at small $\ell$ configuration III is relevant when $\theta>\theta^3$. It should now be evident that configuration II matters in the intermediate regime $\theta^1 \leq \theta \leq \theta^3$.
\subsection{The range $0<\theta<\theta^1$}

\begin{figure}[htbp]
\centering
\includegraphics[width=6.5cm]{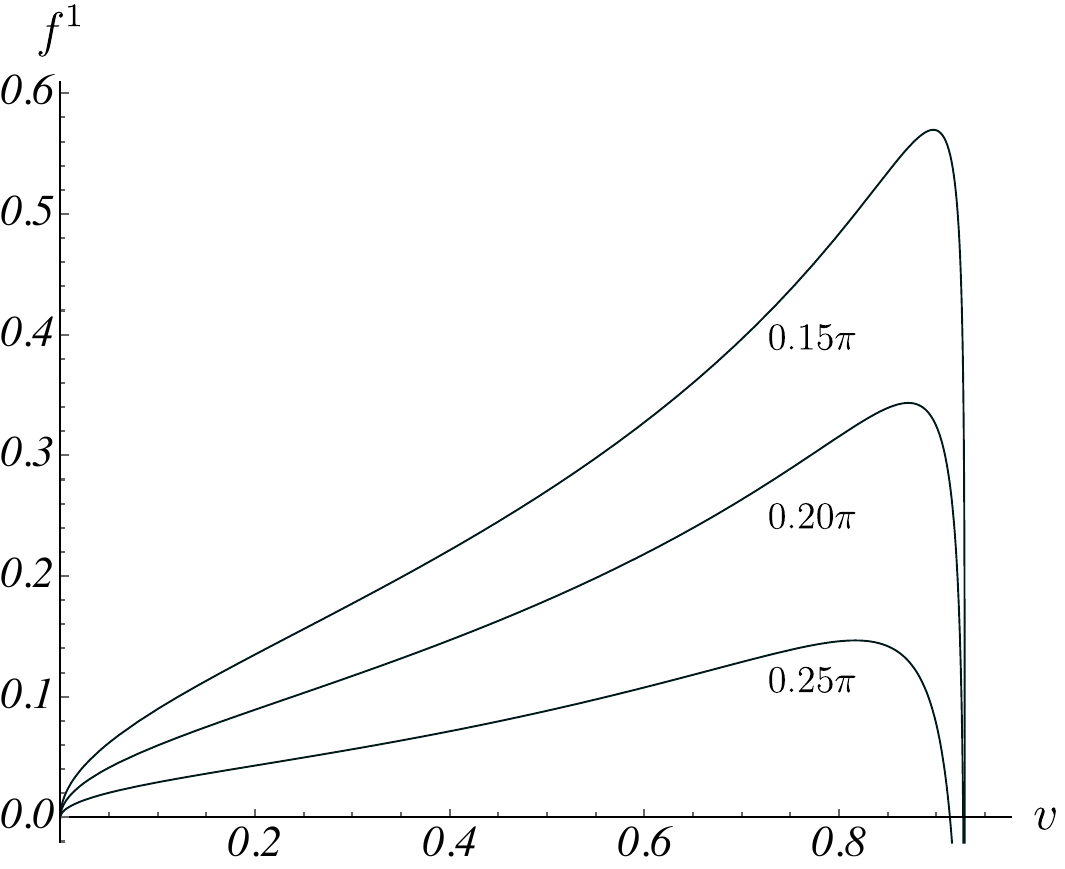}
\hspace{2cm}
\includegraphics[width=7.5cm]{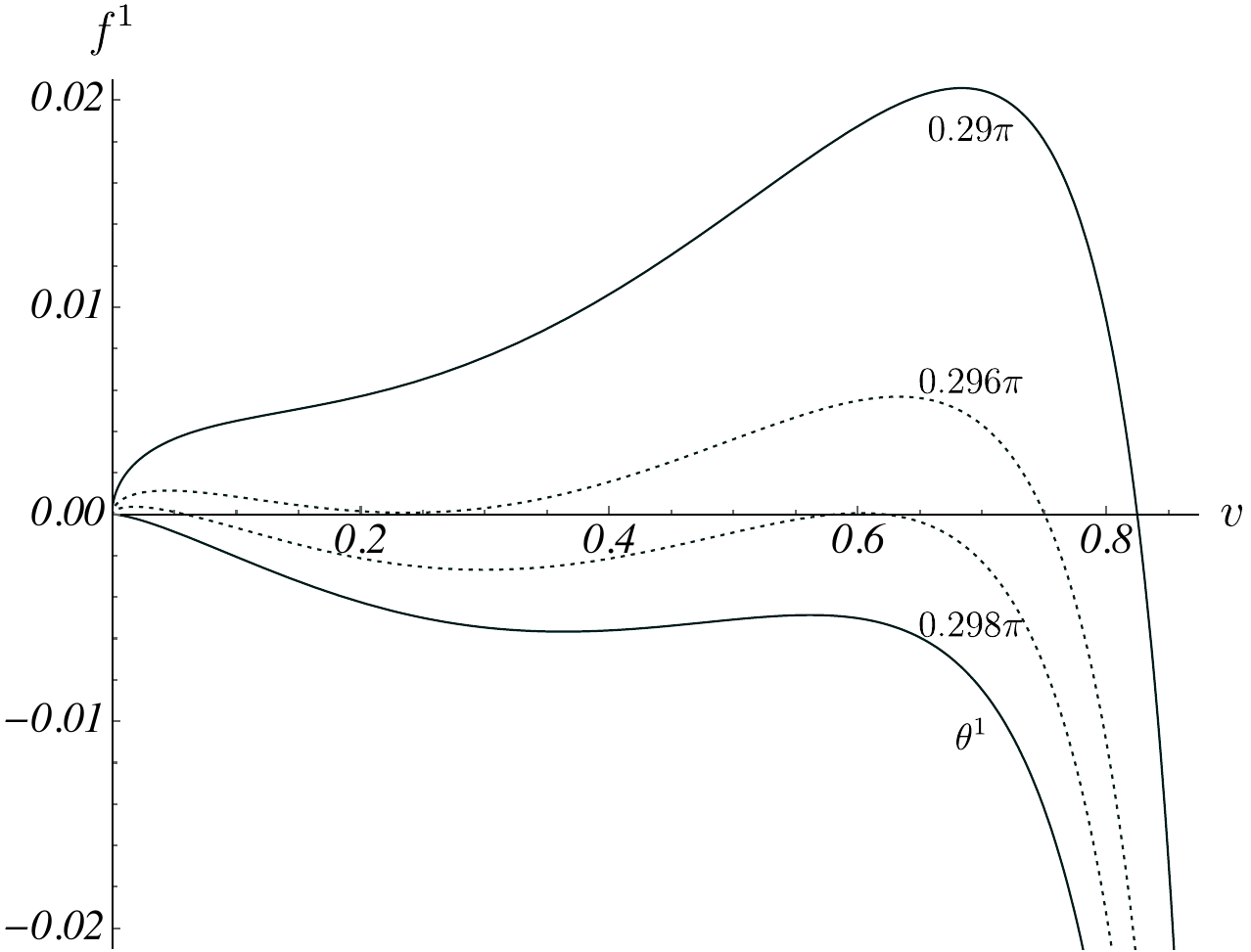}
\caption{{\small The function $f^1$ for several values of the apex angle $\theta$.}}
\label{thetaI}
\end{figure}

Based on the discussion above and in Appendix F, we expect that both configurations I and II are relevant within this range. To examine this in more detail, it is convenient to introduce the function

\begin{equation}
	f^1(\theta,v)=\cos\bigl(\beta+\tfrac{\theta}{2}\bigr){\cal L}^+(0,v)-
\sin\tfrac{\theta}{2}{\cal L}^-(\lambda_3,v)
\,
\end{equation}
whose zeros are solutions to the equation $\alpha_1(v)=0$ if $\beta(v)$ and $\lambda_3(v)$ are defined by the two first equations in \eqref{v01}. We are interested only in the zeros in the interval $0<v<\vo$.\footnote{As explained in Appendix F, the upper bound of $v$ is $\vo$ if $\theta<\frac{2}{3}\pi$.} A numerical analysis shows that this function typically has a single zero, except within the narrow interval $0.296\pi<\theta< 0.298\pi$, where it has three. For illustration, Figure \ref{thetaI} shows $f^1$ plotted as a function of $v$. The zeros 
correspond to transitions between configurations I and II. Accordingly, the potential is described in terms of configurations I and II as follows:

\begin{equation}\label{EL1}
\ell=\begin{cases}
\ell^{\,\i} (v)\,,\,\,&0\,\leq v\leq \vz\,,\\ 
\ell^{\,\ii}(v)\,,\,\,&\vz\leq v\leq \vo\,,
\end{cases}
\qquad
E_{\3Q}=\begin{cases}
E^{\,\i} (v)\,,\,\,&0\,\leq v\leq \vz\,,\\
E^{\,\ii}(v)\,,\,\,&\vz\leq v\leq \vo\,,
\\
\end{cases}
\quad\text{for}\quad 0<\theta\leq 0.296\pi\,,\,0.298\pi\leq\theta<\theta^1\,,
\end{equation}
and 

\begin{equation}\label{EL2}
\ell=\begin{cases}
\ell^{\,\i} (v)\,,\,\,&0\,\leq \, v\leq \,\vzf\,,\\ 
\ell^{\,\ii}(v)\,,\,\,&\vzf\leq v\leq \vzs\,,\\
\ell^{\,\i} (v)\,,\,\,&\vzs\,\leq v\leq \vzt\,,\\ 
\ell^{\,\ii}(v)\,,\,\,&\vzt\leq \,v\leq \,\,\vo\,,
\end{cases}
\qquad
E_{\3Q}=\begin{cases}
E^{\,\i} (v)\,,\,\,&0\,\leq \,\,v\leq \,\vzf\,,\\ 
E^{\,\ii}(v)\,,\,\,&\vzf\leq v\leq \vzs\,,\\
E^{\,\i}(v)\,,\,\,&\vzs\,\leq v\leq \vzt\,,\\ 
E^{\,\ii}(v)\,,\,\,&\vzt\leq \,v\leq \,\,\vo\,,
\end{cases}
\qquad\,\,\,
\text{for}\quad 0.296\pi<\theta<0.298\pi
\,.
\end{equation}
Here $\vz$ and ${\text v}_{\text{\tiny 0\hspace{0.4pt}i}}^{\text{\tiny 1}}$ are the solutions of $\alpha_1(v)=0$. These expressions define a parametric form of $E_{\3Q}(\ell)$ in which the parameter $v$ cannot be easily eliminated.

\begin{figure}[htbp]
\centering
\includegraphics[width=7cm]{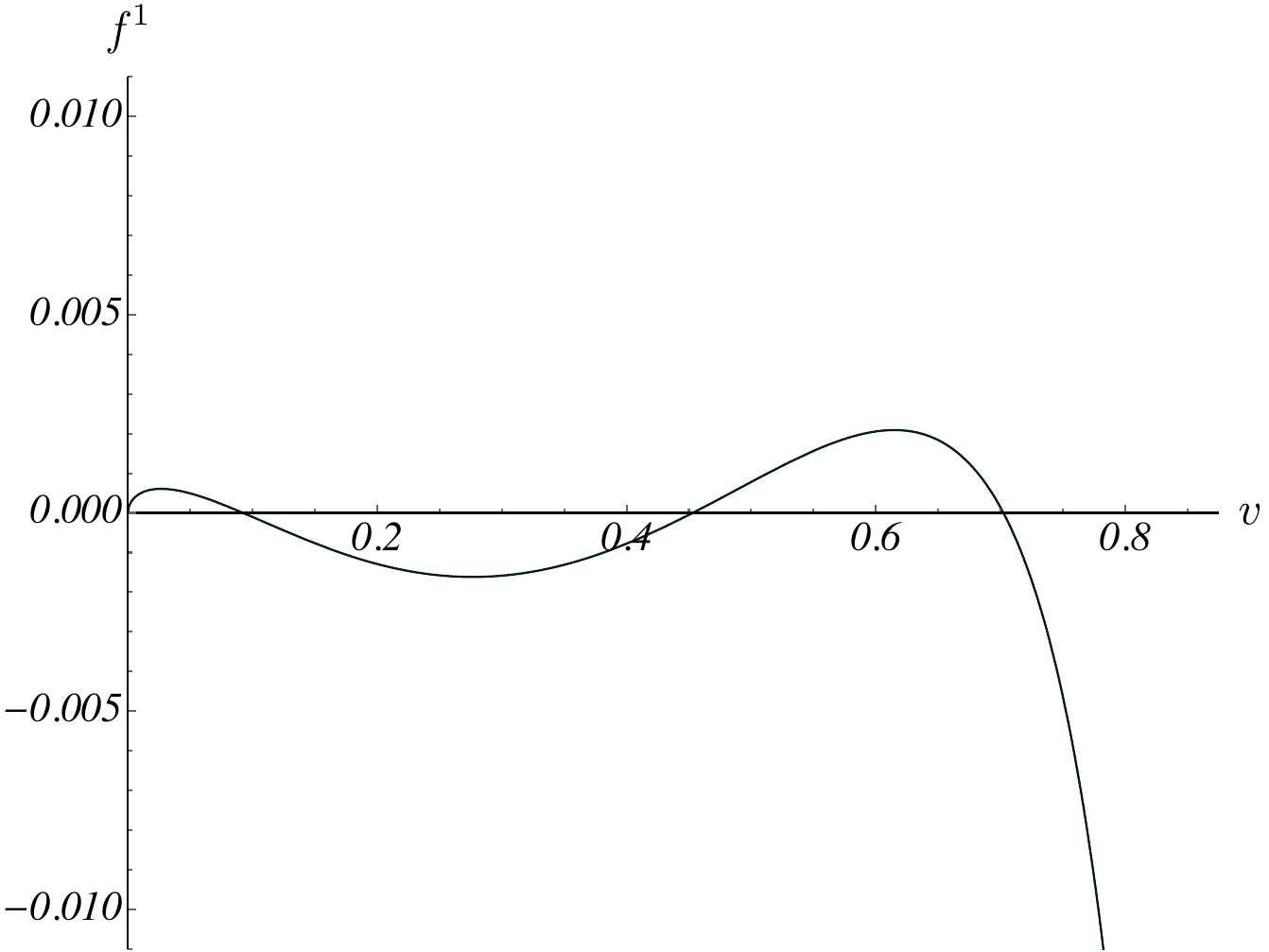}
\hspace{1cm}
\includegraphics[width=7.80cm]{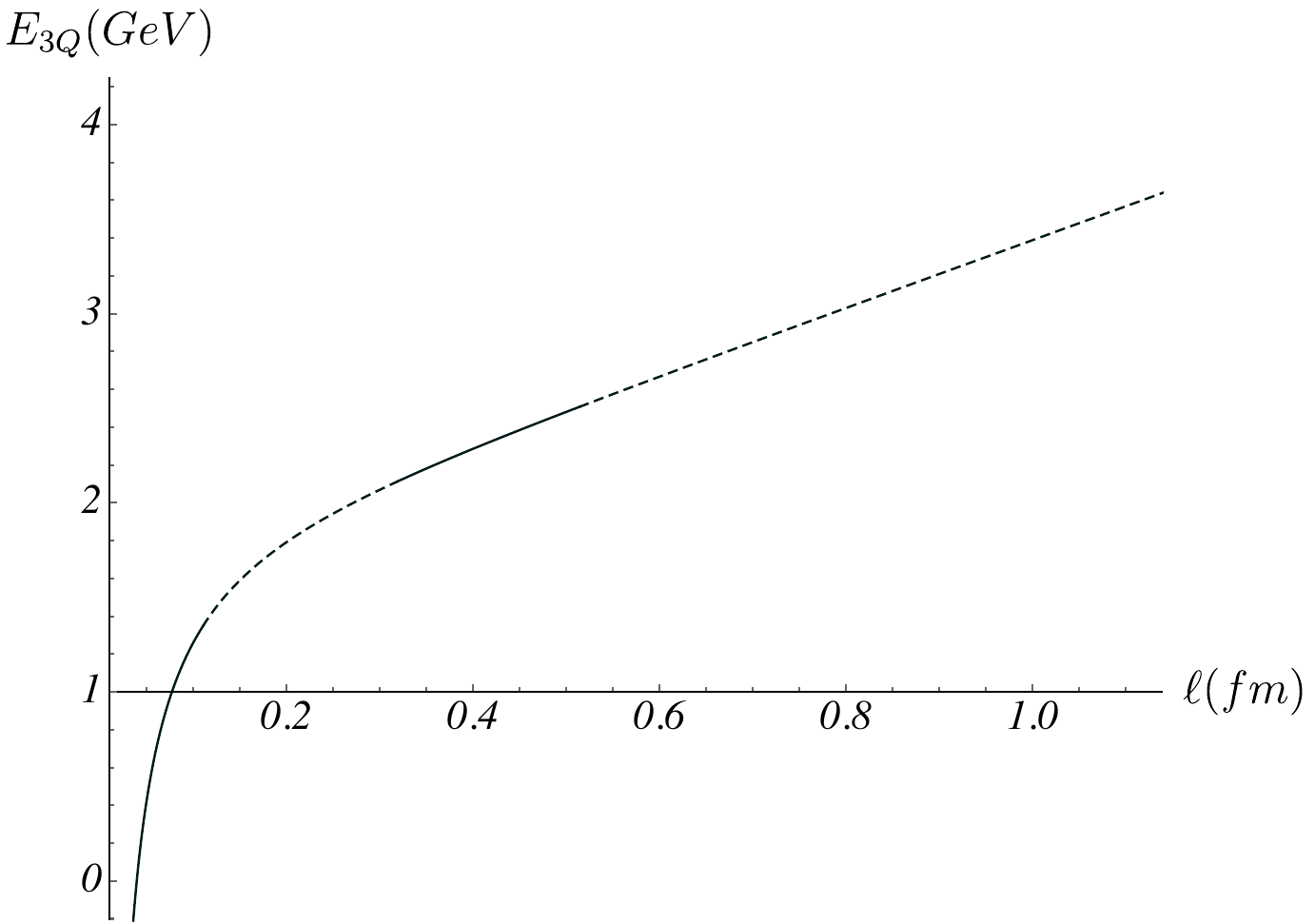}
\caption{{\small The functions $f^1(v)$ and $E_{\3Q}(\ell)$ at $\theta=0.2975\pi$. In the plot of $E_{\3Q}$, the solid and dashed curves correspond respectively to configurations I and II.}}
\label{theta2975}
\end{figure}

The example with $\theta=\frac{\pi}{6}$ discussed in Sec.III is a special case of \eqref{EL1}. Therefore, to illustrate \eqref{EL2}, we consider the case $\theta = 0.2975\pi$. A simple calculation gives $\vzf=0.092$, $\vzs=0.453$, and $\vzt=0.702$. Figure \ref{theta2975} shows both functions $f^1$ and $E_{\3Q}$.

\subsection{The range $\theta^1\leq\theta\leq\theta^3$ }

\begin{figure}[H]
\centering
\includegraphics[width=6.85cm]{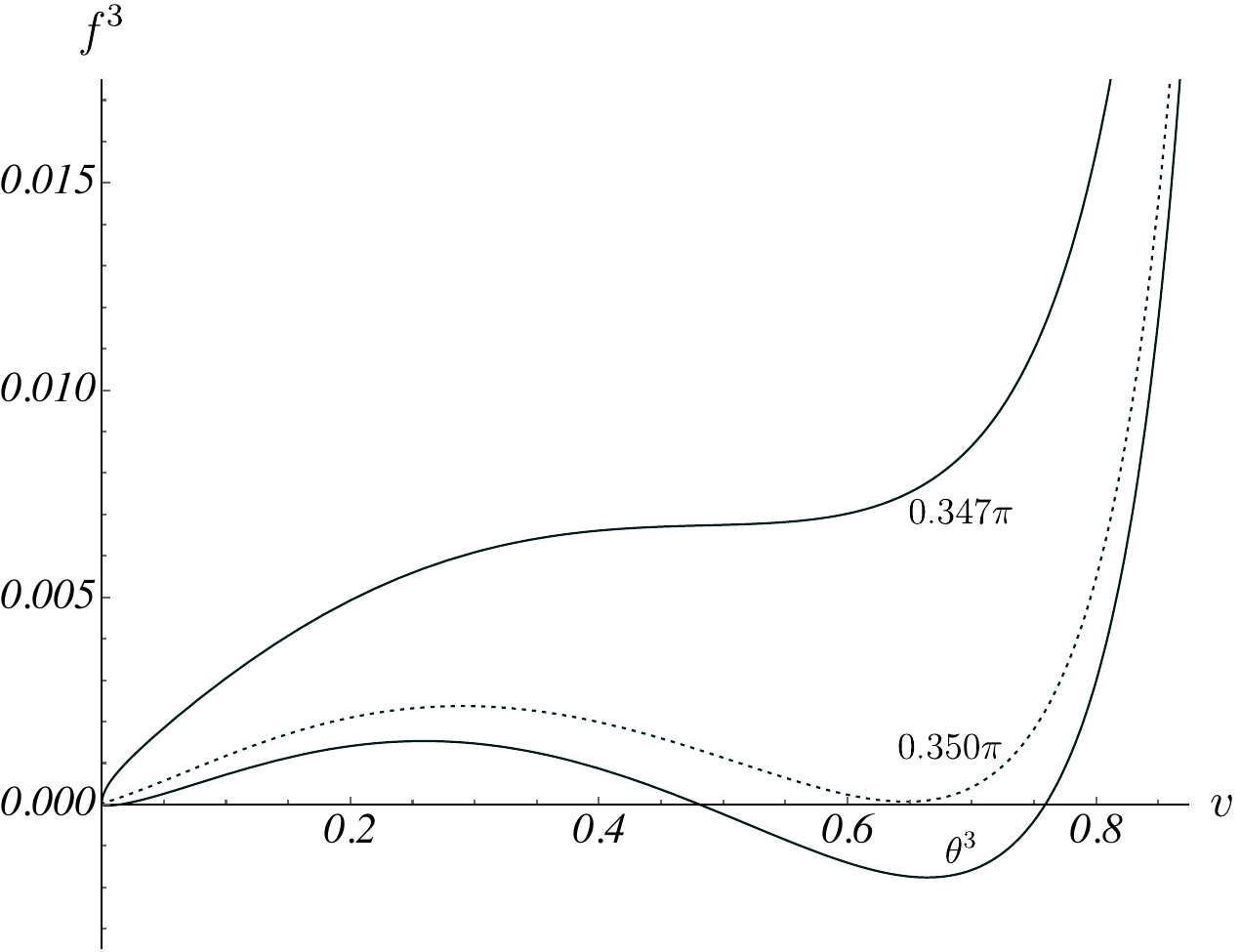}
\caption{{\small The function $f^3$ for several values of the apex angle $\theta$.}}
\label{thetaII}
\end{figure}

In this range, we introduce the function 

\begin{equation}\label{f3}
f^3(\theta,v)=\cos\bigl(\beta+\tfrac{\theta}{2}\bigr){\cal L}^-(\lambda_1,v)-
\sin\tfrac{\theta}{2}{\cal L}^+(0,v)
\,
\end{equation}
whose zeros correspond to the solutions of $\alpha_3(v)=0$ if $\beta(v)$ and $\alpha_1(v)$ are given by the first two equations in \eqref{v03}. As before, we are interested in the zeros within the interval $0<v<\vo$. A numerical analysis of $f^3$ is straightforward and shows two distinct cases: the function has no zeros for $\theta^1 \leq \theta < 0.350\pi$, and has two zeros for $0.350\pi \leq \theta \leq \theta^3$. The former case aligns with the earlier discussion of $\theta = \frac{\pi}{3}$, while the latter implies transitions between configurations II and III. For illustration, we plot $f^3$ as a function of $v$ in Figure \ref{thetaII}. The potential is described in terms of these
\noindent configurations as follows: 

\begin{equation}\label{EL3}
	\ell=\ell^{\,\ii}(v)
	\,,
\,\,
	E_{\3Q}=E^{\,\ii}(v)
	\,,\,\,
	0\leq v\leq\vo\,,
	\quad\text{for}\quad\theta^1\leq\theta\leq 0.350\pi
	\,,
\end{equation}
and 
\begin{equation}\label{EL4}
\ell=\begin{cases}
\ell^{\,\ii} (v)\,,\,\,&0\,\leq \,\, v\leq {\text v}_{\text{\tiny 01}}^{\text{\tiny 3}}\,\,,\\ 
\ell^{\,\iii}(v)\,,\,\,&{\text v}_{\text{\tiny 01}}^{\text{\tiny 3}}\leq v\leq {\text v}_{\text{\tiny 02}}^{\text{\tiny 3}}\,,\\
\ell^{\,\ii}(v)\,,\,\,&{\text v}_{\text{\tiny 02}}^{\text{\tiny 3}}\leq \,v\leq \vo\,\,,
\end{cases}
\qquad
E_{\3Q}=\begin{cases}
E^{\,\ii} (v)\,,\,\,&0\,\leq \,\, v\leq {\text v}_{\text{\tiny 01}}^{\text{\tiny 3}}\,\,,\\  
E^{\,\iii}(v)\,,\,\,&{\text v}_{\text{\tiny 01}}^{\text{\tiny 3}}\leq v\leq {\text v}_{\text{\tiny 02}}^{\text{\tiny 3}}\,,\qquad\text{for}\quad0.350\pi<\theta\leq\theta^3\,,\\
E^{\,\ii}(v)\,,\,\,&{\text v}_{\text{\tiny 02}}^{\text{\tiny 3}}\leq \,v\leq \vo\,\,.
\end{cases}
\end{equation}
Here ${\text v}_{\text{\tiny 0\hspace{0.4pt}i}}^{\text{\tiny 3}}$ are the solutions of the equation $\alpha_3(v)=0$. Again, this parametric representation of $E_{\3Q}(\ell)$ does not allow for an explicit elimination of the parameter $v$.

\begin{figure}[htbp]
\centering
\includegraphics[width=6.8cm]{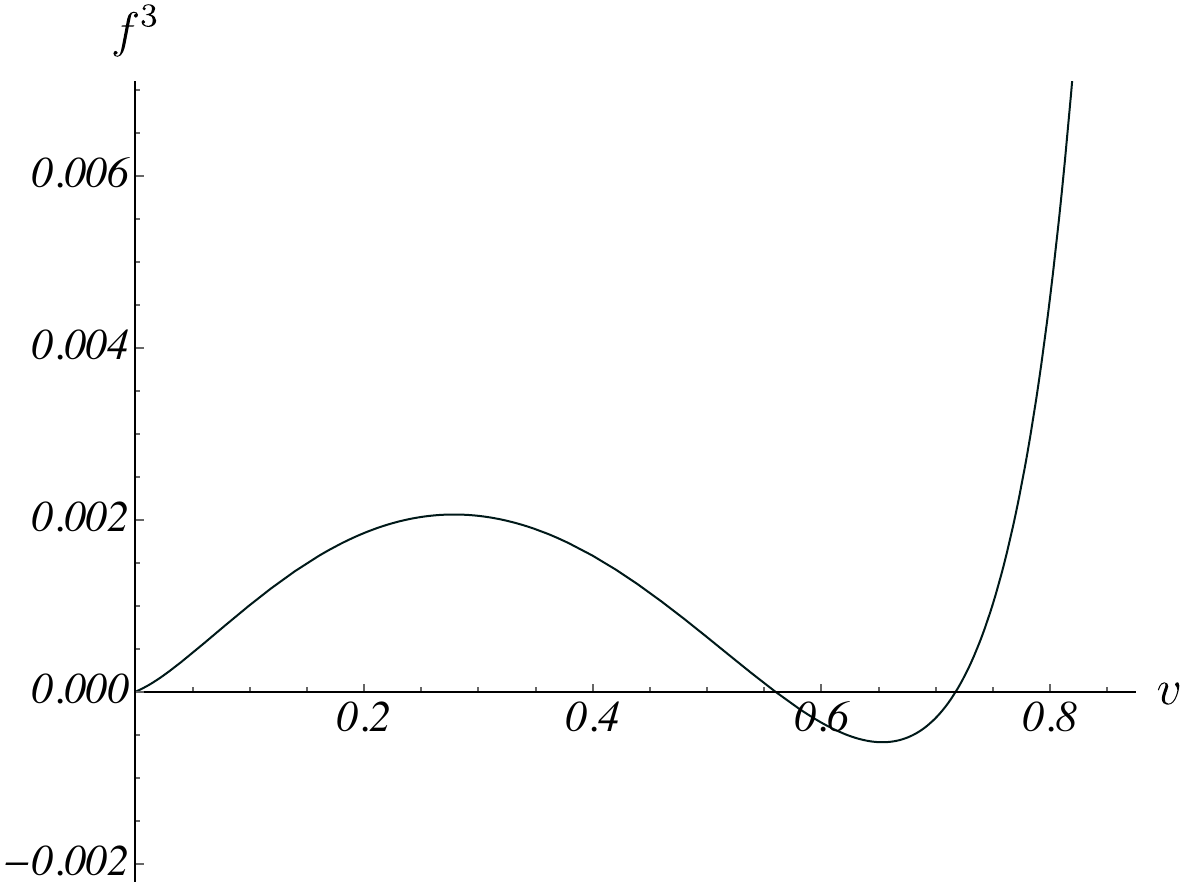}
\hspace{1.3 cm}
\includegraphics[width=8cm]{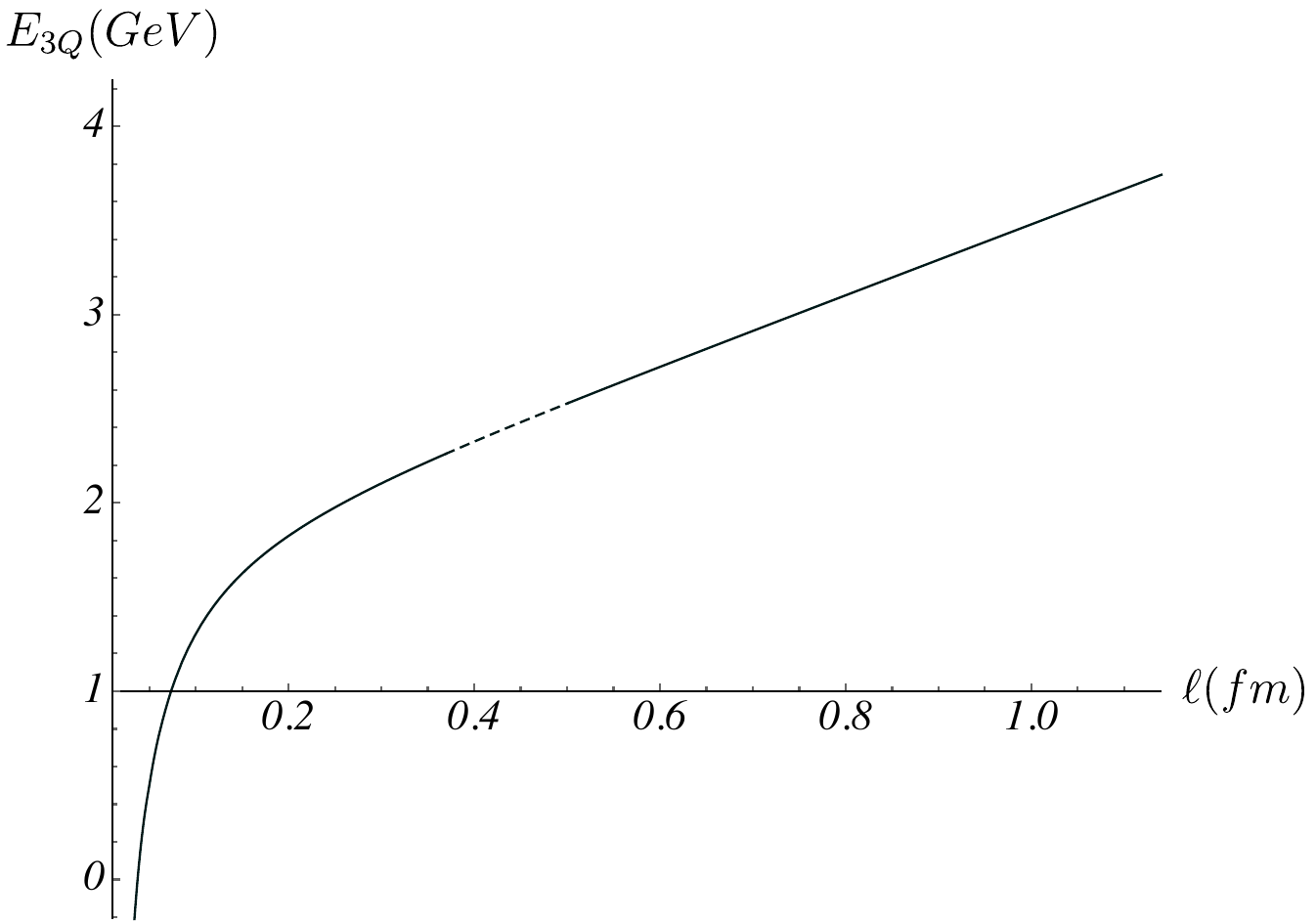}
\caption{{\small The functions $f^3(v)$ and $E_{\3Q}(\ell)$ at $\theta=0.3505\pi$. In the plot of $E_{\3Q}$, the solid and dashed curves correspond respectively to configurations II and III.}}
\label{theta3505}
\end{figure}

Since the case $\theta = \frac{\pi}{3}$ discussed in Sec.III corresponds to \eqref{EL3}, it is natural to consider an example of \eqref{EL4}. Without loss of generality, we take $\theta = 0.3505\pi$. In this case the solutions are ${\text v}_{\text{\tiny 01}}^{\text{\tiny 3}}=0.560$ and ${\text v}_{\text{\tiny 02}}^{\text{\tiny 3}}=0.717$. The corresponding functions $f^3$ and $E_{\3Q}$ are shown in Figure \ref{theta3505}.
\subsection{The range $\theta^3<\theta<0.66638\pi$ }

Before proceeding further, we note that the upper bound of this range arises from the analysis presented in Appendix F. There, it is shown that in the narrow range of $\theta$ values starting from $0.66638\pi$ the potential cannot longer be described in terms of a single string configuration.

\begin{figure}[htbp]
\centering
\includegraphics[width=7.25cm]{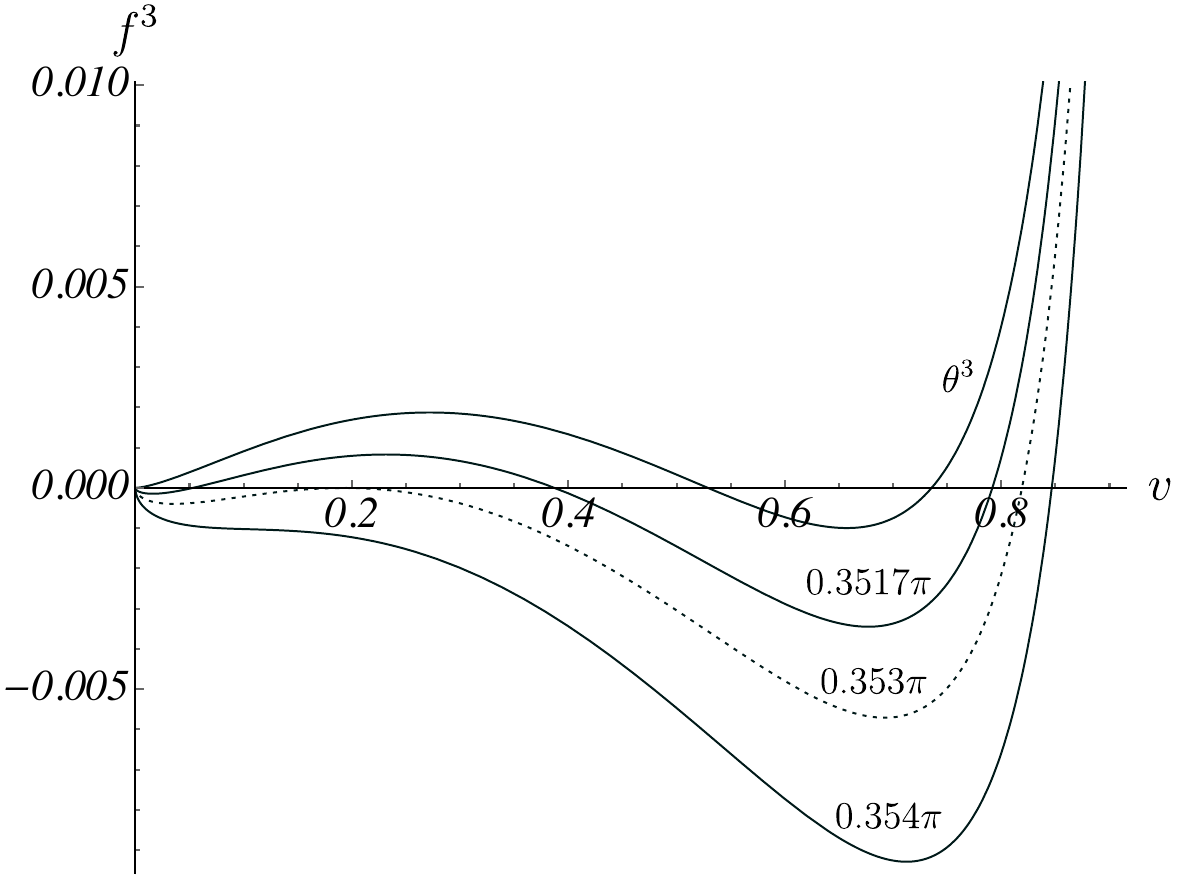}
\caption{{\small The function $f^3$ for several values of $\theta$.}}
\label{thetaIIIa}
\end{figure}

From this point, the analysis proceeds in an obvious way. We look for the zeros of $f^3$ within the interval $0<v<\vo$. A numerical analysis shows that there are three zeros when $\theta^3<\theta\leq 0.353\pi$, and only one zero when $0.353\pi<\theta<0.66638\pi$, as illustrated in Figure \ref{thetaIIIa}. The zeros correspond to transitions between configurations 
II and III. So, the potential is written as 

\begin{equation}\label{EL5}
\ell=\begin{cases}
\ell^{\,\iii} (v)\,,\,\,&0\,\leq \,\, v\leq {\text v}_{\text{\tiny 01}}^{\text{\tiny 3}}\,\,,\\ 
\ell^{\,\ii}(v)\,,\,\,& {\text v}_{\text{\tiny 01}}^{\text{\tiny 3}}\leq v\leq {\text v}_{\text{\tiny 02}}^{\text{\tiny 3}}\,,\\
\ell^{\,\iii}(v)\,,\,\,&{\text v}_{\text{\tiny 02}}^{\text{\tiny 3}}\leq v\leq {\text v}_{\text{\tiny 03}}^{\text{\tiny 3}}\,,\\
\ell^{\,\ii}(v)\,,\,\,&{\text v}_{\text{\tiny 03}}^{\text{\tiny 3}}\leq \,v\leq \vo\,\,,
\end{cases}
\qquad
E_{\3Q}=\begin{cases}
E^{\,\iii} (v)\,,\,\,&0\,\leq \,\, v\leq {\text v}_{\text{\tiny 01}}^{\text{\tiny 3}}\,\,,\\ 
E^{\,\ii}(v)\,,\,\,&{\text v}_{\text{\tiny 01}}^{\text{\tiny 3}}\leq v\leq {\text v}_{\text{\tiny 02}}^{\text{\tiny 3}}\,,\\
E^{\,\iii}(v)\,,\,\,&{\text v}_{\text{\tiny 02}}^{\text{\tiny 3}}\leq v\leq {\text v}_{\text{\tiny 03}}^{\text{\tiny 3}}\,,\\
E^{\,\ii}(v)\,,\,\,&{\text v}_{\text{\tiny 03}}^{\text{\tiny 3}}\leq \,v\leq \vo\,\,,
\end{cases}
\qquad\text{for}\quad \theta^3<\theta\leq 0.353\pi\,,
\end{equation}
and 
\begin{equation}\label{EL6}
\ell=\begin{cases}
\ell^{\,\iii} (v)\,,\,\,&0\,\leq v\leq \vzi\,,\\ 
\ell^{\,\ii}(v)\,,\,\,&\vzi\leq v\leq \vo\,,
\end{cases}
\qquad
E_{\3Q}=\begin{cases}
E^{\,\iii} (v)\,,\,\,&0\,\leq v\leq \vzi\,,
\\
E^{\,\ii}(v)\,,\,\,&\vzi\leq v\leq \vo\,,
\end{cases}
\qquad\text{for}\quad 0.353\pi<\theta<0.66638\pi\,.
\end{equation}
These equations represent the parametric forms of $E_{\3Q}(\ell)$. 
  
\begin{figure}[htbp]
\centering
\includegraphics[width=7.25cm]{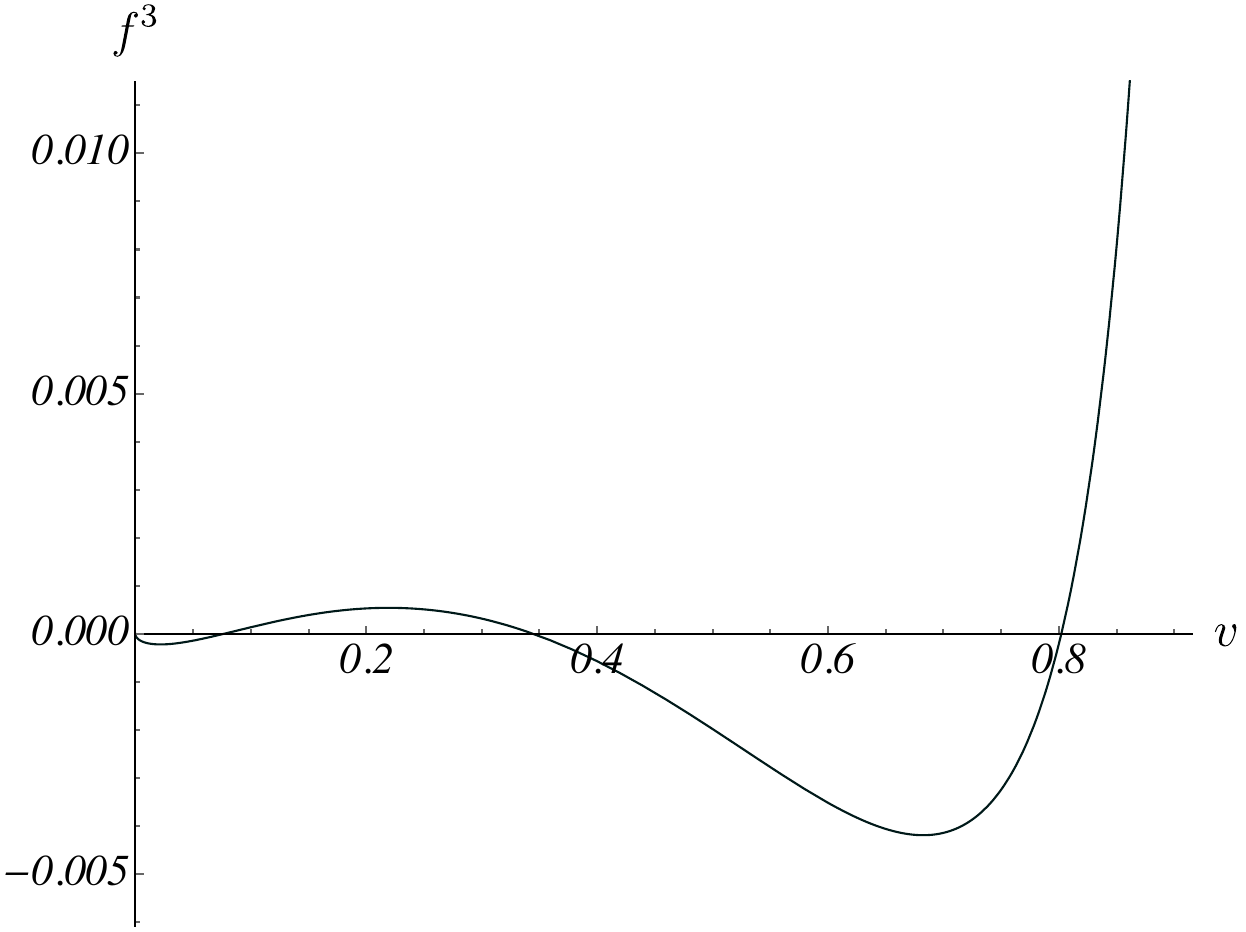}
\hspace{1cm}
\includegraphics[width=8.5cm]{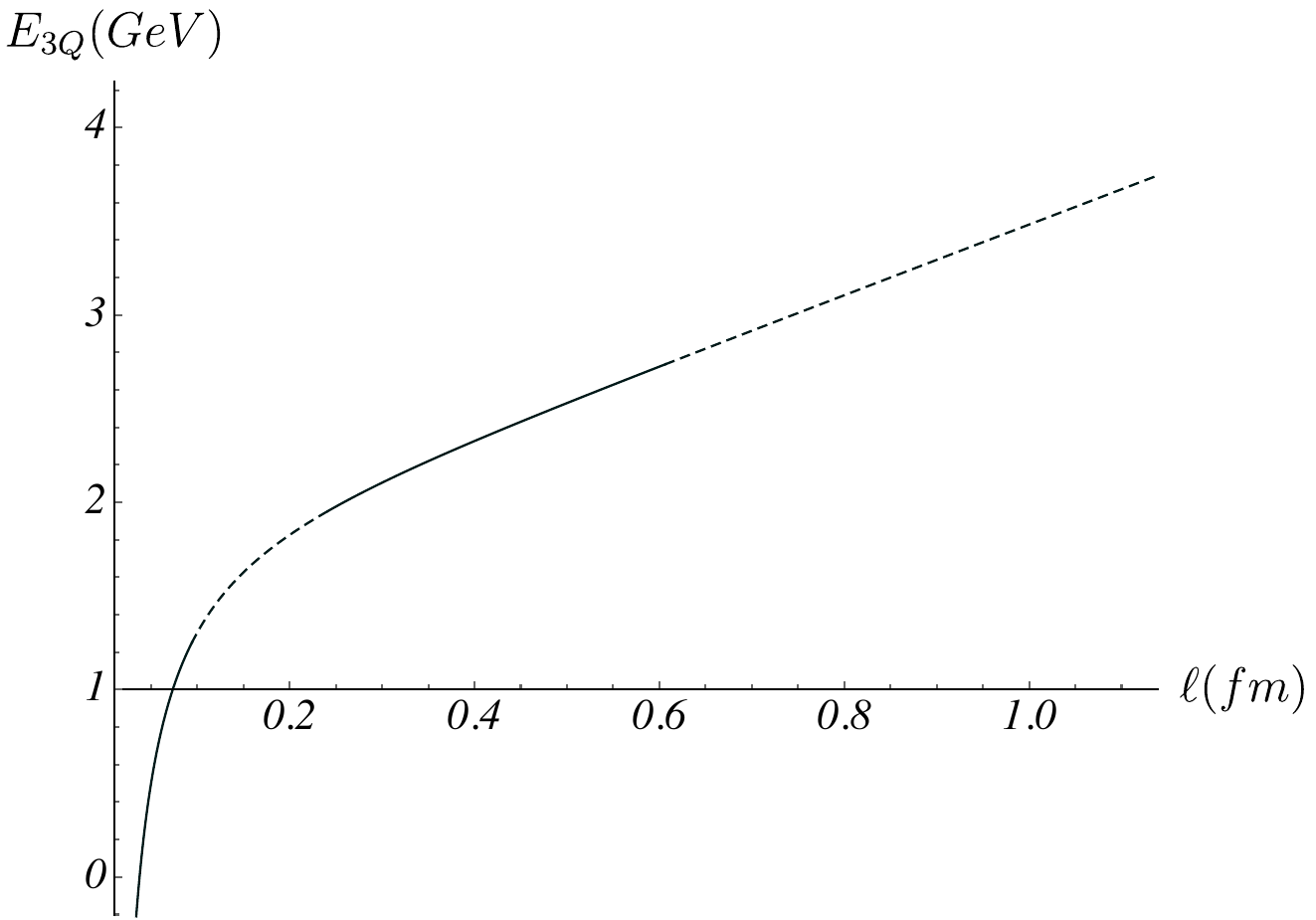}
\caption{{\small The functions $f^3(v)$ and $E_{\3Q}(\ell)$ at $\theta=0.352\pi$. In the plot of $E_{\3Q}$, the dashed and solid curves correspond to configurations II and III, respectively.}}
\label{theta352}
\end{figure}

The example with $\theta=\frac{\pi}{2}$ discussed in Sec.III is a special case of \eqref{EL6}. So, we now consider an example corresponding to \eqref{EL5}, choosing $\theta = 0.352\pi$. In this case, the solutions are ${\text v}_{\text{\tiny 01}}^{\text{\tiny 3}}=0.076$, ${\text v}_{\text{\tiny 02}}^{\text{\tiny 3}}=0.345$, and ${\text v}_{\text{\tiny 03}}^{\text{\tiny 3}}=0.802$. The corresponding functions $f^3$ and $E_{\3Q}$ are shown in Figure \ref{theta352}. 

\subsection{The range $\frac{2}{3}\pi<\theta\leq\pi$ }

\begin{figure}[H]
\centering
\includegraphics[width=6.5cm]{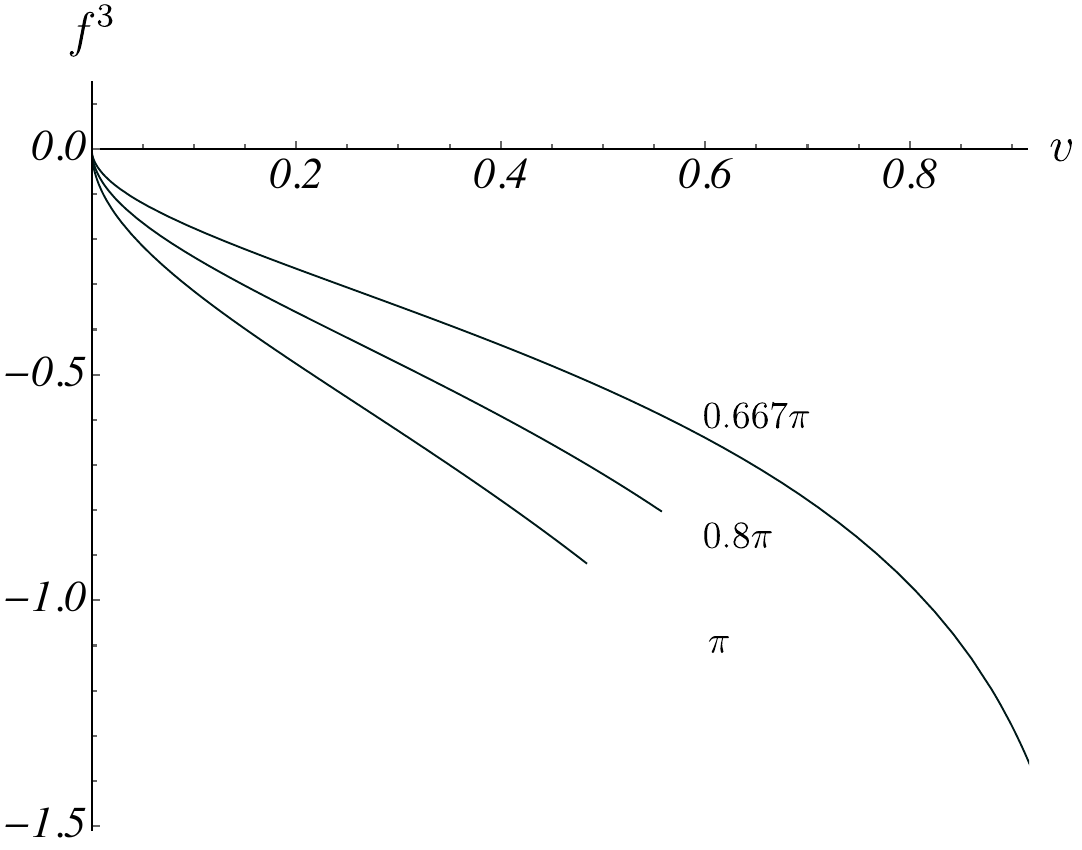}
\caption{{\small The function $f^3$ for  several values  of $\theta$.}}
\label{thetapi}
\end{figure}

After omitting the narrow range $0.66638\pi\leq\theta\leq \frac{2}{3}\pi$, we turn to the last range of $\theta$ values. The analysis proceeds in the same manner as before,  except that we now search for the zeros of $f^3$ in the interval $0<v<\vop$, where $\vop$ depends on $\theta$.\footnote{$\vop$ is a decreasing function of $\theta$, see Figure \ref{v1pm}.} It turns out that $f^3$ has no zeros. It remains strictly negative, as shown in Figure \ref{thetapi}. As a result, 
\noindent the potential is completely determined by configuration III. Explicitly,

\begin{equation}\label{EL7}
	\ell=\ell^{\,\iii}(v)\,,\,\,\,
	E_{\3Q}=E^{\,\iii}(v)\,,\,\,\,
		0\leq v\leq \vop\,,
		\,\,\,\,
	\text{for}
	\,\,\,\,
	\frac{2}{3}\pi<\theta\leq\pi
	\,.
\end{equation} 
The examples discussed in Sec.III are special cases of \eqref{EL7}.

\section{Some details on the three-quark potential for the collinear geometry}
\renewcommand{\theequation}{E.\arabic{equation}}
\setcounter{equation}{0}

In \cite{a3Q2016}, it was shown that for the symmetric collinear geometry ($\eta=1$) the potential is described in terms of configuration A, while in the diquark limit (small $\eta$), it is described by both configurations A and B. Our goal is to understand the transition between these descriptions. 

\begin{figure}[htbp]
\centering
\includegraphics[width=6.75cm]{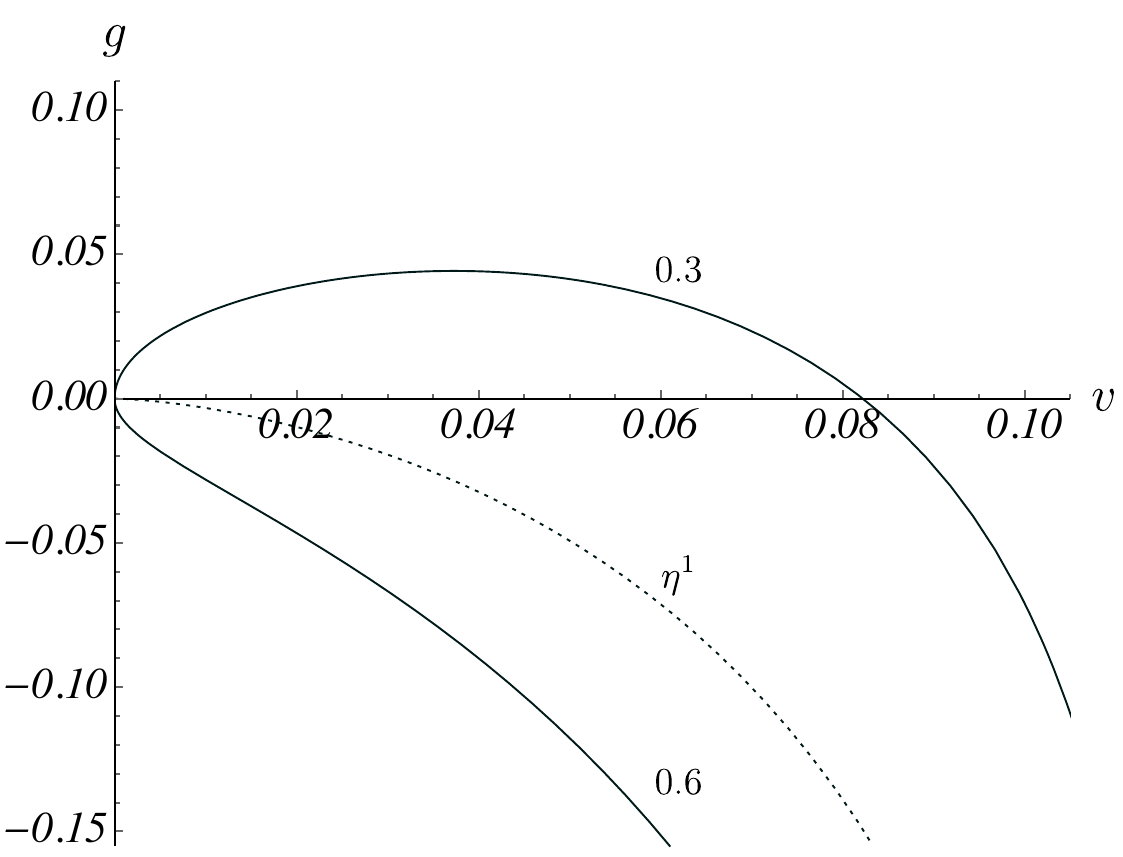}
\caption{{\small The function $g$ for several values of $\eta$.}}
\label{gv0}
\end{figure}

To this end, it is convenient to introduce the function 

\begin{equation}\label{gfun}
g(\eta,v)=
{\cal L}^+(0,v)
-
\eta {\cal L}^-(\lambda_2,v)
+(1+\eta){\cal L}^+(\alpha_3,v)
\,
\end{equation}
which is motivated by the geometrical constraint \eqref{gc-colB}. A key fact is that if $\alpha_2$ and $\alpha_3$ are determined from Eqs.\eqref{fbe-col} at $\alpha_1=0$, then zeros of $g$ coincide with the solutions of the equation $\alpha_1(v)=0$. A numerical analysis shows that there exists a critical value $\eta^1$ such that $g$ has a single zero for $\eta\leq\eta^1$, and no zeros otherwise.\footnote{The zero at $v=\vz$ corresponds to the transition between configurations A and B.} This behavior is illustrated in Figure \ref{gv0}. 

The value of $\eta^1$ can be found by evaluating \eqref{gfun} at $v=0$, yielding

\begin{equation}\label{eta1}
\eta^1=\frac{1+
\cos^{-\oh}\hspace{-.9mm}\alpha_3\,I\bigl(\cos^2\hspace{-.9mm}\alpha_3,\tfrac{3}{4},\oh\bigr)
}
{\cos^{-\oh}\hspace{-.9mm}\alpha_2
\bigl(1+I\bigl(\sin^2\hspace{-.9mm}\alpha_2,\oh,\tfrac{3}{4}\bigr)\bigr)
-
\cos^{-\oh}\hspace{-.9mm}\alpha_3\,I\bigl(\cos^2\hspace{-.9mm}\alpha_3,\tfrac{3}{4},\oh\bigr)}
\,,
\end{equation}
with $\cos\alpha_2=\oh+\frac{3\sqrt{3}}{2}\k\sqrt{\frac{1-3\k^2}{1+9\k^2}}$ and $\cos\alpha_3=\oh-\frac{3\sqrt{3}}{2}\k\sqrt{\frac{1-3\k^2}{1+9\k^2}}$. A straightforward calculation then gives $\eta^1=0.471$. 

Thus, the potential is given parametrically by 

\begin{equation}\label{ELc}
\ell=\begin{cases}
\ell^{\text{\,\tiny B}} (v)\,,\,\,& 0\,\leq v\leq \vz\,\,,\\ 
\ell^{\text{\,\tiny A}}(v)\,,\,\,&\vz\leq v\leq \vop ,
\end{cases}
\quad
E_{\3Q}=\begin{cases}
E^{\,\text{\tiny B}} (v)\,,\,\,&0\,\leq v\leq \vz\,\,,\\
E^{\,\text{\tiny A}}(v)\,,\,\,&\vz\leq v\leq \vop .
\end{cases}
\qquad\text{for}\quad 0<\eta<\eta^1\,
\end{equation}
and 
\begin{equation}\label{ELc2}
	\ell=\ell^{\text{\,\tiny A}}(v)\,,\quad
	E_{\3Q}=E^{\text{\,\tiny A}}(v)\,,\quad 
	0\leq v\leq \vop
	\,,
\qquad\text{for}\quad \eta^1\leq\eta\leq 1\,.
\end{equation} 
A few examples illustrating both cases are given in Sec.III.

\section{More on the IR limit}
\renewcommand{\theequation}{F.\arabic{equation}}
\setcounter{equation}{0}

\subsection{The limiting values of $v$}

As explained in Sec.V, the limiting values of the parameter $v$ denoted as $\vo$, $\vom$, and $\vop$ are the solutions of Eqs.\eqref{v1}, \eqref{v1m}, and \eqref{v1p}, respectively. Here we examine how these solutions behave in the interval $[0, \pi]$.

\begin{figure}[htbp]
\centering
\includegraphics[width=8.1cm]{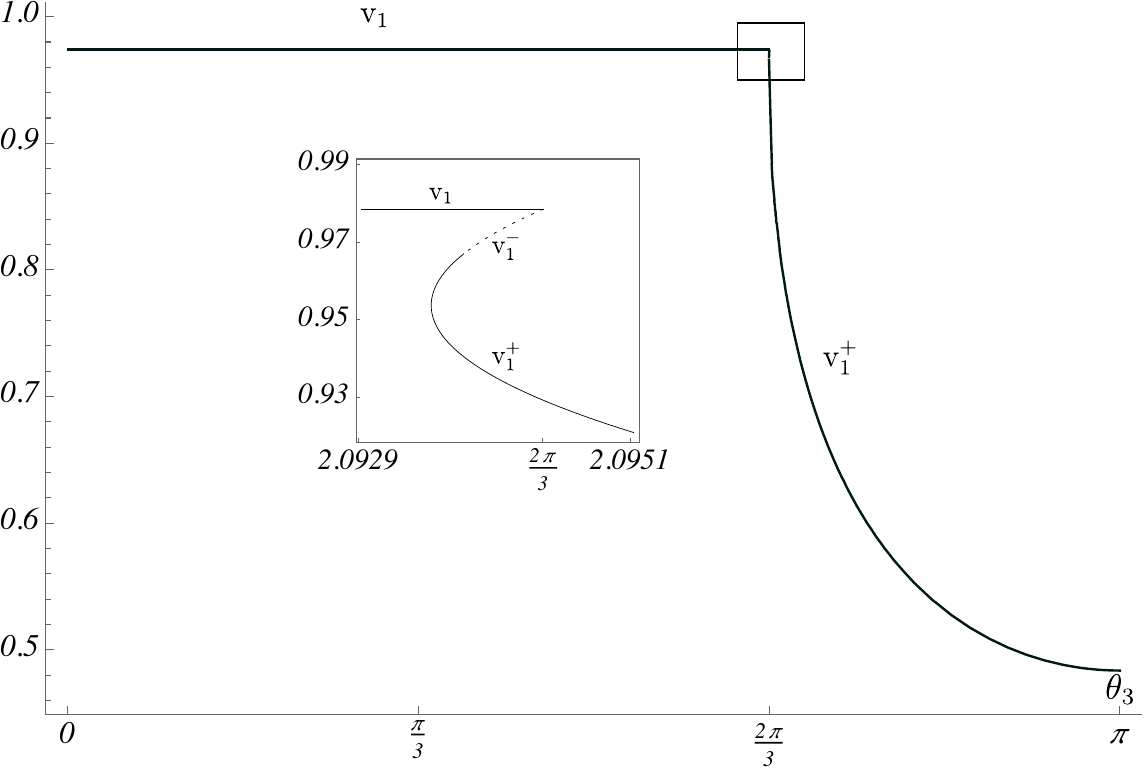}
\hspace{1cm}
\includegraphics[width=8.1cm]{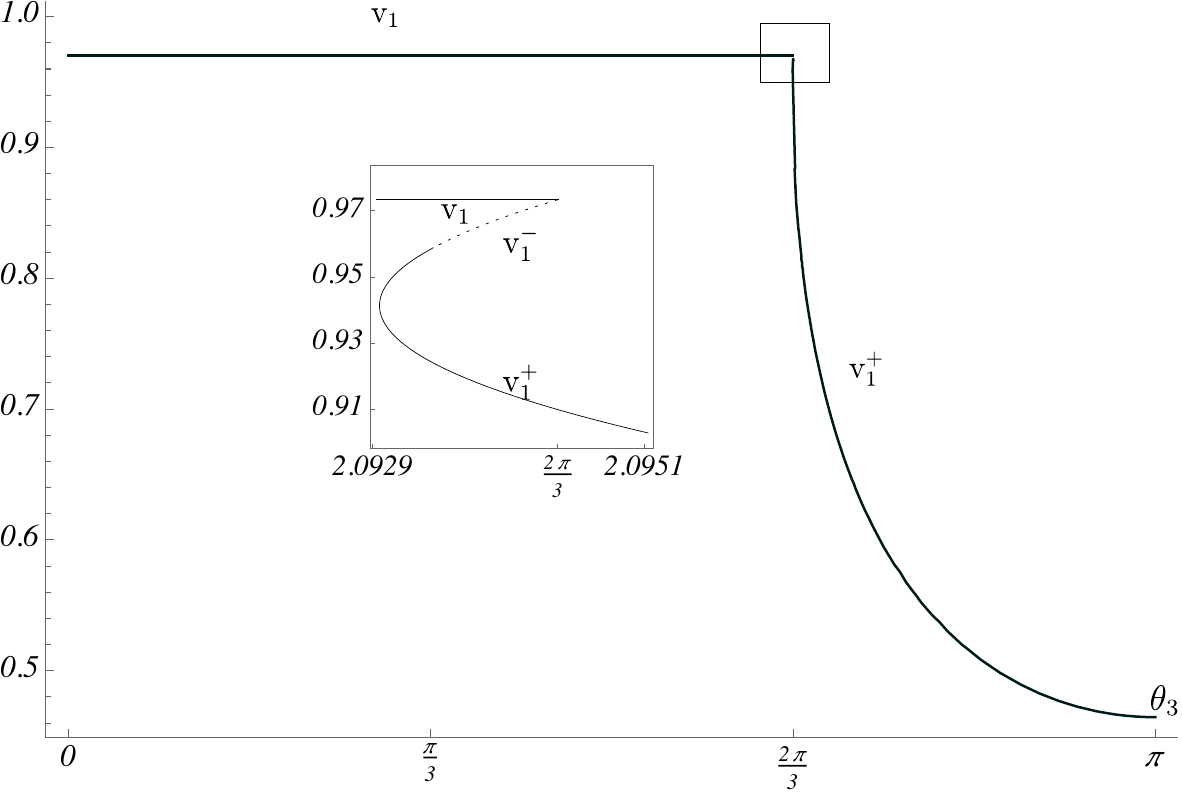}
\caption{{\small The solutions $\vo$, $\vom$, and $\vop$ versus $\theta_3$. The dotted curves correspond to $\vom$. Here $\k=-0.083$ (left) and $\k=-0.102$ (right).}}
\label{v1pm}
\end{figure}

As a shortcut, we perform a numerical analysis under the assumption that the angle $\theta_3$ at $Q_3$ is the largest when obtuse, and that the triangle has no angle greater than $\frac{2}{3}\pi$ when $\theta_3$ is acute. The results are shown in Figure \ref{v1pm} on the left.
The solution $\vo$ exists and equals $0.978$. It is defined over the interval $0\leq\theta_3<\frac{2}{3}\pi$ and, as argued in Sec.V, is meaningful only if the triangle contains no angle greater than $\frac{2}{3}\pi$. In contrast, the solution $\vom$ depends on $\theta_3$. It obeys $\lambda_3(\vom)\leq 1$ and exists only within the narrow interval $0.66646\,\pi\leq\theta_3\leq\frac{2}{3}\pi$. The upper bound corresponds to $\lambda_3=1$, where $\vom=\vo$, and the lower to $\alpha_3=0$, where $\vom=0.967$. Recall that the condition $\alpha_3(v)=0$ describes a transition point between configurations II and III. The analysis of Eq.\eqref{v1p} is slightly more subtle. It shows that this Equation has two solutions within the narrow interval $0.66638\pi<\theta_3\leq 0.66646\pi$, and a single one for $0.66646\pi<\theta_3\leq\pi$. To complete the picture, we repeat the analysis for $\k=-0.102$ \footnote{This value is motivated by phenomenology, specifically by the relation $E_{\QQ} = \tfrac{1}{2} E_{\QQb}$ at small quark separations.} with the results shown in the Figure on the right. Clearly, there are no significant differences between the two cases.

We conclude that the IR limit of the three-quark potential can be described either in terms of configuration II or configuration III, except in the very narrow angular interval near $\frac{2}{3}\pi$.

\subsection{The IR limit and geometrical constraints}

To gain some initial intuition about the IR limit, it is useful to pick a specific path in parameter space. In our case, this can be achieved by imposing geometrical constraints. Below, we present two simple examples.
\subsubsection{Example I}

\begin{figure}[htbp]
\centering
\includegraphics[width=5.5cm]{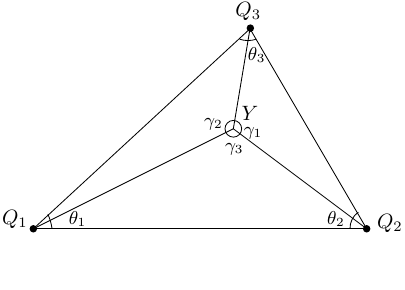}
\caption{{\small A triangle with internal angles $\theta_1$, $\theta_2$, and $\theta_3$. The heavy quarks are at the vertices. $Y$ is the projection of the baryon vertex on the boundary.}}
\label{trian}
\end{figure}

Assume that the triangle shown in Figure \ref{trian} has no internal angle greater than $\frac{2}{3}\pi$. Without loss of generality, we 

\noindent may further assume that $\theta_1<\theta_2<\theta_3$. We impose two constraints by fixing the internal angles and aim to show how two parameters can be express in terms of the remaining one. 

A good starting point is the relations

\begin{equation}\label{sinerule}
\frac{\sin^2\theta_1}{L(2,3\vert 1)}
	=
	\frac{\sin^2\theta_2}{L(1,3\vert 2)}
	=
	\frac{\sin^2\theta_3}{L(1,2\vert 3)}
	\,
\end{equation}
which follow from the sine and cosine rules. Here the notation $L(i,j\vert k) = |YQ_i|^2 + |YQ_j|^2 - 2\cos\gamma_k |YQ_i||YQ_j|$ is used as shorthand for the denominators. We now take the IR limit keeping fixed all the $\theta$'s and letting all $\lambda_i\rightarrow 1$, such that all the strings in question become infinitely long. In this limit, the leading contributions to $|YQ_i|$ and $\gamma_i$ are given by $-\frac{1}{\sqrt{\s}} \ln(1 - \lambda_i)$ and $\frac{2}{3}\pi$, respectively, as discussed in Sec.V. As a result, \eqref{sinerule} reduces to

\begin{equation}\label{sinerule2}
\frac{\sin^2\theta_1}{P(2,3)}
	=
	\frac{\sin^2\theta_2}{P(1,3)}
=
	 \frac{\sin^2\theta_3}{P(1,2)}\,,
\end{equation}
where $P(i,j)=\ln^2(1-\lambda_i)+\ln^2(1-\lambda_j)+\ln(1-\lambda_i)\ln(1-\lambda_j)$.

The above equations allow us to express two of the $\lambda$'s in terms of the remaining one. We look for a solution of the form

\begin{equation}\label{sol-xy}
1-\lambda_2=(1-\lambda_1)^x\,\qquad
\text{and}\qquad
 1-\lambda_3=(1-\lambda_1)^y
\,
\end{equation}
with the positive exponents $x$ and $y$. The exponents are subject to the conditions:

\begin{equation}\label{xy}
\frac{\sin^2\theta_1}{x^2+y^2+xy}
	=
\frac{\sin^2\theta_2}{y^2+y+1}
=
\frac{\sin^2\theta_3}{x^2+x+1}
\,.
\end{equation}

Before discussing the general case, let us consider two special examples. The first is $\theta_1=\theta_2=\theta_3$ that is the equilateral triangle. The solution to Eq.\eqref{xy} is simply $x=y=1$. The second is $\theta_1=\theta_2$ that is the isosceles triangle. In this case, the solution is 

\begin{equation}\label{1y}
x=1\,,\quad
 y=\oh\Bigl(\sqrt{3}\cot\frac{\theta_3}{2}-1\Bigr)
 \,.
 \end{equation}
It can also be derived using the geometrical constraint \eqref{geoc2} as well. Note that $y=0$ at $\theta_3=\frac{2}{3}\pi$, indicating that the IR limit with three infinitely long strings breaks down. 

 A general solution is given by 

\begin{equation}\label{xy2}
	x=\sqrt{\frac{\sin^2(\theta_1+\theta_2)}{\sin^2\theta_2}\bigl(y^2+y+1\bigr)-\frac{3}{4}}-\oh
	\,,\qquad
	y=\frac{1}{2a}\Bigl(\sqrt{b^2-4ac}-b\Bigr)
	\,,
\end{equation}
where $a=\sin^2(2\theta_1+\theta_2)+\sin(\theta_2-\theta_1)\sin(\theta_1+\theta_2)+\sin\theta_2\sin(2\theta_1+\theta_2)$, $b=\sin^2(2\theta_1+\theta_2)+\sin(\theta_1-\theta_2)\sin(\theta_1+\theta_2)+\sin\theta_1\sin(\theta_1+2\theta_2)$, and $c=3\sin^2(2\theta_1+\theta_2)-a-b$. Both exponents remain positive and finite within the domain $\theta_1 < \theta_2 < \theta_3 < \frac{2}{3}\pi$. A notable feature of the solution is that $y=0$ at $\theta_3=\frac{2}{3}\pi$, indicating that the IR limit, as defined above, does not exist. The reason is that the string ending on $Q_3$ cannot be infinitely long.
\subsubsection{Example II}

Now assume that $\theta_3$ is greater than $\frac{2}{3}\pi$. We impose the constraint $\eta=\vert Q_1Q_3\vert/\vert Q_2Q_3\vert$ meaning that the ratio of the side lengths is held fixed. It is convenient for the moment to write it as 

\begin{equation}\label{eta120}
	\eta^2=\frac{L(1,3\vert 2)}{L(2,3\vert 1)}
	\,.
\end{equation}
We take the IR limit  $\lambda_1,\,\lambda_2\rightarrow 1$ keeping $\theta_3$ fixed. In this limit, only the strings ending on $Q_1$ and $Q_2$ become infinitely long. Their leading contributions are proportional to $\ln(1 - \lambda_1)$ and $\ln(1 - \lambda_2)$, respectively. Thus, Eq.\eqref{eta120} simplifies to

\begin{equation}\label{eta1202}
	\eta=\frac{\ln(1-\lambda_1)}{\ln(1-\lambda_2)}
	\,.
\end{equation}
This equation has a simple solution

\begin{equation}\label{eta1203}
	1-\lambda_2=(1-\lambda_1)^{\frac{1}{\eta}}
	\,
\end{equation}
which expresses $\lambda_2$ in terms of $\lambda_1$. The relation used in \eqref{Ecola-large2} is a special case of \eqref{eta1203}, corresponding to $\theta_3=\pi$. 


\end{document}